\documentclass[%
    sigconf,
    balance=false,
    authorversion=true, 
    authordraft=false, 
    natbib=false, 
]{acmart} 

\usepackage[utf8]{inputenc}

\usepackage{amsmath, amssymb, amsfonts}                
\usepackage{algorithmic}
\usepackage{array}
\usepackage{graphics}
\usepackage{textcomp}
\usepackage{xcolor-material} 
\usepackage{tabularx}
\usepackage{makecell}
\usepackage{multirow}
\usepackage{booktabs}
\usepackage{threeparttable}
\usepackage{svg}                    
\usepackage{url}

\usepackage{tikz}                   
\usetikzlibrary{shapes,arrows,fit,matrix,positioning}
\usepackage{hyperref}
\usepackage{etoolbox}
\def\BibTeX{{\rm B\kern-.05em{\sc i\kern-.025em b}\kern-.08em
    T\kern-.1667em\lower.7ex\hbox{E}\kern-.125emX}}
\usepackage{xcolor,colortbl}
\definecolor{Gray}{gray}{0.85}
\newcolumntype{a}{>{\columncolor{Gray}}c}
\usepackage{ragged2e}
\usepackage{placeins}
\usepackage{xtab}
\usepackage{longtable}
\usepackage[
    backend=biber,
    sortcites=true, 
    maxbibnames=99, 
    datamodel=acmdatamodel,
    style=acmnumeric,
    ]{biblatex}

\definecolor{lightgray}{gray}{0.9}

\newbool{isextendedversion}
\setbool{isextendedversion}{true}

\ifbool{isextendedversion}{\newcommand{\extendedversion}[2]{#1}}{\newcommand{\extendedversion}[2]{#2}}

\ifbool{isextendedversion}{\newcommand{\repref}[1]{#1}}{\newcommand{\repref}[1]{\textsuperscript{\ref{replication}}}}

\newcommand{\ie}{i.\,e.}
\newcommand{\eg}{e.\,g.}

\renewcommand{\paragraph}[1]{\textbf{#1.}}

\newcommand{\mfa}{MFA}
\newcommand{\mfas}{MFAs}
\newcommand{\studyone}{deployment evaluation}
\newcommand{\studytwo}{recovery test}

\newcommand\COMMENTX[2]{{\bf({#1}: {#2})}}

\newbool{displaycomments}
\setbool{displaycomments}{false}

\ifbool{displaycomments}{
\newcommand\sabrina[1]{\COMMENTX{sabrina}{{{\color{MaterialRed}#1}}}}
\newcommand\nicolas[1]{\COMMENTX{nicolas}{{{\color{MaterialIndigo}#1}}}}
\newcommand\sandra[1]{\COMMENTX{sandra}{{{\color{MaterialBlue}#1}}}}
\newcommand\alex[1]{\COMMENTX{alex}{{{\color{MaterialPurple}#1}}}}
\newcommand\philip[1]{\COMMENTX{philip}{{{\color{MaterialPurple}#1}}}}
\newcommand\sascha[1]{\COMMENTX{sascha}{{{\color{MaterialOrange}#1}}}}
\newcommand\yas[1]{\COMMENTX{yas}{{{\color{MaterialTeal}#1}}}}
\newcommand\lucy[1]{\COMMENTX{lucy}{{{\color{MaterialBrown}#1}}}}
}
{
\newcommand{\sandra}[1]{}
\newcommand{\sabrina}[1]{}
\newcommand{\yas}[1]{}
\newcommand{\sascha}[1]{}
\newcommand{\nicolas}[1]{}
\newcommand{\philip}[1]{}
\newcommand{\alex}[1]{}
\newcommand{\lucy}[1]{}
}

\newcommand\fixme[1]{{{\color{red}#1}}}

\newcommand\definevar[2]{%
  \expandafter\newcommand\csname var#1var\endcsname{#2}%
}
\newcommand{\var}[1]{\ifcsname var#1var\endcsname%
        \csname var#1var\endcsname%
    \else%
        \fixme{TODO}%
        \GenericWarning{LaTeX Warning: \noexpand Number/Variable {#1} is undefined. on input line \the\inputlineno}%
    \fi%
}

\makeatletter
\newcommand\boldparagraph{%
    \@startsection{boldparagraph}{4}{0\parindent}{4pt plus 2pt minus 1pt}{0pt}
    {\noindent\normalfont\normalsize\bfseries\maybe@addperiod}*
}
\newcommand{\maybe@addperiod}[1]{%
    \let\@period\@empty
    \def\@IEEEsectpunct{}
    #1\@addpunct{.}\enspace%
}
\makeatother

\definevar{PagesNoMFAExcluded}{15}
\definevar{PagesPercentMFA}{71.79}
\definevar{PagesPercentNoMFA}{28.21}

\definevar{AllCountMethSMS}{559}
\definevar{AllPercentMethSMS}{42.90}
\definevar{AllCountMethCall}{156}
\definevar{AllPercentMethCall}{11.97}
\definevar{AllCountMethMail}{205}
\definevar{AllPercentMethMail}{15.73}
\definevar{AllCountMethHWToken}{179}
\definevar{AllPercentMethHWToken}{13.74}
\definevar{AllCountMethSWToken}{1,036}
\definevar{AllPercentMethSWToken}{79.51}
\definevar{AllCountMethOther}{139}
\definevar{AllPercentMethOther}{10.67}
\definevar{AllCountRecCodes}{451}
\definevar{AllPercentRecCodes}{34.61}
\definevar{AllCountRecContactSupport}{491}
\definevar{AllPercentRecContactSupport}{37.68}
\definevar{AllCountRecContactAdmin}{159}
\definevar{AllPercentRecContactAdmin}{12.20}
\definevar{AllCountRecAdd2FA}{150}
\definevar{AllPercentRecAdd2FA}{11.51}
\definevar{AllCountRecTrustedDevice}{37}
\definevar{AllPercentRecTrustedDevice}{2.84}
\definevar{AllCountRecPhone}{78}
\definevar{AllPercentRecPhone}{5.99}
\definevar{AllCountRecMail}{49}
\definevar{AllPercentRecMail}{3.76}
\definevar{AllCountRecRecoveryService}{45}
\definevar{AllPercentRecRecoveryService}{3.45}
\definevar{AllCountRecAccessNotLost}{25}
\definevar{AllPercentRecAccessNotLost}{1.92}
\definevar{AllCountRecPhoto}{30}
\definevar{AllPercentRecPhoto}{2.30}
\definevar{AllCountRecSecret}{44}
\definevar{AllPercentRecSecret}{3.38}
\definevar{AllCountRecPWReset}{13}
\definevar{AllPercentRecPWReset}{1.00}
\definevar{AllCountRecSecQuestion}{16}
\definevar{AllPercentRecSecQuestion}{1.23}
\definevar{AllCountRecInaccesible}{9}
\definevar{AllPercentRecInaccesible}{0.69}
\definevar{AllCountRecOther}{77}
\definevar{AllPercentRecOther}{5.91}
\definevar{AllCountRecNone}{1}
\definevar{AllPercentRecNone}{0.08}
\definevar{AllCountRecNoMention}{321}
\definevar{AllPercentRecNoMention}{24.64}

\definevar{CountRecCodesIfMethSMS}{155}
\definevar{PercentRecCodesIfMethSMS}{\cellcolor{MaterialBlue!27.73}27.73}
\definevar{CountRecContactSupportIfMethSMS}{199}
\definevar{PercentRecContactSupportIfMethSMS}{\cellcolor{MaterialBlue!35.6}35.6}
\definevar{CountRecContactAdminIfMethSMS}{79}
\definevar{PercentRecContactAdminIfMethSMS}{\cellcolor{MaterialBlue!14.13}14.13}
\definevar{CountRecAdd2FAIfMethSMS}{119}
\definevar{PercentRecAdd2FAIfMethSMS}{\cellcolor{MaterialBlue!21.29}21.29}
\definevar{CountRecTrustedDeviceIfMethSMS}{26}
\definevar{PercentRecTrustedDeviceIfMethSMS}{\cellcolor{MaterialBlue!4.65}4.65}
\definevar{CountRecPhoneIfMethSMS}{35}
\definevar{PercentRecPhoneIfMethSMS}{\cellcolor{MaterialBlue!6.26}6.26}
\definevar{CountRecMailIfMethSMS}{23}
\definevar{PercentRecMailIfMethSMS}{\cellcolor{MaterialBlue!4.11}4.11}
\definevar{CountRecRecoveryServiceIfMethSMS}{34}
\definevar{PercentRecRecoveryServiceIfMethSMS}{\cellcolor{MaterialBlue!6.08}6.08}
\definevar{CountRecAccessNotLostIfMethSMS}{14}
\definevar{PercentRecAccessNotLostIfMethSMS}{\cellcolor{MaterialBlue!2.5}2.5}
\definevar{CountRecPhotoIfMethSMS}{10}
\definevar{PercentRecPhotoIfMethSMS}{\cellcolor{MaterialBlue!1.79}1.79}
\definevar{CountRecNoMentionIfMethSMS}{153}
\definevar{PercentRecNoMentionIfMethSMS}{\cellcolor{MaterialBlue!27.37}27.37}
\definevar{CountRecInaccesibleIfMethSMS}{2}
\definevar{PercentRecInaccesibleIfMethSMS}{\cellcolor{MaterialBlue!0.36}0.36}
\definevar{CountRecSecretIfMethSMS}{6}
\definevar{PercentRecSecretIfMethSMS}{\cellcolor{MaterialBlue!1.07}1.07}
\definevar{CountRecPWResetIfMethSMS}{2}
\definevar{PercentRecPWResetIfMethSMS}{\cellcolor{MaterialBlue!0.36}0.36}
\definevar{CountRecSecQuestionIfMethSMS}{9}
\definevar{PercentRecSecQuestionIfMethSMS}{\cellcolor{MaterialBlue!1.61}1.61}
\definevar{CountRecOtherIfMethSMS}{38}
\definevar{PercentRecOtherIfMethSMS}{\cellcolor{MaterialBlue!6.8}6.8}
\definevar{CountRecNoneIfMethSMS}{0}
\definevar{PercentRecNoneIfMethSMS}{\cellcolor{MaterialBlue!0.0}0.0}
\definevar{CountRecCodesIfMethCall}{32}
\definevar{PercentRecCodesIfMethCall}{\cellcolor{MaterialBlue!20.51}20.51}
\definevar{CountRecContactSupportIfMethCall}{32}
\definevar{PercentRecContactSupportIfMethCall}{\cellcolor{MaterialBlue!20.51}20.51}
\definevar{CountRecContactAdminIfMethCall}{41}
\definevar{PercentRecContactAdminIfMethCall}{\cellcolor{MaterialBlue!26.28}26.28}
\definevar{CountRecAdd2FAIfMethCall}{59}
\definevar{PercentRecAdd2FAIfMethCall}{\cellcolor{MaterialBlue!37.82}37.82}
\definevar{CountRecTrustedDeviceIfMethCall}{16}
\definevar{PercentRecTrustedDeviceIfMethCall}{\cellcolor{MaterialBlue!10.26}10.26}
\definevar{CountRecPhoneIfMethCall}{14}
\definevar{PercentRecPhoneIfMethCall}{\cellcolor{MaterialBlue!8.97}8.97}
\definevar{CountRecMailIfMethCall}{4}
\definevar{PercentRecMailIfMethCall}{\cellcolor{MaterialBlue!2.56}2.56}
\definevar{CountRecRecoveryServiceIfMethCall}{20}
\definevar{PercentRecRecoveryServiceIfMethCall}{\cellcolor{MaterialBlue!12.82}12.82}
\definevar{CountRecAccessNotLostIfMethCall}{1}
\definevar{PercentRecAccessNotLostIfMethCall}{\cellcolor{MaterialBlue!0.64}0.64}
\definevar{CountRecPhotoIfMethCall}{3}
\definevar{PercentRecPhotoIfMethCall}{\cellcolor{MaterialBlue!1.92}1.92}
\definevar{CountRecNoMentionIfMethCall}{53}
\definevar{PercentRecNoMentionIfMethCall}{\cellcolor{MaterialBlue!33.97}33.97}
\definevar{CountRecInaccesibleIfMethCall}{1}
\definevar{PercentRecInaccesibleIfMethCall}{\cellcolor{MaterialBlue!0.64}0.64}
\definevar{CountRecSecretIfMethCall}{0}
\definevar{PercentRecSecretIfMethCall}{\cellcolor{MaterialBlue!0.0}0.0}
\definevar{CountRecPWResetIfMethCall}{0}
\definevar{PercentRecPWResetIfMethCall}{\cellcolor{MaterialBlue!0.0}0.0}
\definevar{CountRecSecQuestionIfMethCall}{3}
\definevar{PercentRecSecQuestionIfMethCall}{\cellcolor{MaterialBlue!1.92}1.92}
\definevar{CountRecOtherIfMethCall}{15}
\definevar{PercentRecOtherIfMethCall}{\cellcolor{MaterialBlue!9.62}9.62}
\definevar{CountRecNoneIfMethCall}{0}
\definevar{PercentRecNoneIfMethCall}{\cellcolor{MaterialBlue!0.0}0.0}
\definevar{CountRecCodesIfMethMail}{34}
\definevar{PercentRecCodesIfMethMail}{\cellcolor{MaterialBlue!16.59}16.59}
\definevar{CountRecContactSupportIfMethMail}{72}
\definevar{PercentRecContactSupportIfMethMail}{\cellcolor{MaterialBlue!35.12}35.12}
\definevar{CountRecContactAdminIfMethMail}{13}
\definevar{PercentRecContactAdminIfMethMail}{\cellcolor{MaterialBlue!6.34}6.34}
\definevar{CountRecAdd2FAIfMethMail}{31}
\definevar{PercentRecAdd2FAIfMethMail}{\cellcolor{MaterialBlue!15.12}15.12}
\definevar{CountRecTrustedDeviceIfMethMail}{2}
\definevar{PercentRecTrustedDeviceIfMethMail}{\cellcolor{MaterialBlue!0.98}0.98}
\definevar{CountRecPhoneIfMethMail}{7}
\definevar{PercentRecPhoneIfMethMail}{\cellcolor{MaterialBlue!3.41}3.41}
\definevar{CountRecMailIfMethMail}{7}
\definevar{PercentRecMailIfMethMail}{\cellcolor{MaterialBlue!3.41}3.41}
\definevar{CountRecRecoveryServiceIfMethMail}{3}
\definevar{PercentRecRecoveryServiceIfMethMail}{\cellcolor{MaterialBlue!1.46}1.46}
\definevar{CountRecAccessNotLostIfMethMail}{3}
\definevar{PercentRecAccessNotLostIfMethMail}{\cellcolor{MaterialBlue!1.46}1.46}
\definevar{CountRecPhotoIfMethMail}{2}
\definevar{PercentRecPhotoIfMethMail}{\cellcolor{MaterialBlue!0.98}0.98}
\definevar{CountRecNoMentionIfMethMail}{79}
\definevar{PercentRecNoMentionIfMethMail}{\cellcolor{MaterialBlue!38.54}38.54}
\definevar{CountRecInaccesibleIfMethMail}{1}
\definevar{PercentRecInaccesibleIfMethMail}{\cellcolor{MaterialBlue!0.49}0.49}
\definevar{CountRecSecretIfMethMail}{2}
\definevar{PercentRecSecretIfMethMail}{\cellcolor{MaterialBlue!0.98}0.98}
\definevar{CountRecPWResetIfMethMail}{3}
\definevar{PercentRecPWResetIfMethMail}{\cellcolor{MaterialBlue!1.46}1.46}
\definevar{CountRecSecQuestionIfMethMail}{3}
\definevar{PercentRecSecQuestionIfMethMail}{\cellcolor{MaterialBlue!1.46}1.46}
\definevar{CountRecOtherIfMethMail}{7}
\definevar{PercentRecOtherIfMethMail}{\cellcolor{MaterialBlue!3.41}3.41}
\definevar{CountRecNoneIfMethMail}{0}
\definevar{PercentRecNoneIfMethMail}{\cellcolor{MaterialBlue!0.0}0.0}
\definevar{CountRecCodesIfMethHWToken}{71}
\definevar{PercentRecCodesIfMethHWToken}{\cellcolor{MaterialBlue!39.66}39.66}
\definevar{CountRecContactSupportIfMethHWToken}{57}
\definevar{PercentRecContactSupportIfMethHWToken}{\cellcolor{MaterialBlue!31.84}31.84}
\definevar{CountRecContactAdminIfMethHWToken}{60}
\definevar{PercentRecContactAdminIfMethHWToken}{\cellcolor{MaterialBlue!33.52}33.52}
\definevar{CountRecAdd2FAIfMethHWToken}{70}
\definevar{PercentRecAdd2FAIfMethHWToken}{\cellcolor{MaterialBlue!39.11}39.11}
\definevar{CountRecTrustedDeviceIfMethHWToken}{18}
\definevar{PercentRecTrustedDeviceIfMethHWToken}{\cellcolor{MaterialBlue!10.06}10.06}
\definevar{CountRecPhoneIfMethHWToken}{19}
\definevar{PercentRecPhoneIfMethHWToken}{\cellcolor{MaterialBlue!10.61}10.61}
\definevar{CountRecMailIfMethHWToken}{5}
\definevar{PercentRecMailIfMethHWToken}{\cellcolor{MaterialBlue!2.79}2.79}
\definevar{CountRecRecoveryServiceIfMethHWToken}{19}
\definevar{PercentRecRecoveryServiceIfMethHWToken}{\cellcolor{MaterialBlue!10.61}10.61}
\definevar{CountRecAccessNotLostIfMethHWToken}{9}
\definevar{PercentRecAccessNotLostIfMethHWToken}{\cellcolor{MaterialBlue!5.03}5.03}
\definevar{CountRecPhotoIfMethHWToken}{3}
\definevar{PercentRecPhotoIfMethHWToken}{\cellcolor{MaterialBlue!1.68}1.68}
\definevar{CountRecNoMentionIfMethHWToken}{24}
\definevar{PercentRecNoMentionIfMethHWToken}{\cellcolor{MaterialBlue!13.41}13.41}
\definevar{CountRecInaccesibleIfMethHWToken}{0}
\definevar{PercentRecInaccesibleIfMethHWToken}{\cellcolor{MaterialBlue!0.0}0.0}
\definevar{CountRecSecretIfMethHWToken}{6}
\definevar{PercentRecSecretIfMethHWToken}{\cellcolor{MaterialBlue!3.35}3.35}
\definevar{CountRecPWResetIfMethHWToken}{2}
\definevar{PercentRecPWResetIfMethHWToken}{\cellcolor{MaterialBlue!1.12}1.12}
\definevar{CountRecSecQuestionIfMethHWToken}{0}
\definevar{PercentRecSecQuestionIfMethHWToken}{\cellcolor{MaterialBlue!0.0}0.0}
\definevar{CountRecOtherIfMethHWToken}{27}
\definevar{PercentRecOtherIfMethHWToken}{\cellcolor{MaterialBlue!15.08}15.08}
\definevar{CountRecNoneIfMethHWToken}{0}
\definevar{PercentRecNoneIfMethHWToken}{\cellcolor{MaterialBlue!0.0}0.0}
\definevar{CountRecCodesIfMethSWToken}{435}
\definevar{PercentRecCodesIfMethSWToken}{\cellcolor{MaterialBlue!41.99}41.99}
\definevar{CountRecContactSupportIfMethSWToken}{413}
\definevar{PercentRecContactSupportIfMethSWToken}{\cellcolor{MaterialBlue!39.86}39.86}
\definevar{CountRecContactAdminIfMethSWToken}{152}
\definevar{PercentRecContactAdminIfMethSWToken}{\cellcolor{MaterialBlue!14.67}14.67}
\definevar{CountRecAdd2FAIfMethSWToken}{134}
\definevar{PercentRecAdd2FAIfMethSWToken}{\cellcolor{MaterialBlue!12.93}12.93}
\definevar{CountRecTrustedDeviceIfMethSWToken}{33}
\definevar{PercentRecTrustedDeviceIfMethSWToken}{\cellcolor{MaterialBlue!3.19}3.19}
\definevar{CountRecPhoneIfMethSWToken}{74}
\definevar{PercentRecPhoneIfMethSWToken}{\cellcolor{MaterialBlue!7.14}7.14}
\definevar{CountRecMailIfMethSWToken}{41}
\definevar{PercentRecMailIfMethSWToken}{\cellcolor{MaterialBlue!3.96}3.96}
\definevar{CountRecRecoveryServiceIfMethSWToken}{39}
\definevar{PercentRecRecoveryServiceIfMethSWToken}{\cellcolor{MaterialBlue!3.76}3.76}
\definevar{CountRecAccessNotLostIfMethSWToken}{23}
\definevar{PercentRecAccessNotLostIfMethSWToken}{\cellcolor{MaterialBlue!2.22}2.22}
\definevar{CountRecPhotoIfMethSWToken}{30}
\definevar{PercentRecPhotoIfMethSWToken}{\cellcolor{MaterialBlue!2.9}2.9}
\definevar{CountRecNoMentionIfMethSWToken}{180}
\definevar{PercentRecNoMentionIfMethSWToken}{\cellcolor{MaterialBlue!17.37}17.37}
\definevar{CountRecInaccesibleIfMethSWToken}{6}
\definevar{PercentRecInaccesibleIfMethSWToken}{\cellcolor{MaterialBlue!0.58}0.58}
\definevar{CountRecSecretIfMethSWToken}{44}
\definevar{PercentRecSecretIfMethSWToken}{\cellcolor{MaterialBlue!4.25}4.25}
\definevar{CountRecPWResetIfMethSWToken}{11}
\definevar{PercentRecPWResetIfMethSWToken}{\cellcolor{MaterialBlue!1.06}1.06}
\definevar{CountRecSecQuestionIfMethSWToken}{10}
\definevar{PercentRecSecQuestionIfMethSWToken}{\cellcolor{MaterialBlue!0.97}0.97}
\definevar{CountRecOtherIfMethSWToken}{62}
\definevar{PercentRecOtherIfMethSWToken}{\cellcolor{MaterialBlue!5.98}5.98}
\definevar{CountRecNoneIfMethSWToken}{1}
\definevar{PercentRecNoneIfMethSWToken}{\cellcolor{MaterialBlue!0.1}0.1}
\definevar{CountRecCodesIfMethOther}{17}
\definevar{PercentRecCodesIfMethOther}{\cellcolor{MaterialBlue!12.23}12.23}
\definevar{CountRecContactSupportIfMethOther}{35}
\definevar{PercentRecContactSupportIfMethOther}{\cellcolor{MaterialBlue!25.18}25.18}
\definevar{CountRecContactAdminIfMethOther}{12}
\definevar{PercentRecContactAdminIfMethOther}{\cellcolor{MaterialBlue!8.63}8.63}
\definevar{CountRecAdd2FAIfMethOther}{15}
\definevar{PercentRecAdd2FAIfMethOther}{\cellcolor{MaterialBlue!10.79}10.79}
\definevar{CountRecTrustedDeviceIfMethOther}{4}
\definevar{PercentRecTrustedDeviceIfMethOther}{\cellcolor{MaterialBlue!2.88}2.88}
\definevar{CountRecPhoneIfMethOther}{3}
\definevar{PercentRecPhoneIfMethOther}{\cellcolor{MaterialBlue!2.16}2.16}
\definevar{CountRecMailIfMethOther}{4}
\definevar{PercentRecMailIfMethOther}{\cellcolor{MaterialBlue!2.88}2.88}
\definevar{CountRecRecoveryServiceIfMethOther}{3}
\definevar{PercentRecRecoveryServiceIfMethOther}{\cellcolor{MaterialBlue!2.16}2.16}
\definevar{CountRecAccessNotLostIfMethOther}{3}
\definevar{PercentRecAccessNotLostIfMethOther}{\cellcolor{MaterialBlue!2.16}2.16}
\definevar{CountRecPhotoIfMethOther}{1}
\definevar{PercentRecPhotoIfMethOther}{\cellcolor{MaterialBlue!0.72}0.72}
\definevar{CountRecNoMentionIfMethOther}{65}
\definevar{PercentRecNoMentionIfMethOther}{\cellcolor{MaterialBlue!46.76}46.76}
\definevar{CountRecInaccesibleIfMethOther}{3}
\definevar{PercentRecInaccesibleIfMethOther}{\cellcolor{MaterialBlue!2.16}2.16}
\definevar{CountRecSecretIfMethOther}{1}
\definevar{PercentRecSecretIfMethOther}{\cellcolor{MaterialBlue!0.72}0.72}
\definevar{CountRecPWResetIfMethOther}{1}
\definevar{PercentRecPWResetIfMethOther}{\cellcolor{MaterialBlue!0.72}0.72}
\definevar{CountRecSecQuestionIfMethOther}{1}
\definevar{PercentRecSecQuestionIfMethOther}{\cellcolor{MaterialBlue!0.72}0.72}
\definevar{CountRecOtherIfMethOther}{16}
\definevar{PercentRecOtherIfMethOther}{\cellcolor{MaterialBlue!11.51}11.51}
\definevar{CountRecNoneIfMethOther}{0}
\definevar{PercentRecNoneIfMethOther}{\cellcolor{MaterialBlue!0.0}0.0}
\definevar{AllMethMean}{1.75}
\definevar{AllMethMedian}{1.0}
\definevar{AllRecMean}{1.28}
\definevar{AllRecMedian}{1.0}
\definevar{CatCountMFABackup+Sync}{30}
\definevar{CatCountMFABanking}{136}
\definevar{CatCountMFABetting}{7}
\definevar{CatCountMFACloudComputing}{35}
\definevar{CatCountMFACommunication}{50}
\definevar{CatCountMFACreativity}{8}
\definevar{CatCountMFACrowdfunding}{5}
\definevar{CatCountMFACryptocurrencies}{110}
\definevar{CatCountMFADevelopers}{101}
\definevar{CatCountMFADomains}{59}
\definevar{CatCountMFAEducation}{12}
\definevar{CatCountMFAEmail}{28}
\definevar{CatCountMFAEntertainment}{9}
\definevar{CatCountMFAFinance}{54}
\definevar{CatCountMFAFood}{5}
\definevar{CatCountMFAGaming}{53}
\definevar{CatCountMFAGovernment}{17}
\definevar{CatCountMFAHealth}{21}
\definevar{CatCountMFAHosting+VPS}{74}
\definevar{CatCountMFAHotels+Accomodations}{5}
\definevar{CatCountMFAIdentityManagement}{20}
\definevar{CatCountMFAInvesting}{51}
\definevar{CatCountMFAIoT}{17}
\definevar{CatCountMFALegal}{8}
\definevar{CatCountMFAMarketing+Analytics}{59}
\definevar{CatCountMFAOther}{82}
\definevar{CatCountMFAPayment}{34}
\definevar{CatCountMFAPostandShipping}{3}
\definevar{CatCountMFARemoteAccess}{16}
\definevar{CatCountMFARetail}{25}
\definevar{CatCountMFASecurity}{38}
\definevar{CatCountMFASocial}{28}
\definevar{CatCountMFATaskManagement}{20}
\definevar{CatCountMFATickets+Events}{1}
\definevar{CatCountMFATransport}{6}
\definevar{CatCountMFAUniversities}{41}
\definevar{CatCountMFAUtilities}{26}
\definevar{CatCountMFAVPNProviders}{9}

\definevar{numberCategories}{38}
\definevar{numberTechCategories}{568}
\definevar{percentTechCategories}{43.59}

\definevar{CatPercentBackup+SyncMFA}{2.30}
\definevar{CatPercentBankingMFA}{10.44}
\definevar{CatPercentBettingMFA}{0.54}
\definevar{CatPercentCloudComputingMFA}{2.69}
\definevar{CatPercentCommunicationMFA}{3.84}
\definevar{CatPercentCreativityMFA}{0.61}
\definevar{CatPercentCrowdfundingMFA}{0.38}
\definevar{CatPercentCryptocurrenciesMFA}{8.44}
\definevar{CatPercentDevelopersMFA}{7.75}
\definevar{CatPercentDomainsMFA}{4.53}
\definevar{CatPercentEducationMFA}{0.92}
\definevar{CatPercentEmailMFA}{2.15}
\definevar{CatPercentEntertainmentMFA}{0.69}
\definevar{CatPercentFinanceMFA}{4.14}
\definevar{CatPercentFoodMFA}{0.38}
\definevar{CatPercentGamingMFA}{4.07}
\definevar{CatPercentGovernmentMFA}{1.30}
\definevar{CatPercentHealthMFA}{1.61}
\definevar{CatPercentHosting+VPSMFA}{5.68}
\definevar{CatPercentHotels+AccomodationsMFA}{0.38}
\definevar{CatPercentIdentityManagementMFA}{1.53}
\definevar{CatPercentInvestingMFA}{3.91}
\definevar{CatPercentIoTMFA}{1.30}
\definevar{CatPercentLegalMFA}{0.61}
\definevar{CatPercentMarketing+AnalyticsMFA}{4.53}
\definevar{CatPercentOtherMFA}{6.29}
\definevar{CatPercentPaymentMFA}{2.61}
\definevar{CatPercentPostandShippingMFA}{0.23}
\definevar{CatPercentRemoteAccessMFA}{1.23}
\definevar{CatPercentRetailMFA}{1.92}
\definevar{CatPercentSecurityMFA}{2.92}
\definevar{CatPercentSocialMFA}{2.15}
\definevar{CatPercentTaskManagementMFA}{1.53}
\definevar{CatPercentTickets+EventsMFA}{0.08}
\definevar{CatPercentTransportMFA}{0.46}
\definevar{CatPercentUniversitiesMFA}{3.15}
\definevar{CatPercentUtilitiesMFA}{2.00}
\definevar{CatPercentVPNProvidersMFA}{0.69}

\definevar{CatRelBackup+SyncMethSMS}{56.7}
\definevar{CatRelBackup+SyncMethCall}{16.7}
\definevar{CatRelBackup+SyncMethMail}{23.3}
\definevar{CatRelBackup+SyncMethHWToken}{13.3}
\definevar{CatRelBackup+SyncMethSWToken}{90.0}
\definevar{CatRelBackup+SyncMethOther}{6.7}
\definevar{CatRelBankingMethSMS}{70.6}
\definevar{CatRelBankingMethCall}{31.6}
\definevar{CatRelBankingMethMail}{21.3}
\definevar{CatRelBankingMethHWToken}{22.1}
\definevar{CatRelBankingMethSWToken}{52.9}
\definevar{CatRelBankingMethOther}{27.9}
\definevar{CatRelCloudComputingMethSMS}{31.4}
\definevar{CatRelCloudComputingMethCall}{8.6}
\definevar{CatRelCloudComputingMethMail}{8.6}
\definevar{CatRelCloudComputingMethHWToken}{25.7}
\definevar{CatRelCloudComputingMethSWToken}{97.1}
\definevar{CatRelCloudComputingMethOther}{5.7}
\definevar{CatRelCommunicationMethSMS}{44.0}
\definevar{CatRelCommunicationMethCall}{8.0}
\definevar{CatRelCommunicationMethMail}{12.0}
\definevar{CatRelCommunicationMethHWToken}{16.0}
\definevar{CatRelCommunicationMethSWToken}{82.0}
\definevar{CatRelCommunicationMethOther}{6.0}
\definevar{CatRelCryptocurrenciesMethSMS}{29.1}
\definevar{CatRelCryptocurrenciesMethCall}{0.9}
\definevar{CatRelCryptocurrenciesMethMail}{7.3}
\definevar{CatRelCryptocurrenciesMethHWToken}{8.2}
\definevar{CatRelCryptocurrenciesMethSWToken}{96.4}
\definevar{CatRelCryptocurrenciesMethOther}{4.5}
\definevar{CatRelDevelopersMethSMS}{27.7}
\definevar{CatRelDevelopersMethCall}{5.9}
\definevar{CatRelDevelopersMethMail}{5.0}
\definevar{CatRelDevelopersMethHWToken}{10.9}
\definevar{CatRelDevelopersMethSWToken}{94.1}
\definevar{CatRelDevelopersMethOther}{3.0}
\definevar{CatRelDomainsMethSMS}{30.5}
\definevar{CatRelDomainsMethCall}{3.4}
\definevar{CatRelDomainsMethMail}{8.5}
\definevar{CatRelDomainsMethHWToken}{13.6}
\definevar{CatRelDomainsMethSWToken}{91.5}
\definevar{CatRelDomainsMethOther}{3.4}
\definevar{CatRelEducationMethSMS}{25.0}
\definevar{CatRelEducationMethCall}{0.0}
\definevar{CatRelEducationMethMail}{25.0}
\definevar{CatRelEducationMethHWToken}{0.0}
\definevar{CatRelEducationMethSWToken}{50.0}
\definevar{CatRelEducationMethOther}{33.3}
\definevar{CatRelEmailMethSMS}{35.7}
\definevar{CatRelEmailMethCall}{14.3}
\definevar{CatRelEmailMethMail}{10.7}
\definevar{CatRelEmailMethHWToken}{42.9}
\definevar{CatRelEmailMethSWToken}{100.0}
\definevar{CatRelEmailMethOther}{0.0}
\definevar{CatRelFinanceMethSMS}{53.7}
\definevar{CatRelFinanceMethCall}{25.9}
\definevar{CatRelFinanceMethMail}{24.1}
\definevar{CatRelFinanceMethHWToken}{1.9}
\definevar{CatRelFinanceMethSWToken}{59.3}
\definevar{CatRelFinanceMethOther}{20.4}
\definevar{CatRelGamingMethSMS}{18.9}
\definevar{CatRelGamingMethCall}{3.8}
\definevar{CatRelGamingMethMail}{37.7}
\definevar{CatRelGamingMethHWToken}{3.8}
\definevar{CatRelGamingMethSWToken}{86.8}
\definevar{CatRelGamingMethOther}{1.9}
\definevar{CatRelGovernmentMethSMS}{70.6}
\definevar{CatRelGovernmentMethCall}{35.3}
\definevar{CatRelGovernmentMethMail}{23.5}
\definevar{CatRelGovernmentMethHWToken}{23.5}
\definevar{CatRelGovernmentMethSWToken}{52.9}
\definevar{CatRelGovernmentMethOther}{52.9}
\definevar{CatRelHealthMethSMS}{52.4}
\definevar{CatRelHealthMethCall}{4.8}
\definevar{CatRelHealthMethMail}{33.3}
\definevar{CatRelHealthMethHWToken}{4.8}
\definevar{CatRelHealthMethSWToken}{47.6}
\definevar{CatRelHealthMethOther}{4.8}
\definevar{CatRelHosting+VPSMethSMS}{17.6}
\definevar{CatRelHosting+VPSMethCall}{0.0}
\definevar{CatRelHosting+VPSMethMail}{8.1}
\definevar{CatRelHosting+VPSMethHWToken}{5.4}
\definevar{CatRelHosting+VPSMethSWToken}{95.9}
\definevar{CatRelHosting+VPSMethOther}{2.7}
\definevar{CatRelIdentityManagementMethSMS}{45.0}
\definevar{CatRelIdentityManagementMethCall}{20.0}
\definevar{CatRelIdentityManagementMethMail}{35.0}
\definevar{CatRelIdentityManagementMethHWToken}{60.0}
\definevar{CatRelIdentityManagementMethSWToken}{90.0}
\definevar{CatRelIdentityManagementMethOther}{35.0}
\definevar{CatRelInvestingMethSMS}{54.9}
\definevar{CatRelInvestingMethCall}{23.5}
\definevar{CatRelInvestingMethMail}{11.8}
\definevar{CatRelInvestingMethHWToken}{11.8}
\definevar{CatRelInvestingMethSWToken}{62.7}
\definevar{CatRelInvestingMethOther}{17.6}
\definevar{CatRelIoTMethSMS}{64.7}
\definevar{CatRelIoTMethCall}{0.0}
\definevar{CatRelIoTMethMail}{29.4}
\definevar{CatRelIoTMethHWToken}{0.0}
\definevar{CatRelIoTMethSWToken}{70.6}
\definevar{CatRelIoTMethOther}{5.9}
\definevar{CatRelMarketing+AnalyticsMethSMS}{40.7}
\definevar{CatRelMarketing+AnalyticsMethCall}{0.0}
\definevar{CatRelMarketing+AnalyticsMethMail}{11.9}
\definevar{CatRelMarketing+AnalyticsMethHWToken}{1.7}
\definevar{CatRelMarketing+AnalyticsMethSWToken}{88.1}
\definevar{CatRelMarketing+AnalyticsMethOther}{3.4}
\definevar{CatRelOtherMethSMS}{30.5}
\definevar{CatRelOtherMethCall}{3.7}
\definevar{CatRelOtherMethMail}{15.9}
\definevar{CatRelOtherMethHWToken}{8.5}
\definevar{CatRelOtherMethSWToken}{81.7}
\definevar{CatRelOtherMethOther}{7.3}
\definevar{CatRelPaymentMethSMS}{64.7}
\definevar{CatRelPaymentMethCall}{8.8}
\definevar{CatRelPaymentMethMail}{20.6}
\definevar{CatRelPaymentMethHWToken}{8.8}
\definevar{CatRelPaymentMethSWToken}{73.5}
\definevar{CatRelPaymentMethOther}{5.9}
\definevar{CatRelRemoteAccessMethSMS}{25.0}
\definevar{CatRelRemoteAccessMethCall}{6.2}
\definevar{CatRelRemoteAccessMethMail}{31.2}
\definevar{CatRelRemoteAccessMethHWToken}{18.8}
\definevar{CatRelRemoteAccessMethSWToken}{87.5}
\definevar{CatRelRemoteAccessMethOther}{12.5}
\definevar{CatRelRetailMethSMS}{56.0}
\definevar{CatRelRetailMethCall}{8.0}
\definevar{CatRelRetailMethMail}{20.0}
\definevar{CatRelRetailMethHWToken}{4.0}
\definevar{CatRelRetailMethSWToken}{52.0}
\definevar{CatRelRetailMethOther}{32.0}
\definevar{CatRelSecurityMethSMS}{13.2}
\definevar{CatRelSecurityMethCall}{2.6}
\definevar{CatRelSecurityMethMail}{5.3}
\definevar{CatRelSecurityMethHWToken}{13.2}
\definevar{CatRelSecurityMethSWToken}{94.7}
\definevar{CatRelSecurityMethOther}{2.6}
\definevar{CatRelSocialMethSMS}{64.3}
\definevar{CatRelSocialMethCall}{0.0}
\definevar{CatRelSocialMethMail}{10.7}
\definevar{CatRelSocialMethHWToken}{10.7}
\definevar{CatRelSocialMethSWToken}{78.6}
\definevar{CatRelSocialMethOther}{0.0}
\definevar{CatRelTaskManagementMethSMS}{35.0}
\definevar{CatRelTaskManagementMethCall}{0.0}
\definevar{CatRelTaskManagementMethMail}{5.0}
\definevar{CatRelTaskManagementMethHWToken}{0.0}
\definevar{CatRelTaskManagementMethSWToken}{95.0}
\definevar{CatRelTaskManagementMethOther}{5.0}
\definevar{CatRelUniversitiesMethSMS}{63.4}
\definevar{CatRelUniversitiesMethCall}{61.0}
\definevar{CatRelUniversitiesMethMail}{2.4}
\definevar{CatRelUniversitiesMethHWToken}{68.3}
\definevar{CatRelUniversitiesMethSWToken}{95.1}
\definevar{CatRelUniversitiesMethOther}{12.2}
\definevar{CatRelUtilitiesMethSMS}{65.4}
\definevar{CatRelUtilitiesMethCall}{26.9}
\definevar{CatRelUtilitiesMethMail}{23.1}
\definevar{CatRelUtilitiesMethHWToken}{7.7}
\definevar{CatRelUtilitiesMethSWToken}{53.8}
\definevar{CatRelUtilitiesMethOther}{30.8}

\definevar{CatRelBackup+SyncBackupCodes}{53.3}
\definevar{CatRelBackup+SyncContactService/Website}{20.0}
\definevar{CatRelBackup+SyncContactLocalAdmin}{23.3}
\definevar{CatRelBackup+SyncAdditionalMFAMethod}{10.0}
\definevar{CatRelBackup+SyncBackupSMS/PhoneCall}{16.7}
\definevar{CatRelBackup+SyncTrustedDevice}{0.0}
\definevar{CatRelBackup+SyncBackupEmail}{10.0}
\definevar{CatRelBackup+SyncAccountRecoveryService}{6.7}
\definevar{CatRelBackup+SyncAccessNotLost}{0.0}
\definevar{CatRelBackup+SyncContactwithPhotoProof}{0.0}
\definevar{CatRelBackup+SyncNothingMentioned}{20.0}
\definevar{CatRelBackup+SyncHelpNotAccesible}{0.0}
\definevar{CatRelBackup+SyncInitialSecret}{3.3}
\definevar{CatRelBackup+SyncPasswordReset}{0.0}
\definevar{CatRelBackup+SyncSecurityQuestions}{3.3}
\definevar{CatRelBackup+SyncOtherRecovery}{10.0}
\definevar{CatRelBackup+SyncNoRecovery}{0.0}
\definevar{CatRelBankingBackupCodes}{2.2}
\definevar{CatRelBankingContactService/Website}{34.6}
\definevar{CatRelBankingContactLocalAdmin}{0.0}
\definevar{CatRelBankingAdditionalMFAMethod}{6.6}
\definevar{CatRelBankingBackupSMS/PhoneCall}{1.5}
\definevar{CatRelBankingTrustedDevice}{0.7}
\definevar{CatRelBankingBackupEmail}{1.5}
\definevar{CatRelBankingAccountRecoveryService}{0.0}
\definevar{CatRelBankingAccessNotLost}{5.1}
\definevar{CatRelBankingContactwithPhotoProof}{0.0}
\definevar{CatRelBankingNothingMentioned}{51.5}
\definevar{CatRelBankingHelpNotAccesible}{0.0}
\definevar{CatRelBankingInitialSecret}{0.0}
\definevar{CatRelBankingPasswordReset}{0.7}
\definevar{CatRelBankingSecurityQuestions}{0.0}
\definevar{CatRelBankingOtherRecovery}{11.0}
\definevar{CatRelBankingNoRecovery}{0.0}
\definevar{CatRelCloudComputingBackupCodes}{51.4}
\definevar{CatRelCloudComputingContactService/Website}{62.9}
\definevar{CatRelCloudComputingContactLocalAdmin}{22.9}
\definevar{CatRelCloudComputingAdditionalMFAMethod}{17.1}
\definevar{CatRelCloudComputingBackupSMS/PhoneCall}{2.9}
\definevar{CatRelCloudComputingTrustedDevice}{8.6}
\definevar{CatRelCloudComputingBackupEmail}{5.7}
\definevar{CatRelCloudComputingAccountRecoveryService}{2.9}
\definevar{CatRelCloudComputingAccessNotLost}{5.7}
\definevar{CatRelCloudComputingContactwithPhotoProof}{0.0}
\definevar{CatRelCloudComputingNothingMentioned}{8.6}
\definevar{CatRelCloudComputingHelpNotAccesible}{0.0}
\definevar{CatRelCloudComputingInitialSecret}{0.0}
\definevar{CatRelCloudComputingPasswordReset}{8.6}
\definevar{CatRelCloudComputingSecurityQuestions}{0.0}
\definevar{CatRelCloudComputingOtherRecovery}{2.9}
\definevar{CatRelCloudComputingNoRecovery}{0.0}
\definevar{CatRelCommunicationBackupCodes}{50.0}
\definevar{CatRelCommunicationContactService/Website}{32.0}
\definevar{CatRelCommunicationContactLocalAdmin}{32.0}
\definevar{CatRelCommunicationAdditionalMFAMethod}{12.0}
\definevar{CatRelCommunicationBackupSMS/PhoneCall}{6.0}
\definevar{CatRelCommunicationTrustedDevice}{14.0}
\definevar{CatRelCommunicationBackupEmail}{6.0}
\definevar{CatRelCommunicationAccountRecoveryService}{6.0}
\definevar{CatRelCommunicationAccessNotLost}{6.0}
\definevar{CatRelCommunicationContactwithPhotoProof}{0.0}
\definevar{CatRelCommunicationNothingMentioned}{10.0}
\definevar{CatRelCommunicationHelpNotAccesible}{0.0}
\definevar{CatRelCommunicationInitialSecret}{0.0}
\definevar{CatRelCommunicationPasswordReset}{2.0}
\definevar{CatRelCommunicationSecurityQuestions}{2.0}
\definevar{CatRelCommunicationOtherRecovery}{2.0}
\definevar{CatRelCommunicationNoRecovery}{0.0}
\definevar{CatRelCryptocurrenciesBackupCodes}{24.5}
\definevar{CatRelCryptocurrenciesContactService/Website}{44.5}
\definevar{CatRelCryptocurrenciesContactLocalAdmin}{0.0}
\definevar{CatRelCryptocurrenciesAdditionalMFAMethod}{4.5}
\definevar{CatRelCryptocurrenciesBackupSMS/PhoneCall}{0.0}
\definevar{CatRelCryptocurrenciesTrustedDevice}{2.7}
\definevar{CatRelCryptocurrenciesBackupEmail}{1.8}
\definevar{CatRelCryptocurrenciesAccountRecoveryService}{10.0}
\definevar{CatRelCryptocurrenciesAccessNotLost}{1.8}
\definevar{CatRelCryptocurrenciesContactwithPhotoProof}{20.0}
\definevar{CatRelCryptocurrenciesNothingMentioned}{12.7}
\definevar{CatRelCryptocurrenciesHelpNotAccesible}{4.5}
\definevar{CatRelCryptocurrenciesInitialSecret}{22.7}
\definevar{CatRelCryptocurrenciesPasswordReset}{0.0}
\definevar{CatRelCryptocurrenciesSecurityQuestions}{0.9}
\definevar{CatRelCryptocurrenciesOtherRecovery}{6.4}
\definevar{CatRelCryptocurrenciesNoRecovery}{0.0}
\definevar{CatRelDevelopersBackupCodes}{61.4}
\definevar{CatRelDevelopersContactService/Website}{25.7}
\definevar{CatRelDevelopersContactLocalAdmin}{10.9}
\definevar{CatRelDevelopersAdditionalMFAMethod}{8.9}
\definevar{CatRelDevelopersBackupSMS/PhoneCall}{3.0}
\definevar{CatRelDevelopersTrustedDevice}{3.0}
\definevar{CatRelDevelopersBackupEmail}{1.0}
\definevar{CatRelDevelopersAccountRecoveryService}{0.0}
\definevar{CatRelDevelopersAccessNotLost}{0.0}
\definevar{CatRelDevelopersContactwithPhotoProof}{0.0}
\definevar{CatRelDevelopersNothingMentioned}{27.7}
\definevar{CatRelDevelopersHelpNotAccesible}{0.0}
\definevar{CatRelDevelopersInitialSecret}{0.0}
\definevar{CatRelDevelopersPasswordReset}{1.0}
\definevar{CatRelDevelopersSecurityQuestions}{0.0}
\definevar{CatRelDevelopersOtherRecovery}{3.0}
\definevar{CatRelDevelopersNoRecovery}{0.0}
\definevar{CatRelDomainsBackupCodes}{44.1}
\definevar{CatRelDomainsContactService/Website}{54.2}
\definevar{CatRelDomainsContactLocalAdmin}{5.1}
\definevar{CatRelDomainsAdditionalMFAMethod}{6.8}
\definevar{CatRelDomainsBackupSMS/PhoneCall}{1.7}
\definevar{CatRelDomainsTrustedDevice}{8.5}
\definevar{CatRelDomainsBackupEmail}{0.0}
\definevar{CatRelDomainsAccountRecoveryService}{5.1}
\definevar{CatRelDomainsAccessNotLost}{1.7}
\definevar{CatRelDomainsContactwithPhotoProof}{0.0}
\definevar{CatRelDomainsNothingMentioned}{13.6}
\definevar{CatRelDomainsHelpNotAccesible}{0.0}
\definevar{CatRelDomainsInitialSecret}{10.2}
\definevar{CatRelDomainsPasswordReset}{0.0}
\definevar{CatRelDomainsSecurityQuestions}{0.0}
\definevar{CatRelDomainsOtherRecovery}{6.8}
\definevar{CatRelDomainsNoRecovery}{0.0}
\definevar{CatRelEducationBackupCodes}{8.3}
\definevar{CatRelEducationContactService/Website}{50.0}
\definevar{CatRelEducationContactLocalAdmin}{0.0}
\definevar{CatRelEducationAdditionalMFAMethod}{0.0}
\definevar{CatRelEducationBackupSMS/PhoneCall}{0.0}
\definevar{CatRelEducationTrustedDevice}{0.0}
\definevar{CatRelEducationBackupEmail}{8.3}
\definevar{CatRelEducationAccountRecoveryService}{0.0}
\definevar{CatRelEducationAccessNotLost}{0.0}
\definevar{CatRelEducationContactwithPhotoProof}{0.0}
\definevar{CatRelEducationNothingMentioned}{41.7}
\definevar{CatRelEducationHelpNotAccesible}{0.0}
\definevar{CatRelEducationInitialSecret}{0.0}
\definevar{CatRelEducationPasswordReset}{0.0}
\definevar{CatRelEducationSecurityQuestions}{0.0}
\definevar{CatRelEducationOtherRecovery}{0.0}
\definevar{CatRelEducationNoRecovery}{0.0}
\definevar{CatRelEmailBackupCodes}{60.7}
\definevar{CatRelEmailContactService/Website}{35.7}
\definevar{CatRelEmailContactLocalAdmin}{17.9}
\definevar{CatRelEmailAdditionalMFAMethod}{14.3}
\definevar{CatRelEmailBackupSMS/PhoneCall}{3.6}
\definevar{CatRelEmailTrustedDevice}{17.9}
\definevar{CatRelEmailBackupEmail}{3.6}
\definevar{CatRelEmailAccountRecoveryService}{7.1}
\definevar{CatRelEmailAccessNotLost}{0.0}
\definevar{CatRelEmailContactwithPhotoProof}{0.0}
\definevar{CatRelEmailNothingMentioned}{7.1}
\definevar{CatRelEmailHelpNotAccesible}{0.0}
\definevar{CatRelEmailInitialSecret}{3.6}
\definevar{CatRelEmailPasswordReset}{3.6}
\definevar{CatRelEmailSecurityQuestions}{0.0}
\definevar{CatRelEmailOtherRecovery}{10.7}
\definevar{CatRelEmailNoRecovery}{0.0}
\definevar{CatRelFinanceBackupCodes}{18.5}
\definevar{CatRelFinanceContactService/Website}{33.3}
\definevar{CatRelFinanceContactLocalAdmin}{11.1}
\definevar{CatRelFinanceAdditionalMFAMethod}{13.0}
\definevar{CatRelFinanceBackupSMS/PhoneCall}{0.0}
\definevar{CatRelFinanceTrustedDevice}{1.9}
\definevar{CatRelFinanceBackupEmail}{7.4}
\definevar{CatRelFinanceAccountRecoveryService}{5.6}
\definevar{CatRelFinanceAccessNotLost}{1.9}
\definevar{CatRelFinanceContactwithPhotoProof}{7.4}
\definevar{CatRelFinanceNothingMentioned}{31.5}
\definevar{CatRelFinanceHelpNotAccesible}{1.9}
\definevar{CatRelFinanceInitialSecret}{0.0}
\definevar{CatRelFinancePasswordReset}{0.0}
\definevar{CatRelFinanceSecurityQuestions}{1.9}
\definevar{CatRelFinanceOtherRecovery}{3.7}
\definevar{CatRelFinanceNoRecovery}{0.0}
\definevar{CatRelGamingBackupCodes}{37.7}
\definevar{CatRelGamingContactService/Website}{52.8}
\definevar{CatRelGamingContactLocalAdmin}{1.9}
\definevar{CatRelGamingAdditionalMFAMethod}{5.7}
\definevar{CatRelGamingBackupSMS/PhoneCall}{0.0}
\definevar{CatRelGamingTrustedDevice}{11.3}
\definevar{CatRelGamingBackupEmail}{3.8}
\definevar{CatRelGamingAccountRecoveryService}{1.9}
\definevar{CatRelGamingAccessNotLost}{0.0}
\definevar{CatRelGamingContactwithPhotoProof}{0.0}
\definevar{CatRelGamingNothingMentioned}{18.9}
\definevar{CatRelGamingHelpNotAccesible}{0.0}
\definevar{CatRelGamingInitialSecret}{1.9}
\definevar{CatRelGamingPasswordReset}{1.9}
\definevar{CatRelGamingSecurityQuestions}{0.0}
\definevar{CatRelGamingOtherRecovery}{7.5}
\definevar{CatRelGamingNoRecovery}{1.9}
\definevar{CatRelGovernmentBackupCodes}{11.8}
\definevar{CatRelGovernmentContactService/Website}{35.3}
\definevar{CatRelGovernmentContactLocalAdmin}{0.0}
\definevar{CatRelGovernmentAdditionalMFAMethod}{17.6}
\definevar{CatRelGovernmentBackupSMS/PhoneCall}{0.0}
\definevar{CatRelGovernmentTrustedDevice}{0.0}
\definevar{CatRelGovernmentBackupEmail}{0.0}
\definevar{CatRelGovernmentAccountRecoveryService}{0.0}
\definevar{CatRelGovernmentAccessNotLost}{5.9}
\definevar{CatRelGovernmentContactwithPhotoProof}{0.0}
\definevar{CatRelGovernmentNothingMentioned}{41.2}
\definevar{CatRelGovernmentHelpNotAccesible}{0.0}
\definevar{CatRelGovernmentInitialSecret}{0.0}
\definevar{CatRelGovernmentPasswordReset}{0.0}
\definevar{CatRelGovernmentSecurityQuestions}{0.0}
\definevar{CatRelGovernmentOtherRecovery}{0.0}
\definevar{CatRelGovernmentNoRecovery}{0.0}
\definevar{CatRelHealthBackupCodes}{19.0}
\definevar{CatRelHealthContactService/Website}{42.9}
\definevar{CatRelHealthContactLocalAdmin}{14.3}
\definevar{CatRelHealthAdditionalMFAMethod}{9.5}
\definevar{CatRelHealthBackupSMS/PhoneCall}{4.8}
\definevar{CatRelHealthTrustedDevice}{0.0}
\definevar{CatRelHealthBackupEmail}{19.0}
\definevar{CatRelHealthAccountRecoveryService}{4.8}
\definevar{CatRelHealthAccessNotLost}{0.0}
\definevar{CatRelHealthContactwithPhotoProof}{0.0}
\definevar{CatRelHealthNothingMentioned}{23.8}
\definevar{CatRelHealthHelpNotAccesible}{0.0}
\definevar{CatRelHealthInitialSecret}{0.0}
\definevar{CatRelHealthPasswordReset}{0.0}
\definevar{CatRelHealthSecurityQuestions}{0.0}
\definevar{CatRelHealthOtherRecovery}{9.5}
\definevar{CatRelHealthNoRecovery}{0.0}
\definevar{CatRelHosting+VPSBackupCodes}{43.2}
\definevar{CatRelHosting+VPSContactService/Website}{50.0}
\definevar{CatRelHosting+VPSContactLocalAdmin}{2.7}
\definevar{CatRelHosting+VPSAdditionalMFAMethod}{9.5}
\definevar{CatRelHosting+VPSBackupSMS/PhoneCall}{0.0}
\definevar{CatRelHosting+VPSTrustedDevice}{13.5}
\definevar{CatRelHosting+VPSBackupEmail}{2.7}
\definevar{CatRelHosting+VPSAccountRecoveryService}{1.4}
\definevar{CatRelHosting+VPSAccessNotLost}{1.4}
\definevar{CatRelHosting+VPSContactwithPhotoProof}{0.0}
\definevar{CatRelHosting+VPSNothingMentioned}{21.6}
\definevar{CatRelHosting+VPSHelpNotAccesible}{0.0}
\definevar{CatRelHosting+VPSInitialSecret}{6.8}
\definevar{CatRelHosting+VPSPasswordReset}{0.0}
\definevar{CatRelHosting+VPSSecurityQuestions}{4.1}
\definevar{CatRelHosting+VPSOtherRecovery}{6.8}
\definevar{CatRelHosting+VPSNoRecovery}{0.0}
\definevar{CatRelIdentityManagementBackupCodes}{45.0}
\definevar{CatRelIdentityManagementContactService/Website}{15.0}
\definevar{CatRelIdentityManagementContactLocalAdmin}{20.0}
\definevar{CatRelIdentityManagementAdditionalMFAMethod}{40.0}
\definevar{CatRelIdentityManagementBackupSMS/PhoneCall}{10.0}
\definevar{CatRelIdentityManagementTrustedDevice}{15.0}
\definevar{CatRelIdentityManagementBackupEmail}{0.0}
\definevar{CatRelIdentityManagementAccountRecoveryService}{0.0}
\definevar{CatRelIdentityManagementAccessNotLost}{5.0}
\definevar{CatRelIdentityManagementContactwithPhotoProof}{0.0}
\definevar{CatRelIdentityManagementNothingMentioned}{20.0}
\definevar{CatRelIdentityManagementHelpNotAccesible}{0.0}
\definevar{CatRelIdentityManagementInitialSecret}{0.0}
\definevar{CatRelIdentityManagementPasswordReset}{0.0}
\definevar{CatRelIdentityManagementSecurityQuestions}{0.0}
\definevar{CatRelIdentityManagementOtherRecovery}{5.0}
\definevar{CatRelIdentityManagementNoRecovery}{0.0}
\definevar{CatRelInvestingBackupCodes}{25.5}
\definevar{CatRelInvestingContactService/Website}{51.0}
\definevar{CatRelInvestingContactLocalAdmin}{0.0}
\definevar{CatRelInvestingAdditionalMFAMethod}{9.8}
\definevar{CatRelInvestingBackupSMS/PhoneCall}{0.0}
\definevar{CatRelInvestingTrustedDevice}{3.9}
\definevar{CatRelInvestingBackupEmail}{2.0}
\definevar{CatRelInvestingAccountRecoveryService}{0.0}
\definevar{CatRelInvestingAccessNotLost}{2.0}
\definevar{CatRelInvestingContactwithPhotoProof}{3.9}
\definevar{CatRelInvestingNothingMentioned}{31.4}
\definevar{CatRelInvestingHelpNotAccesible}{0.0}
\definevar{CatRelInvestingInitialSecret}{0.0}
\definevar{CatRelInvestingPasswordReset}{3.9}
\definevar{CatRelInvestingSecurityQuestions}{2.0}
\definevar{CatRelInvestingOtherRecovery}{2.0}
\definevar{CatRelInvestingNoRecovery}{0.0}
\definevar{CatRelIoTBackupCodes}{47.1}
\definevar{CatRelIoTContactService/Website}{41.2}
\definevar{CatRelIoTContactLocalAdmin}{5.9}
\definevar{CatRelIoTAdditionalMFAMethod}{5.9}
\definevar{CatRelIoTBackupSMS/PhoneCall}{0.0}
\definevar{CatRelIoTTrustedDevice}{5.9}
\definevar{CatRelIoTBackupEmail}{5.9}
\definevar{CatRelIoTAccountRecoveryService}{0.0}
\definevar{CatRelIoTAccessNotLost}{0.0}
\definevar{CatRelIoTContactwithPhotoProof}{0.0}
\definevar{CatRelIoTNothingMentioned}{17.6}
\definevar{CatRelIoTHelpNotAccesible}{0.0}
\definevar{CatRelIoTInitialSecret}{0.0}
\definevar{CatRelIoTPasswordReset}{0.0}
\definevar{CatRelIoTSecurityQuestions}{0.0}
\definevar{CatRelIoTOtherRecovery}{5.9}
\definevar{CatRelIoTNoRecovery}{0.0}
\definevar{CatRelMarketing+AnalyticsBackupCodes}{42.4}
\definevar{CatRelMarketing+AnalyticsContactService/Website}{47.5}
\definevar{CatRelMarketing+AnalyticsContactLocalAdmin}{25.4}
\definevar{CatRelMarketing+AnalyticsAdditionalMFAMethod}{8.5}
\definevar{CatRelMarketing+AnalyticsBackupSMS/PhoneCall}{0.0}
\definevar{CatRelMarketing+AnalyticsTrustedDevice}{6.8}
\definevar{CatRelMarketing+AnalyticsBackupEmail}{5.1}
\definevar{CatRelMarketing+AnalyticsAccountRecoveryService}{0.0}
\definevar{CatRelMarketing+AnalyticsAccessNotLost}{0.0}
\definevar{CatRelMarketing+AnalyticsContactwithPhotoProof}{0.0}
\definevar{CatRelMarketing+AnalyticsNothingMentioned}{22.0}
\definevar{CatRelMarketing+AnalyticsHelpNotAccesible}{0.0}
\definevar{CatRelMarketing+AnalyticsInitialSecret}{0.0}
\definevar{CatRelMarketing+AnalyticsPasswordReset}{0.0}
\definevar{CatRelMarketing+AnalyticsSecurityQuestions}{0.0}
\definevar{CatRelMarketing+AnalyticsOtherRecovery}{3.4}
\definevar{CatRelMarketing+AnalyticsNoRecovery}{0.0}
\definevar{CatRelOtherBackupCodes}{39.0}
\definevar{CatRelOtherContactService/Website}{42.7}
\definevar{CatRelOtherContactLocalAdmin}{14.6}
\definevar{CatRelOtherAdditionalMFAMethod}{13.4}
\definevar{CatRelOtherBackupSMS/PhoneCall}{2.4}
\definevar{CatRelOtherTrustedDevice}{9.8}
\definevar{CatRelOtherBackupEmail}{4.9}
\definevar{CatRelOtherAccountRecoveryService}{3.7}
\definevar{CatRelOtherAccessNotLost}{1.2}
\definevar{CatRelOtherContactwithPhotoProof}{1.2}
\definevar{CatRelOtherNothingMentioned}{17.1}
\definevar{CatRelOtherHelpNotAccesible}{0.0}
\definevar{CatRelOtherInitialSecret}{2.4}
\definevar{CatRelOtherPasswordReset}{2.4}
\definevar{CatRelOtherSecurityQuestions}{2.4}
\definevar{CatRelOtherOtherRecovery}{1.2}
\definevar{CatRelOtherNoRecovery}{0.0}
\definevar{CatRelPaymentBackupCodes}{26.5}
\definevar{CatRelPaymentContactService/Website}{44.1}
\definevar{CatRelPaymentContactLocalAdmin}{17.6}
\definevar{CatRelPaymentAdditionalMFAMethod}{17.6}
\definevar{CatRelPaymentBackupSMS/PhoneCall}{11.8}
\definevar{CatRelPaymentTrustedDevice}{5.9}
\definevar{CatRelPaymentBackupEmail}{2.9}
\definevar{CatRelPaymentAccountRecoveryService}{11.8}
\definevar{CatRelPaymentAccessNotLost}{5.9}
\definevar{CatRelPaymentContactwithPhotoProof}{0.0}
\definevar{CatRelPaymentNothingMentioned}{32.4}
\definevar{CatRelPaymentHelpNotAccesible}{0.0}
\definevar{CatRelPaymentInitialSecret}{0.0}
\definevar{CatRelPaymentPasswordReset}{0.0}
\definevar{CatRelPaymentSecurityQuestions}{0.0}
\definevar{CatRelPaymentOtherRecovery}{14.7}
\definevar{CatRelPaymentNoRecovery}{0.0}
\definevar{CatRelRemoteAccessBackupCodes}{56.2}
\definevar{CatRelRemoteAccessContactService/Website}{18.8}
\definevar{CatRelRemoteAccessContactLocalAdmin}{12.5}
\definevar{CatRelRemoteAccessAdditionalMFAMethod}{6.2}
\definevar{CatRelRemoteAccessBackupSMS/PhoneCall}{0.0}
\definevar{CatRelRemoteAccessTrustedDevice}{6.2}
\definevar{CatRelRemoteAccessBackupEmail}{6.2}
\definevar{CatRelRemoteAccessAccountRecoveryService}{0.0}
\definevar{CatRelRemoteAccessAccessNotLost}{0.0}
\definevar{CatRelRemoteAccessContactwithPhotoProof}{0.0}
\definevar{CatRelRemoteAccessNothingMentioned}{31.2}
\definevar{CatRelRemoteAccessHelpNotAccesible}{0.0}
\definevar{CatRelRemoteAccessInitialSecret}{0.0}
\definevar{CatRelRemoteAccessPasswordReset}{0.0}
\definevar{CatRelRemoteAccessSecurityQuestions}{0.0}
\definevar{CatRelRemoteAccessOtherRecovery}{6.2}
\definevar{CatRelRemoteAccessNoRecovery}{0.0}
\definevar{CatRelRetailBackupCodes}{40.0}
\definevar{CatRelRetailContactService/Website}{44.0}
\definevar{CatRelRetailContactLocalAdmin}{0.0}
\definevar{CatRelRetailAdditionalMFAMethod}{20.0}
\definevar{CatRelRetailBackupSMS/PhoneCall}{8.0}
\definevar{CatRelRetailTrustedDevice}{4.0}
\definevar{CatRelRetailBackupEmail}{0.0}
\definevar{CatRelRetailAccountRecoveryService}{8.0}
\definevar{CatRelRetailAccessNotLost}{0.0}
\definevar{CatRelRetailContactwithPhotoProof}{4.0}
\definevar{CatRelRetailNothingMentioned}{28.0}
\definevar{CatRelRetailHelpNotAccesible}{0.0}
\definevar{CatRelRetailInitialSecret}{0.0}
\definevar{CatRelRetailPasswordReset}{0.0}
\definevar{CatRelRetailSecurityQuestions}{0.0}
\definevar{CatRelRetailOtherRecovery}{4.0}
\definevar{CatRelRetailNoRecovery}{0.0}
\definevar{CatRelSecurityBackupCodes}{39.5}
\definevar{CatRelSecurityContactService/Website}{28.9}
\definevar{CatRelSecurityContactLocalAdmin}{23.7}
\definevar{CatRelSecurityAdditionalMFAMethod}{7.9}
\definevar{CatRelSecurityBackupSMS/PhoneCall}{2.6}
\definevar{CatRelSecurityTrustedDevice}{5.3}
\definevar{CatRelSecurityBackupEmail}{5.3}
\definevar{CatRelSecurityAccountRecoveryService}{0.0}
\definevar{CatRelSecurityAccessNotLost}{0.0}
\definevar{CatRelSecurityContactwithPhotoProof}{0.0}
\definevar{CatRelSecurityNothingMentioned}{15.8}
\definevar{CatRelSecurityHelpNotAccesible}{0.0}
\definevar{CatRelSecurityInitialSecret}{5.3}
\definevar{CatRelSecurityPasswordReset}{0.0}
\definevar{CatRelSecuritySecurityQuestions}{5.3}
\definevar{CatRelSecurityOtherRecovery}{2.6}
\definevar{CatRelSecurityNoRecovery}{0.0}
\definevar{CatRelSocialBackupCodes}{53.6}
\definevar{CatRelSocialContactService/Website}{28.6}
\definevar{CatRelSocialContactLocalAdmin}{3.6}
\definevar{CatRelSocialAdditionalMFAMethod}{10.7}
\definevar{CatRelSocialBackupSMS/PhoneCall}{10.7}
\definevar{CatRelSocialTrustedDevice}{3.6}
\definevar{CatRelSocialBackupEmail}{7.1}
\definevar{CatRelSocialAccountRecoveryService}{3.6}
\definevar{CatRelSocialAccessNotLost}{0.0}
\definevar{CatRelSocialContactwithPhotoProof}{0.0}
\definevar{CatRelSocialNothingMentioned}{25.0}
\definevar{CatRelSocialHelpNotAccesible}{0.0}
\definevar{CatRelSocialInitialSecret}{0.0}
\definevar{CatRelSocialPasswordReset}{0.0}
\definevar{CatRelSocialSecurityQuestions}{0.0}
\definevar{CatRelSocialOtherRecovery}{7.1}
\definevar{CatRelSocialNoRecovery}{0.0}
\definevar{CatRelTaskManagementBackupCodes}{45.0}
\definevar{CatRelTaskManagementContactService/Website}{10.0}
\definevar{CatRelTaskManagementContactLocalAdmin}{20.0}
\definevar{CatRelTaskManagementAdditionalMFAMethod}{0.0}
\definevar{CatRelTaskManagementBackupSMS/PhoneCall}{0.0}
\definevar{CatRelTaskManagementTrustedDevice}{10.0}
\definevar{CatRelTaskManagementBackupEmail}{0.0}
\definevar{CatRelTaskManagementAccountRecoveryService}{0.0}
\definevar{CatRelTaskManagementAccessNotLost}{0.0}
\definevar{CatRelTaskManagementContactwithPhotoProof}{0.0}
\definevar{CatRelTaskManagementNothingMentioned}{25.0}
\definevar{CatRelTaskManagementHelpNotAccesible}{5.0}
\definevar{CatRelTaskManagementInitialSecret}{0.0}
\definevar{CatRelTaskManagementPasswordReset}{0.0}
\definevar{CatRelTaskManagementSecurityQuestions}{0.0}
\definevar{CatRelTaskManagementOtherRecovery}{0.0}
\definevar{CatRelTaskManagementNoRecovery}{0.0}
\definevar{CatRelUniversitiesBackupCodes}{14.6}
\definevar{CatRelUniversitiesContactService/Website}{0.0}
\definevar{CatRelUniversitiesContactLocalAdmin}{82.9}
\definevar{CatRelUniversitiesAdditionalMFAMethod}{56.1}
\definevar{CatRelUniversitiesBackupSMS/PhoneCall}{0.0}
\definevar{CatRelUniversitiesTrustedDevice}{4.9}
\definevar{CatRelUniversitiesBackupEmail}{2.4}
\definevar{CatRelUniversitiesAccountRecoveryService}{4.9}
\definevar{CatRelUniversitiesAccessNotLost}{0.0}
\definevar{CatRelUniversitiesContactwithPhotoProof}{0.0}
\definevar{CatRelUniversitiesNothingMentioned}{2.4}
\definevar{CatRelUniversitiesHelpNotAccesible}{2.4}
\definevar{CatRelUniversitiesInitialSecret}{0.0}
\definevar{CatRelUniversitiesPasswordReset}{0.0}
\definevar{CatRelUniversitiesSecurityQuestions}{2.4}
\definevar{CatRelUniversitiesOtherRecovery}{22.0}
\definevar{CatRelUniversitiesNoRecovery}{0.0}
\definevar{CatRelUtilitiesBackupCodes}{15.4}
\definevar{CatRelUtilitiesContactService/Website}{23.1}
\definevar{CatRelUtilitiesContactLocalAdmin}{11.5}
\definevar{CatRelUtilitiesAdditionalMFAMethod}{7.7}
\definevar{CatRelUtilitiesBackupSMS/PhoneCall}{7.7}
\definevar{CatRelUtilitiesTrustedDevice}{11.5}
\definevar{CatRelUtilitiesBackupEmail}{7.7}
\definevar{CatRelUtilitiesAccountRecoveryService}{7.7}
\definevar{CatRelUtilitiesAccessNotLost}{0.0}
\definevar{CatRelUtilitiesContactwithPhotoProof}{0.0}
\definevar{CatRelUtilitiesNothingMentioned}{53.8}
\definevar{CatRelUtilitiesHelpNotAccesible}{3.8}
\definevar{CatRelUtilitiesInitialSecret}{3.8}
\definevar{CatRelUtilitiesPasswordReset}{0.0}
\definevar{CatRelUtilitiesSecurityQuestions}{0.0}
\definevar{CatRelUtilitiesOtherRecovery}{0.0}
\definevar{CatRelUtilitiesNoRecovery}{0.0}

\definevar{MethBackup+SyncMean}{2.07}
\definevar{MethBackup+SyncMedian}{2.00}
\definevar{MethBackup+SyncMode}{1.00}
\definevar{MethBankingMean}{2.26}
\definevar{MethBankingMedian}{2.00}
\definevar{MethBankingMode}{2.00}
\definevar{MethBettingMean}{1.43}
\definevar{MethBettingMedian}{1.00}
\definevar{MethBettingMode}{1.00}
\definevar{MethCloudComputingMean}{1.77}
\definevar{MethCloudComputingMedian}{1.00}
\definevar{MethCloudComputingMode}{1.00}
\definevar{MethCommunicationMean}{1.68}
\definevar{MethCommunicationMedian}{1.00}
\definevar{MethCommunicationMode}{1.00}
\definevar{MethCreativityMean}{1.75}
\definevar{MethCreativityMedian}{2.00}
\definevar{MethCreativityMode}{2.00}
\definevar{MethCrowdfundingMean}{1.40}
\definevar{MethCrowdfundingMedian}{1.00}
\definevar{MethCrowdfundingMode}{1.00}
\definevar{MethCryptocurrenciesMean}{1.46}
\definevar{MethCryptocurrenciesMedian}{1.00}
\definevar{MethCryptocurrenciesMode}{1.00}
\definevar{MethDevelopersMean}{1.47}
\definevar{MethDevelopersMedian}{1.00}
\definevar{MethDevelopersMode}{1.00}
\definevar{MethDomainsMean}{1.51}
\definevar{MethDomainsMedian}{1.00}
\definevar{MethDomainsMode}{1.00}
\definevar{MethEducationMean}{1.33}
\definevar{MethEducationMedian}{1.00}
\definevar{MethEducationMode}{1.00}
\definevar{MethEmailMean}{2.04}
\definevar{MethEmailMedian}{2.00}
\definevar{MethEmailMode}{1.00}
\definevar{MethEntertainmentMean}{1.78}
\definevar{MethEntertainmentMedian}{1.00}
\definevar{MethEntertainmentMode}{1.00}
\definevar{MethFinanceMean}{1.85}
\definevar{MethFinanceMedian}{2.00}
\definevar{MethFinanceMode}{1.00}
\definevar{MethFoodMean}{1.80}
\definevar{MethFoodMedian}{2.00}
\definevar{MethFoodMode}{2.00}
\definevar{MethGamingMean}{1.53}
\definevar{MethGamingMedian}{1.00}
\definevar{MethGamingMode}{1.00}
\definevar{MethGovernmentMean}{2.59}
\definevar{MethGovernmentMedian}{3.00}
\definevar{MethGovernmentMode}{3.00}
\definevar{MethHealthMean}{1.63}
\definevar{MethHealthMedian}{1.00}
\definevar{MethHealthMode}{1.00}
\definevar{MethHosting+VPSMean}{1.30}
\definevar{MethHosting+VPSMedian}{1.00}
\definevar{MethHosting+VPSMode}{1.00}
\definevar{MethHotels+AccomodationsMean}{2.00}
\definevar{MethHotels+AccomodationsMedian}{2.00}
\definevar{MethHotels+AccomodationsMode}{1.00}
\definevar{MethIdentityManagementMean}{2.85}
\definevar{MethIdentityManagementMedian}{3.00}
\definevar{MethIdentityManagementMode}{1.00}
\definevar{MethInvestingMean}{1.82}
\definevar{MethInvestingMedian}{1.00}
\definevar{MethInvestingMode}{1.00}
\definevar{MethIoTMean}{1.71}
\definevar{MethIoTMedian}{2.00}
\definevar{MethIoTMode}{2.00}
\definevar{MethLegalMean}{1.88}
\definevar{MethLegalMedian}{1.50}
\definevar{MethLegalMode}{1.00}
\definevar{MethMarketing+AnalyticsMean}{1.46}
\definevar{MethMarketing+AnalyticsMedian}{1.00}
\definevar{MethMarketing+AnalyticsMode}{1.00}
\definevar{MethOtherMean}{1.48}
\definevar{MethOtherMedian}{1.00}
\definevar{MethOtherMode}{1.00}
\definevar{MethPaymentMean}{1.82}
\definevar{MethPaymentMedian}{2.00}
\definevar{MethPaymentMode}{1.00}
\definevar{MethPostandShippingMean}{1.67}
\definevar{MethPostandShippingMedian}{2.00}
\definevar{MethPostandShippingMode}{2.00}
\definevar{MethRemoteAccessMean}{1.81}
\definevar{MethRemoteAccessMedian}{1.00}
\definevar{MethRemoteAccessMode}{1.00}
\definevar{MethRetailMean}{1.72}
\definevar{MethRetailMedian}{1.00}
\definevar{MethRetailMode}{1.00}
\definevar{MethSecurityMean}{1.32}
\definevar{MethSecurityMedian}{1.00}
\definevar{MethSecurityMode}{1.00}
\definevar{MethSocialMean}{1.64}
\definevar{MethSocialMedian}{2.00}
\definevar{MethSocialMode}{1.00}
\definevar{MethTaskManagementMean}{1.40}
\definevar{MethTaskManagementMedian}{1.00}
\definevar{MethTaskManagementMode}{1.00}
\definevar{MethTransportMean}{2.00}
\definevar{MethTransportMedian}{2.00}
\definevar{MethTransportMode}{2.00}
\definevar{MethUniversitiesMean}{3.02}
\definevar{MethUniversitiesMedian}{3.00}
\definevar{MethUniversitiesMode}{3.00}
\definevar{MethUtilitiesMean}{2.08}
\definevar{MethUtilitiesMedian}{2.00}
\definevar{MethUtilitiesMode}{1.00}
\definevar{MethVPNProvidersMean}{1.44}
\definevar{MethVPNProvidersMedian}{1.00}
\definevar{MethVPNProvidersMode}{1.00}

\definevar{RecBackup+SyncMean}{1.53}
\definevar{RecBackup+SyncMedian}{1.00}
\definevar{RecBackup+SyncMode}{1.00}
\definevar{RecBankingMean}{0.63}
\definevar{RecBankingMedian}{0.00}
\definevar{RecBankingMode}{0.00}
\definevar{RecBettingMean}{0.86}
\definevar{RecBettingMedian}{1.00}
\definevar{RecBettingMode}{0.00}
\definevar{RecCloudComputingMean}{1.91}
\definevar{RecCloudComputingMedian}{2.00}
\definevar{RecCloudComputingMode}{2.00}
\definevar{RecCommunicationMean}{1.70}
\definevar{RecCommunicationMedian}{1.00}
\definevar{RecCommunicationMode}{1.00}
\definevar{RecCreativityMean}{1.62}
\definevar{RecCreativityMedian}{2.00}
\definevar{RecCreativityMode}{2.00}
\definevar{RecCrowdfundingMean}{0.60}
\definevar{RecCrowdfundingMedian}{0.00}
\definevar{RecCrowdfundingMode}{0.00}
\definevar{RecCryptocurrenciesMean}{1.40}
\definevar{RecCryptocurrenciesMedian}{1.00}
\definevar{RecCryptocurrenciesMode}{1.00}
\definevar{RecDevelopersMean}{1.17}
\definevar{RecDevelopersMedian}{1.00}
\definevar{RecDevelopersMode}{1.00}
\definevar{RecDomainsMean}{1.44}
\definevar{RecDomainsMedian}{1.00}
\definevar{RecDomainsMode}{1.00}
\definevar{RecEducationMean}{0.67}
\definevar{RecEducationMedian}{1.00}
\definevar{RecEducationMode}{1.00}
\definevar{RecEmailMean}{1.79}
\definevar{RecEmailMedian}{2.00}
\definevar{RecEmailMode}{2.00}
\definevar{RecEntertainmentMean}{2.11}
\definevar{RecEntertainmentMedian}{2.00}
\definevar{RecEntertainmentMode}{2.00}
\definevar{RecFinanceMean}{1.06}
\definevar{RecFinanceMedian}{1.00}
\definevar{RecFinanceMode}{0.00}
\definevar{RecFoodMean}{1.00}
\definevar{RecFoodMedian}{1.00}
\definevar{RecFoodMode}{1.00}
\definevar{RecGamingMean}{1.25}
\definevar{RecGamingMedian}{1.00}
\definevar{RecGamingMode}{1.00}
\definevar{RecGovernmentMean}{0.71}
\definevar{RecGovernmentMedian}{1.00}
\definevar{RecGovernmentMode}{0.00}
\definevar{RecHealthMean}{1.24}
\definevar{RecHealthMedian}{1.00}
\definevar{RecHealthMode}{0.00}
\definevar{RecHosting+VPSMean}{1.39}
\definevar{RecHosting+VPSMedian}{1.00}
\definevar{RecHosting+VPSMode}{1.00}
\definevar{RecHotels+AccomodationsMean}{1.00}
\definevar{RecHotels+AccomodationsMedian}{1.00}
\definevar{RecHotels+AccomodationsMode}{1.00}
\definevar{RecIdentityManagementMean}{1.55}
\definevar{RecIdentityManagementMedian}{2.00}
\definevar{RecIdentityManagementMode}{2.00}
\definevar{RecInvestingMean}{1.06}
\definevar{RecInvestingMedian}{1.00}
\definevar{RecInvestingMode}{1.00}
\definevar{RecIoTMean}{1.18}
\definevar{RecIoTMedian}{1.00}
\definevar{RecIoTMode}{1.00}
\definevar{RecLegalMean}{1.38}
\definevar{RecLegalMedian}{1.00}
\definevar{RecLegalMode}{1.00}
\definevar{RecMarketing+AnalyticsMean}{1.39}
\definevar{RecMarketing+AnalyticsMedian}{1.00}
\definevar{RecMarketing+AnalyticsMode}{1.00}
\definevar{RecOtherMean}{1.41}
\definevar{RecOtherMedian}{1.00}
\definevar{RecOtherMode}{1.00}
\definevar{RecPaymentMean}{1.59}
\definevar{RecPaymentMedian}{2.00}
\definevar{RecPaymentMode}{0.00}
\definevar{RecPostandShippingMean}{2.33}
\definevar{RecPostandShippingMedian}{3.00}
\definevar{RecPostandShippingMode}{0.00}
\definevar{RecRemoteAccessMean}{1.12}
\definevar{RecRemoteAccessMedian}{1.00}
\definevar{RecRemoteAccessMode}{1.00}
\definevar{RecRetailMean}{1.32}
\definevar{RecRetailMedian}{1.00}
\definevar{RecRetailMode}{2.00}
\definevar{RecSecurityMean}{1.26}
\definevar{RecSecurityMedian}{1.00}
\definevar{RecSecurityMode}{1.00}
\definevar{RecSocialMean}{1.29}
\definevar{RecSocialMedian}{1.00}
\definevar{RecSocialMode}{1.00}
\definevar{RecTaskManagementMean}{0.85}
\definevar{RecTaskManagementMedian}{1.00}
\definevar{RecTaskManagementMode}{0.00}
\definevar{RecTickets+EventsMean}{2.00}
\definevar{RecTickets+EventsMedian}{2.00}
\definevar{RecTickets+EventsMode}{2.00}
\definevar{RecTransportMean}{1.00}
\definevar{RecTransportMedian}{1.00}
\definevar{RecTransportMode}{0.00}
\definevar{RecUniversitiesMean}{1.90}
\definevar{RecUniversitiesMedian}{2.00}
\definevar{RecUniversitiesMode}{2.00}
\definevar{RecUtilitiesMean}{0.96}
\definevar{RecUtilitiesMedian}{0.00}
\definevar{RecUtilitiesMode}{0.00}
\definevar{RecVPNProvidersMean}{0.78}
\definevar{RecVPNProvidersMedian}{1.00}
\definevar{RecVPNProvidersMode}{1.00}

\definevar{TLDCount_ca}{15}
\definevar{TLDCount_ch}{13}
\definevar{TLDCount_co.uk}{21}
\definevar{TLDCount_com}{944}
\definevar{TLDCount_com.au}{22}
\definevar{TLDCount_de}{48}
\definevar{TLDCount_edu}{32}
\definevar{TLDCount_gov}{10}
\definevar{TLDCount_io}{52}
\definevar{TLDCount_net}{32}
\definevar{TLDCount_org}{41}
\definevar{TLDCount_other}{73}

\definevar{TLDPercent_ca}{1.15}
\definevar{TLDPercent_ch}{1.00}
\definevar{TLDPercent_co.uk}{1.61}
\definevar{TLDPercent_com}{72.45}
\definevar{TLDPercent_com.au}{1.69}
\definevar{TLDPercent_de}{3.68}
\definevar{TLDPercent_edu}{2.46}
\definevar{TLDPercent_gov}{0.77}
\definevar{TLDPercent_io}{3.99}
\definevar{TLDPercent_net}{2.46}
\definevar{TLDPercent_org}{3.15}
\definevar{TLDPercent_other}{5.60}

\definevar{TLDMethcaMean}{2.20}
\definevar{TLDMethcaMedian}{2.00}
\definevar{TLDMethcaMode}{2.00}
\definevar{TLDMethchMean}{1.31}
\definevar{TLDMethchMedian}{1.00}
\definevar{TLDMethchMode}{1.00}
\definevar{TLDMethco.ukMean}{2.05}
\definevar{TLDMethco.ukMedian}{2.00}
\definevar{TLDMethco.ukMode}{1.00}
\definevar{TLDMethcomMean}{1.73}
\definevar{TLDMethcomMedian}{1.00}
\definevar{TLDMethcomMode}{1.00}
\definevar{TLDMethcom.auMean}{1.73}
\definevar{TLDMethcom.auMedian}{1.50}
\definevar{TLDMethcom.auMode}{1.00}
\definevar{TLDMethdeMean}{1.96}
\definevar{TLDMethdeMedian}{2.00}
\definevar{TLDMethdeMode}{2.00}
\definevar{TLDMetheduMean}{3.22}
\definevar{TLDMetheduMedian}{3.00}
\definevar{TLDMetheduMode}{3.00}
\definevar{TLDMethgovMean}{2.90}
\definevar{TLDMethgovMedian}{3.00}
\definevar{TLDMethgovMode}{3.00}
\definevar{TLDMethioMean}{1.21}
\definevar{TLDMethioMedian}{1.00}
\definevar{TLDMethioMode}{1.00}
\definevar{TLDMethnetMean}{1.31}
\definevar{TLDMethnetMedian}{1.00}
\definevar{TLDMethnetMode}{1.00}
\definevar{TLDMethorgMean}{1.93}
\definevar{TLDMethorgMedian}{2.00}
\definevar{TLDMethorgMode}{1.00}
\definevar{TLDMethotherMean}{1.40}
\definevar{TLDMethotherMedian}{1.00}
\definevar{TLDMethotherMode}{1.00}

\definevar{TLDReccaMean}{1.67}
\definevar{TLDReccaMedian}{2.00}
\definevar{TLDReccaMode}{1.00}
\definevar{TLDRecchMean}{0.92}
\definevar{TLDRecchMedian}{1.00}
\definevar{TLDRecchMode}{0.00}
\definevar{TLDRecco.ukMean}{0.81}
\definevar{TLDRecco.ukMedian}{1.00}
\definevar{TLDRecco.ukMode}{1.00}
\definevar{TLDReccomMean}{1.33}
\definevar{TLDReccomMedian}{1.00}
\definevar{TLDReccomMode}{1.00}
\definevar{TLDReccom.auMean}{1.23}
\definevar{TLDReccom.auMedian}{1.00}
\definevar{TLDReccom.auMode}{1.00}
\definevar{TLDRecdeMean}{0.85}
\definevar{TLDRecdeMedian}{1.00}
\definevar{TLDRecdeMode}{0.00}
\definevar{TLDReceduMean}{1.91}
\definevar{TLDReceduMedian}{2.00}
\definevar{TLDReceduMode}{2.00}
\definevar{TLDRecgovMean}{0.70}
\definevar{TLDRecgovMedian}{1.00}
\definevar{TLDRecgovMode}{1.00}
\definevar{TLDRecioMean}{1.12}
\definevar{TLDRecioMedian}{1.00}
\definevar{TLDRecioMode}{1.00}
\definevar{TLDRecnetMean}{1.34}
\definevar{TLDRecnetMedian}{1.00}
\definevar{TLDRecnetMode}{1.00}
\definevar{TLDRecorgMean}{0.80}
\definevar{TLDRecorgMedian}{0.00}
\definevar{TLDRecorgMode}{0.00}
\definevar{TLDRecotherMean}{1.04}
\definevar{TLDRecotherMedian}{1.00}
\definevar{TLDRecotherMode}{1.00}

\definevar{TLDRel_caRecCodes}{26.7}
\definevar{TLDRel_caRecContactSupport}{46.7}
\definevar{TLDRel_caRecContactAdmin}{26.7}
\definevar{TLDRel_caRecAdd2FA}{26.7}
\definevar{TLDRel_caRecTrustedDevice}{0.0}
\definevar{TLDRel_caRecPhone}{6.7}
\definevar{TLDRel_caRecMail}{13.3}
\definevar{TLDRel_caRecRecoveryService}{0.0}
\definevar{TLDRel_caRecAccessNotLost}{6.7}
\definevar{TLDRel_caRecPhoto}{0.0}
\definevar{TLDRel_caRecNoMention}{20.0}
\definevar{TLDRel_caRecInaccesible}{0.0}
\definevar{TLDRel_caRecSecret}{0.0}
\definevar{TLDRel_caRecPWReset}{0.0}
\definevar{TLDRel_caRecSecQuestion}{6.7}
\definevar{TLDRel_caRecOther}{6.7}
\definevar{TLDRel_caRecNone}{0.0}
\definevar{TLDRel_chRecCodes}{23.1}
\definevar{TLDRel_chRecContactSupport}{38.5}
\definevar{TLDRel_chRecContactAdmin}{0.0}
\definevar{TLDRel_chRecAdd2FA}{7.7}
\definevar{TLDRel_chRecTrustedDevice}{0.0}
\definevar{TLDRel_chRecPhone}{7.7}
\definevar{TLDRel_chRecMail}{0.0}
\definevar{TLDRel_chRecRecoveryService}{0.0}
\definevar{TLDRel_chRecAccessNotLost}{0.0}
\definevar{TLDRel_chRecPhoto}{0.0}
\definevar{TLDRel_chRecNoMention}{38.5}
\definevar{TLDRel_chRecInaccesible}{7.7}
\definevar{TLDRel_chRecSecret}{0.0}
\definevar{TLDRel_chRecPWReset}{0.0}
\definevar{TLDRel_chRecSecQuestion}{7.7}
\definevar{TLDRel_chRecOther}{7.7}
\definevar{TLDRel_chRecNone}{0.0}
\definevar{TLDRel_co.ukRecCodes}{23.8}
\definevar{TLDRel_co.ukRecContactSupport}{38.1}
\definevar{TLDRel_co.ukRecContactAdmin}{4.8}
\definevar{TLDRel_co.ukRecAdd2FA}{9.5}
\definevar{TLDRel_co.ukRecTrustedDevice}{0.0}
\definevar{TLDRel_co.ukRecPhone}{0.0}
\definevar{TLDRel_co.ukRecMail}{0.0}
\definevar{TLDRel_co.ukRecRecoveryService}{0.0}
\definevar{TLDRel_co.ukRecAccessNotLost}{0.0}
\definevar{TLDRel_co.ukRecPhoto}{0.0}
\definevar{TLDRel_co.ukRecNoMention}{33.3}
\definevar{TLDRel_co.ukRecInaccesible}{0.0}
\definevar{TLDRel_co.ukRecSecret}{4.8}
\definevar{TLDRel_co.ukRecPWReset}{0.0}
\definevar{TLDRel_co.ukRecSecQuestion}{0.0}
\definevar{TLDRel_co.ukRecOther}{0.0}
\definevar{TLDRel_co.ukRecNone}{0.0}
\definevar{TLDRel_comRecCodes}{37.4}
\definevar{TLDRel_comRecContactSupport}{39.0}
\definevar{TLDRel_comRecContactAdmin}{12.1}
\definevar{TLDRel_comRecAdd2FA}{11.3}
\definevar{TLDRel_comRecTrustedDevice}{3.8}
\definevar{TLDRel_comRecPhone}{7.1}
\definevar{TLDRel_comRecMail}{4.4}
\definevar{TLDRel_comRecRecoveryService}{4.0}
\definevar{TLDRel_comRecAccessNotLost}{2.0}
\definevar{TLDRel_comRecPhoto}{2.3}
\definevar{TLDRel_comRecNoMention}{23.3}
\definevar{TLDRel_comRecInaccesible}{0.6}
\definevar{TLDRel_comRecSecret}{3.4}
\definevar{TLDRel_comRecPWReset}{0.8}
\definevar{TLDRel_comRecSecQuestion}{1.3}
\definevar{TLDRel_comRecOther}{4.9}
\definevar{TLDRel_comRecNone}{0.1}
\definevar{TLDRel_com.auRecCodes}{18.2}
\definevar{TLDRel_com.auRecContactSupport}{68.2}
\definevar{TLDRel_com.auRecContactAdmin}{0.0}
\definevar{TLDRel_com.auRecAdd2FA}{4.5}
\definevar{TLDRel_com.auRecTrustedDevice}{0.0}
\definevar{TLDRel_com.auRecPhone}{9.1}
\definevar{TLDRel_com.auRecMail}{0.0}
\definevar{TLDRel_com.auRecRecoveryService}{0.0}
\definevar{TLDRel_com.auRecAccessNotLost}{4.5}
\definevar{TLDRel_com.auRecPhoto}{4.5}
\definevar{TLDRel_com.auRecNoMention}{13.6}
\definevar{TLDRel_com.auRecInaccesible}{0.0}
\definevar{TLDRel_com.auRecSecret}{9.1}
\definevar{TLDRel_com.auRecPWReset}{0.0}
\definevar{TLDRel_com.auRecSecQuestion}{0.0}
\definevar{TLDRel_com.auRecOther}{4.5}
\definevar{TLDRel_com.auRecNone}{0.0}
\definevar{TLDRel_deRecCodes}{6.2}
\definevar{TLDRel_deRecContactSupport}{39.6}
\definevar{TLDRel_deRecContactAdmin}{4.2}
\definevar{TLDRel_deRecAdd2FA}{6.2}
\definevar{TLDRel_deRecTrustedDevice}{2.1}
\definevar{TLDRel_deRecPhone}{2.1}
\definevar{TLDRel_deRecMail}{0.0}
\definevar{TLDRel_deRecRecoveryService}{0.0}
\definevar{TLDRel_deRecAccessNotLost}{4.2}
\definevar{TLDRel_deRecPhoto}{0.0}
\definevar{TLDRel_deRecNoMention}{45.8}
\definevar{TLDRel_deRecInaccesible}{0.0}
\definevar{TLDRel_deRecSecret}{4.2}
\definevar{TLDRel_deRecPWReset}{0.0}
\definevar{TLDRel_deRecSecQuestion}{0.0}
\definevar{TLDRel_deRecOther}{18.8}
\definevar{TLDRel_deRecNone}{0.0}
\definevar{TLDRel_eduRecCodes}{15.6}
\definevar{TLDRel_eduRecContactSupport}{0.0}
\definevar{TLDRel_eduRecContactAdmin}{78.1}
\definevar{TLDRel_eduRecAdd2FA}{59.4}
\definevar{TLDRel_eduRecTrustedDevice}{0.0}
\definevar{TLDRel_eduRecPhone}{3.1}
\definevar{TLDRel_eduRecMail}{0.0}
\definevar{TLDRel_eduRecRecoveryService}{6.2}
\definevar{TLDRel_eduRecAccessNotLost}{0.0}
\definevar{TLDRel_eduRecPhoto}{0.0}
\definevar{TLDRel_eduRecNoMention}{3.1}
\definevar{TLDRel_eduRecInaccesible}{3.1}
\definevar{TLDRel_eduRecSecret}{0.0}
\definevar{TLDRel_eduRecPWReset}{0.0}
\definevar{TLDRel_eduRecSecQuestion}{0.0}
\definevar{TLDRel_eduRecOther}{28.1}
\definevar{TLDRel_eduRecNone}{0.0}
\definevar{TLDRel_govRecCodes}{20.0}
\definevar{TLDRel_govRecContactSupport}{30.0}
\definevar{TLDRel_govRecContactAdmin}{0.0}
\definevar{TLDRel_govRecAdd2FA}{10.0}
\definevar{TLDRel_govRecTrustedDevice}{0.0}
\definevar{TLDRel_govRecPhone}{0.0}
\definevar{TLDRel_govRecMail}{0.0}
\definevar{TLDRel_govRecRecoveryService}{0.0}
\definevar{TLDRel_govRecAccessNotLost}{10.0}
\definevar{TLDRel_govRecPhoto}{0.0}
\definevar{TLDRel_govRecNoMention}{20.0}
\definevar{TLDRel_govRecInaccesible}{0.0}
\definevar{TLDRel_govRecSecret}{0.0}
\definevar{TLDRel_govRecPWReset}{0.0}
\definevar{TLDRel_govRecSecQuestion}{0.0}
\definevar{TLDRel_govRecOther}{0.0}
\definevar{TLDRel_govRecNone}{0.0}
\definevar{TLDRel_ioRecCodes}{44.2}
\definevar{TLDRel_ioRecContactSupport}{32.7}
\definevar{TLDRel_ioRecContactAdmin}{11.5}
\definevar{TLDRel_ioRecAdd2FA}{0.0}
\definevar{TLDRel_ioRecTrustedDevice}{0.0}
\definevar{TLDRel_ioRecPhone}{1.9}
\definevar{TLDRel_ioRecMail}{0.0}
\definevar{TLDRel_ioRecRecoveryService}{3.8}
\definevar{TLDRel_ioRecAccessNotLost}{0.0}
\definevar{TLDRel_ioRecPhoto}{7.7}
\definevar{TLDRel_ioRecNoMention}{21.2}
\definevar{TLDRel_ioRecInaccesible}{0.0}
\definevar{TLDRel_ioRecSecret}{1.9}
\definevar{TLDRel_ioRecPWReset}{3.8}
\definevar{TLDRel_ioRecSecQuestion}{0.0}
\definevar{TLDRel_ioRecOther}{3.8}
\definevar{TLDRel_ioRecNone}{0.0}
\definevar{TLDRel_netRecCodes}{56.2}
\definevar{TLDRel_netRecContactSupport}{46.9}
\definevar{TLDRel_netRecContactAdmin}{0.0}
\definevar{TLDRel_netRecAdd2FA}{3.1}
\definevar{TLDRel_netRecTrustedDevice}{0.0}
\definevar{TLDRel_netRecPhone}{9.4}
\definevar{TLDRel_netRecMail}{3.1}
\definevar{TLDRel_netRecRecoveryService}{3.1}
\definevar{TLDRel_netRecAccessNotLost}{0.0}
\definevar{TLDRel_netRecPhoto}{0.0}
\definevar{TLDRel_netRecNoMention}{21.9}
\definevar{TLDRel_netRecInaccesible}{0.0}
\definevar{TLDRel_netRecSecret}{0.0}
\definevar{TLDRel_netRecPWReset}{0.0}
\definevar{TLDRel_netRecSecQuestion}{6.2}
\definevar{TLDRel_netRecOther}{6.2}
\definevar{TLDRel_netRecNone}{0.0}
\definevar{TLDRel_orgRecCodes}{24.4}
\definevar{TLDRel_orgRecContactSupport}{19.5}
\definevar{TLDRel_orgRecContactAdmin}{2.4}
\definevar{TLDRel_orgRecAdd2FA}{4.9}
\definevar{TLDRel_orgRecTrustedDevice}{0.0}
\definevar{TLDRel_orgRecPhone}{0.0}
\definevar{TLDRel_orgRecMail}{4.9}
\definevar{TLDRel_orgRecRecoveryService}{2.4}
\definevar{TLDRel_orgRecAccessNotLost}{0.0}
\definevar{TLDRel_orgRecPhoto}{0.0}
\definevar{TLDRel_orgRecNoMention}{51.2}
\definevar{TLDRel_orgRecInaccesible}{0.0}
\definevar{TLDRel_orgRecSecret}{4.9}
\definevar{TLDRel_orgRecPWReset}{7.3}
\definevar{TLDRel_orgRecSecQuestion}{0.0}
\definevar{TLDRel_orgRecOther}{9.8}
\definevar{TLDRel_orgRecNone}{0.0}

\definevar{TrancoCount_1}{692}
\definevar{TrancoCount_20001}{165}
\definevar{TrancoCount_40001}{95}
\definevar{TrancoCount_60001}{55}
\definevar{TrancoCount_80001}{42}
\definevar{TrancoCount_100001}{34}
\definevar{TrancoCount_120001}{37}
\definevar{TrancoCount_140001}{23}
\definevar{TrancoCount_160001}{11}
\definevar{TrancoCount_180001}{19}
\definevar{TrancoCount_200001}{130}

\definevar{TrancoPercent_1}{53.11}
\definevar{TrancoPercent_20001}{12.66}
\definevar{TrancoPercent_40001}{7.29}
\definevar{TrancoPercent_60001}{4.22}
\definevar{TrancoPercent_80001}{3.22}
\definevar{TrancoPercent_100001}{2.61}
\definevar{TrancoPercent_120001}{2.84}
\definevar{TrancoPercent_140001}{1.77}
\definevar{TrancoPercent_160001}{0.84}
\definevar{TrancoPercent_180001}{1.46}
\definevar{TrancoPercent_200001}{9.98}
\definevar{PagesInTop200K}{1,173}
\definevar{PercentInTop200K}{90.02}

\definevar{TrancoMeth1Mean}{1.92}
\definevar{TrancoMeth1Median}{2.00}
\definevar{TrancoMeth1Mode}{1.00}
\definevar{TrancoMeth20001Mean}{1.62}
\definevar{TrancoMeth20001Median}{1.00}
\definevar{TrancoMeth20001Mode}{1.00}
\definevar{TrancoMeth40001Mean}{1.68}
\definevar{TrancoMeth40001Median}{1.00}
\definevar{TrancoMeth40001Mode}{1.00}
\definevar{TrancoMeth60001Mean}{1.56}
\definevar{TrancoMeth60001Median}{1.00}
\definevar{TrancoMeth60001Mode}{1.00}
\definevar{TrancoMeth80001Mean}{1.52}
\definevar{TrancoMeth80001Median}{1.00}
\definevar{TrancoMeth80001Mode}{1.00}
\definevar{TrancoMeth100001Mean}{1.59}
\definevar{TrancoMeth100001Median}{1.00}
\definevar{TrancoMeth100001Mode}{1.00}
\definevar{TrancoMeth120001Mean}{1.54}
\definevar{TrancoMeth120001Median}{1.00}
\definevar{TrancoMeth120001Mode}{1.00}
\definevar{TrancoMeth140001Mean}{1.48}
\definevar{TrancoMeth140001Median}{1.00}
\definevar{TrancoMeth140001Mode}{1.00}
\definevar{TrancoMeth160001Mean}{1.27}
\definevar{TrancoMeth160001Median}{1.00}
\definevar{TrancoMeth160001Mode}{1.00}
\definevar{TrancoMeth180001Mean}{1.42}
\definevar{TrancoMeth180001Median}{1.00}
\definevar{TrancoMeth180001Mode}{1.00}
\definevar{TrancoMeth200001Mean}{1.45}
\definevar{TrancoMeth200001Median}{1.00}
\definevar{TrancoMeth200001Mode}{1.00}

\definevar{TrancoRec1Mean}{1.48}
\definevar{TrancoRec1Median}{1.00}
\definevar{TrancoRec1Mode}{1.00}
\definevar{TrancoRec20001Mean}{1.12}
\definevar{TrancoRec20001Median}{1.00}
\definevar{TrancoRec20001Mode}{0.00}
\definevar{TrancoRec40001Mean}{1.00}
\definevar{TrancoRec40001Median}{1.00}
\definevar{TrancoRec40001Mode}{1.00}
\definevar{TrancoRec60001Mean}{1.09}
\definevar{TrancoRec60001Median}{1.00}
\definevar{TrancoRec60001Mode}{1.00}
\definevar{TrancoRec80001Mean}{1.17}
\definevar{TrancoRec80001Median}{1.00}
\definevar{TrancoRec80001Mode}{1.00}
\definevar{TrancoRec100001Mean}{0.85}
\definevar{TrancoRec100001Median}{1.00}
\definevar{TrancoRec100001Mode}{1.00}
\definevar{TrancoRec120001Mean}{1.11}
\definevar{TrancoRec120001Median}{1.00}
\definevar{TrancoRec120001Mode}{1.00}
\definevar{TrancoRec140001Mean}{1.00}
\definevar{TrancoRec140001Median}{1.00}
\definevar{TrancoRec140001Mode}{1.00}
\definevar{TrancoRec160001Mean}{0.64}
\definevar{TrancoRec160001Median}{0.00}
\definevar{TrancoRec160001Mode}{0.00}
\definevar{TrancoRec180001Mean}{1.32}
\definevar{TrancoRec180001Median}{1.00}
\definevar{TrancoRec180001Mode}{1.00}
\definevar{TrancoRec200001Mean}{0.92}
\definevar{TrancoRec200001Median}{1.00}
\definevar{TrancoRec200001Mode}{0.00}

\definevar{TrancoRel_1MethSMS}{50.3}
\definevar{TrancoRel_1MethCall}{14.0}
\definevar{TrancoRel_1MethMail}{18.4}
\definevar{TrancoRel_1MethHWToken}{18.2}
\definevar{TrancoRel_1MethSWToken}{79.8}
\definevar{TrancoRel_1MethNone}{0.0}
\definevar{TrancoRel_1MethOther}{10.4}
\definevar{TrancoRel_20001MethSMS}{41.8}
\definevar{TrancoRel_20001MethCall}{12.7}
\definevar{TrancoRel_20001MethMail}{12.1}
\definevar{TrancoRel_20001MethHWToken}{7.3}
\definevar{TrancoRel_20001MethSWToken}{73.9}
\definevar{TrancoRel_20001MethNone}{0.0}
\definevar{TrancoRel_20001MethOther}{13.9}
\definevar{TrancoRel_40001MethSMS}{35.8}
\definevar{TrancoRel_40001MethCall}{13.7}
\definevar{TrancoRel_40001MethMail}{20.0}
\definevar{TrancoRel_40001MethHWToken}{12.6}
\definevar{TrancoRel_40001MethSWToken}{78.9}
\definevar{TrancoRel_40001MethNone}{0.0}
\definevar{TrancoRel_40001MethOther}{7.4}
\definevar{TrancoRel_60001MethSMS}{29.1}
\definevar{TrancoRel_60001MethCall}{7.3}
\definevar{TrancoRel_60001MethMail}{12.7}
\definevar{TrancoRel_60001MethHWToken}{10.9}
\definevar{TrancoRel_60001MethSWToken}{80.0}
\definevar{TrancoRel_60001MethNone}{0.0}
\definevar{TrancoRel_60001MethOther}{16.4}
\definevar{TrancoRel_80001MethSMS}{35.7}
\definevar{TrancoRel_80001MethCall}{4.8}
\definevar{TrancoRel_80001MethMail}{11.9}
\definevar{TrancoRel_80001MethHWToken}{2.4}
\definevar{TrancoRel_80001MethSWToken}{90.5}
\definevar{TrancoRel_80001MethNone}{0.0}
\definevar{TrancoRel_80001MethOther}{7.1}
\definevar{TrancoRel_100001MethSMS}{35.3}
\definevar{TrancoRel_100001MethCall}{8.8}
\definevar{TrancoRel_100001MethMail}{14.7}
\definevar{TrancoRel_100001MethHWToken}{14.7}
\definevar{TrancoRel_100001MethSWToken}{70.6}
\definevar{TrancoRel_100001MethNone}{0.0}
\definevar{TrancoRel_100001MethOther}{14.7}
\definevar{TrancoRel_120001MethSMS}{37.8}
\definevar{TrancoRel_120001MethCall}{8.1}
\definevar{TrancoRel_120001MethMail}{16.2}
\definevar{TrancoRel_120001MethHWToken}{8.1}
\definevar{TrancoRel_120001MethSWToken}{81.1}
\definevar{TrancoRel_120001MethNone}{0.0}
\definevar{TrancoRel_120001MethOther}{2.7}
\definevar{TrancoRel_140001MethSMS}{39.1}
\definevar{TrancoRel_140001MethCall}{8.7}
\definevar{TrancoRel_140001MethMail}{4.3}
\definevar{TrancoRel_140001MethHWToken}{0.0}
\definevar{TrancoRel_140001MethSWToken}{78.3}
\definevar{TrancoRel_140001MethNone}{0.0}
\definevar{TrancoRel_140001MethOther}{17.4}
\definevar{TrancoRel_160001MethSMS}{18.2}
\definevar{TrancoRel_160001MethCall}{0.0}
\definevar{TrancoRel_160001MethMail}{18.2}
\definevar{TrancoRel_160001MethHWToken}{0.0}
\definevar{TrancoRel_160001MethSWToken}{90.9}
\definevar{TrancoRel_160001MethNone}{0.0}
\definevar{TrancoRel_160001MethOther}{0.0}
\definevar{TrancoRel_180001MethSMS}{26.3}
\definevar{TrancoRel_180001MethCall}{0.0}
\definevar{TrancoRel_180001MethMail}{0.0}
\definevar{TrancoRel_180001MethHWToken}{15.8}
\definevar{TrancoRel_180001MethSWToken}{89.5}
\definevar{TrancoRel_180001MethNone}{0.0}
\definevar{TrancoRel_180001MethOther}{10.5}
\definevar{TrancoRel_200001MethSMS}{26.9}
\definevar{TrancoRel_200001MethCall}{8.5}
\definevar{TrancoRel_200001MethMail}{10.0}
\definevar{TrancoRel_200001MethHWToken}{8.5}
\definevar{TrancoRel_200001MethSWToken}{81.5}
\definevar{TrancoRel_200001MethNone}{0.0}
\definevar{TrancoRel_200001MethOther}{10.0}

\definevar{TrancoRel_1RecCodes}{41.0}
\definevar{TrancoRel_1RecContactSupport}{39.9}
\definevar{TrancoRel_1RecContactAdmin}{15.9}
\definevar{TrancoRel_1RecAdd2FA}{14.9}
\definevar{TrancoRel_1RecTrustedDevice}{4.5}
\definevar{TrancoRel_1RecPhone}{7.7}
\definevar{TrancoRel_1RecMail}{4.9}
\definevar{TrancoRel_1RecRecoveryService}{5.3}
\definevar{TrancoRel_1RecAccessNotLost}{2.2}
\definevar{TrancoRel_1RecPhoto}{1.4}
\definevar{TrancoRel_1RecNoMention}{18.9}
\definevar{TrancoRel_1RecInaccesible}{0.1}
\definevar{TrancoRel_1RecSecret}{2.0}
\definevar{TrancoRel_1RecPWReset}{0.7}
\definevar{TrancoRel_1RecSecQuestion}{1.6}
\definevar{TrancoRel_1RecOther}{6.8}
\definevar{TrancoRel_1RecNone}{0.1}
\definevar{TrancoRel_20001RecCodes}{28.5}
\definevar{TrancoRel_20001RecContactSupport}{31.5}
\definevar{TrancoRel_20001RecContactAdmin}{9.1}
\definevar{TrancoRel_20001RecAdd2FA}{11.5}
\definevar{TrancoRel_20001RecTrustedDevice}{0.6}
\definevar{TrancoRel_20001RecPhone}{4.8}
\definevar{TrancoRel_20001RecMail}{3.6}
\definevar{TrancoRel_20001RecRecoveryService}{3.0}
\definevar{TrancoRel_20001RecAccessNotLost}{0.6}
\definevar{TrancoRel_20001RecPhoto}{3.6}
\definevar{TrancoRel_20001RecNoMention}{35.8}
\definevar{TrancoRel_20001RecInaccesible}{0.0}
\definevar{TrancoRel_20001RecSecret}{6.7}
\definevar{TrancoRel_20001RecPWReset}{0.6}
\definevar{TrancoRel_20001RecSecQuestion}{0.0}
\definevar{TrancoRel_20001RecOther}{7.3}
\definevar{TrancoRel_20001RecNone}{0.0}
\definevar{TrancoRel_40001RecCodes}{26.3}
\definevar{TrancoRel_40001RecContactSupport}{36.8}
\definevar{TrancoRel_40001RecContactAdmin}{9.5}
\definevar{TrancoRel_40001RecAdd2FA}{8.4}
\definevar{TrancoRel_40001RecTrustedDevice}{2.1}
\definevar{TrancoRel_40001RecPhone}{3.2}
\definevar{TrancoRel_40001RecMail}{0.0}
\definevar{TrancoRel_40001RecRecoveryService}{0.0}
\definevar{TrancoRel_40001RecAccessNotLost}{2.1}
\definevar{TrancoRel_40001RecPhoto}{2.1}
\definevar{TrancoRel_40001RecNoMention}{31.6}
\definevar{TrancoRel_40001RecInaccesible}{0.0}
\definevar{TrancoRel_40001RecSecret}{3.2}
\definevar{TrancoRel_40001RecPWReset}{2.1}
\definevar{TrancoRel_40001RecSecQuestion}{2.1}
\definevar{TrancoRel_40001RecOther}{3.2}
\definevar{TrancoRel_40001RecNone}{0.0}
\definevar{TrancoRel_60001RecCodes}{30.9}
\definevar{TrancoRel_60001RecContactSupport}{38.2}
\definevar{TrancoRel_60001RecContactAdmin}{10.9}
\definevar{TrancoRel_60001RecAdd2FA}{9.1}
\definevar{TrancoRel_60001RecTrustedDevice}{1.8}
\definevar{TrancoRel_60001RecPhone}{1.8}
\definevar{TrancoRel_60001RecMail}{3.6}
\definevar{TrancoRel_60001RecRecoveryService}{1.8}
\definevar{TrancoRel_60001RecAccessNotLost}{0.0}
\definevar{TrancoRel_60001RecPhoto}{1.8}
\definevar{TrancoRel_60001RecNoMention}{21.8}
\definevar{TrancoRel_60001RecInaccesible}{1.8}
\definevar{TrancoRel_60001RecSecret}{1.8}
\definevar{TrancoRel_60001RecPWReset}{3.6}
\definevar{TrancoRel_60001RecSecQuestion}{0.0}
\definevar{TrancoRel_60001RecOther}{3.6}
\definevar{TrancoRel_60001RecNone}{0.0}
\definevar{TrancoRel_80001RecCodes}{31.0}
\definevar{TrancoRel_80001RecContactSupport}{38.1}
\definevar{TrancoRel_80001RecContactAdmin}{14.3}
\definevar{TrancoRel_80001RecAdd2FA}{4.8}
\definevar{TrancoRel_80001RecTrustedDevice}{2.4}
\definevar{TrancoRel_80001RecPhone}{7.1}
\definevar{TrancoRel_80001RecMail}{2.4}
\definevar{TrancoRel_80001RecRecoveryService}{2.4}
\definevar{TrancoRel_80001RecAccessNotLost}{2.4}
\definevar{TrancoRel_80001RecPhoto}{2.4}
\definevar{TrancoRel_80001RecNoMention}{28.6}
\definevar{TrancoRel_80001RecInaccesible}{0.0}
\definevar{TrancoRel_80001RecSecret}{4.8}
\definevar{TrancoRel_80001RecPWReset}{0.0}
\definevar{TrancoRel_80001RecSecQuestion}{0.0}
\definevar{TrancoRel_80001RecOther}{4.8}
\definevar{TrancoRel_80001RecNone}{0.0}
\definevar{TrancoRel_100001RecCodes}{23.5}
\definevar{TrancoRel_100001RecContactSupport}{26.5}
\definevar{TrancoRel_100001RecContactAdmin}{0.0}
\definevar{TrancoRel_100001RecAdd2FA}{8.8}
\definevar{TrancoRel_100001RecTrustedDevice}{0.0}
\definevar{TrancoRel_100001RecPhone}{5.9}
\definevar{TrancoRel_100001RecMail}{2.9}
\definevar{TrancoRel_100001RecRecoveryService}{0.0}
\definevar{TrancoRel_100001RecAccessNotLost}{2.9}
\definevar{TrancoRel_100001RecPhoto}{5.9}
\definevar{TrancoRel_100001RecNoMention}{32.4}
\definevar{TrancoRel_100001RecInaccesible}{2.9}
\definevar{TrancoRel_100001RecSecret}{5.9}
\definevar{TrancoRel_100001RecPWReset}{0.0}
\definevar{TrancoRel_100001RecSecQuestion}{0.0}
\definevar{TrancoRel_100001RecOther}{5.9}
\definevar{TrancoRel_100001RecNone}{0.0}
\definevar{TrancoRel_120001RecCodes}{29.7}
\definevar{TrancoRel_120001RecContactSupport}{45.9}
\definevar{TrancoRel_120001RecContactAdmin}{2.7}
\definevar{TrancoRel_120001RecAdd2FA}{5.4}
\definevar{TrancoRel_120001RecTrustedDevice}{0.0}
\definevar{TrancoRel_120001RecPhone}{0.0}
\definevar{TrancoRel_120001RecMail}{5.4}
\definevar{TrancoRel_120001RecRecoveryService}{0.0}
\definevar{TrancoRel_120001RecAccessNotLost}{2.7}
\definevar{TrancoRel_120001RecPhoto}{2.7}
\definevar{TrancoRel_120001RecNoMention}{24.3}
\definevar{TrancoRel_120001RecInaccesible}{0.0}
\definevar{TrancoRel_120001RecSecret}{5.4}
\definevar{TrancoRel_120001RecPWReset}{0.0}
\definevar{TrancoRel_120001RecSecQuestion}{0.0}
\definevar{TrancoRel_120001RecOther}{10.8}
\definevar{TrancoRel_120001RecNone}{0.0}
\definevar{TrancoRel_140001RecCodes}{21.7}
\definevar{TrancoRel_140001RecContactSupport}{43.5}
\definevar{TrancoRel_140001RecContactAdmin}{4.3}
\definevar{TrancoRel_140001RecAdd2FA}{0.0}
\definevar{TrancoRel_140001RecTrustedDevice}{0.0}
\definevar{TrancoRel_140001RecPhone}{8.7}
\definevar{TrancoRel_140001RecMail}{0.0}
\definevar{TrancoRel_140001RecRecoveryService}{0.0}
\definevar{TrancoRel_140001RecAccessNotLost}{4.3}
\definevar{TrancoRel_140001RecPhoto}{4.3}
\definevar{TrancoRel_140001RecNoMention}{21.7}
\definevar{TrancoRel_140001RecInaccesible}{0.0}
\definevar{TrancoRel_140001RecSecret}{4.3}
\definevar{TrancoRel_140001RecPWReset}{0.0}
\definevar{TrancoRel_140001RecSecQuestion}{4.3}
\definevar{TrancoRel_140001RecOther}{4.3}
\definevar{TrancoRel_140001RecNone}{0.0}
\definevar{TrancoRel_160001RecCodes}{18.2}
\definevar{TrancoRel_160001RecContactSupport}{18.2}
\definevar{TrancoRel_160001RecContactAdmin}{0.0}
\definevar{TrancoRel_160001RecAdd2FA}{18.2}
\definevar{TrancoRel_160001RecTrustedDevice}{0.0}
\definevar{TrancoRel_160001RecPhone}{0.0}
\definevar{TrancoRel_160001RecMail}{0.0}
\definevar{TrancoRel_160001RecRecoveryService}{0.0}
\definevar{TrancoRel_160001RecAccessNotLost}{0.0}
\definevar{TrancoRel_160001RecPhoto}{0.0}
\definevar{TrancoRel_160001RecNoMention}{54.5}
\definevar{TrancoRel_160001RecInaccesible}{0.0}
\definevar{TrancoRel_160001RecSecret}{9.1}
\definevar{TrancoRel_160001RecPWReset}{0.0}
\definevar{TrancoRel_160001RecSecQuestion}{0.0}
\definevar{TrancoRel_160001RecOther}{0.0}
\definevar{TrancoRel_160001RecNone}{0.0}
\definevar{TrancoRel_180001RecCodes}{36.8}
\definevar{TrancoRel_180001RecContactSupport}{47.4}
\definevar{TrancoRel_180001RecContactAdmin}{15.8}
\definevar{TrancoRel_180001RecAdd2FA}{5.3}
\definevar{TrancoRel_180001RecTrustedDevice}{0.0}
\definevar{TrancoRel_180001RecPhone}{5.3}
\definevar{TrancoRel_180001RecMail}{5.3}
\definevar{TrancoRel_180001RecRecoveryService}{0.0}
\definevar{TrancoRel_180001RecAccessNotLost}{0.0}
\definevar{TrancoRel_180001RecPhoto}{0.0}
\definevar{TrancoRel_180001RecNoMention}{15.8}
\definevar{TrancoRel_180001RecInaccesible}{0.0}
\definevar{TrancoRel_180001RecSecret}{5.3}
\definevar{TrancoRel_180001RecPWReset}{5.3}
\definevar{TrancoRel_180001RecSecQuestion}{0.0}
\definevar{TrancoRel_180001RecOther}{5.3}
\definevar{TrancoRel_180001RecNone}{0.0}
\definevar{TrancoRel_200001RecCodes}{24.6}
\definevar{TrancoRel_200001RecContactSupport}{33.8}
\definevar{TrancoRel_200001RecContactAdmin}{6.2}
\definevar{TrancoRel_200001RecAdd2FA}{3.8}
\definevar{TrancoRel_200001RecTrustedDevice}{0.8}
\definevar{TrancoRel_200001RecPhone}{3.8}
\definevar{TrancoRel_200001RecMail}{1.5}
\definevar{TrancoRel_200001RecRecoveryService}{0.8}
\definevar{TrancoRel_200001RecAccessNotLost}{2.3}
\definevar{TrancoRel_200001RecPhoto}{4.6}
\definevar{TrancoRel_200001RecNoMention}{33.1}
\definevar{TrancoRel_200001RecInaccesible}{4.6}
\definevar{TrancoRel_200001RecSecret}{4.6}
\definevar{TrancoRel_200001RecPWReset}{1.5}
\definevar{TrancoRel_200001RecSecQuestion}{1.5}
\definevar{TrancoRel_200001RecOther}{2.3}
\definevar{TrancoRel_200001RecNone}{0.0}

\definevar{MFAEmailSuccessCount}{6}
\definevar{MFAHardwareTokenSuccessCount}{1}
\definevar{MFAMobileAppLostCount}{21}
\definevar{MFAMobileAppWaitingCount}{3}
\definevar{MFAMobileAppSuccessCount}{25}
\definevar{MFASMSLostCount}{9}
\definevar{MFASMSWaitingCount}{1}
\definevar{MFASMSSuccessCount}{5}
\definevar{MFAEmailRegainedPercent}{100.0}
\definevar{MFAEmailLostPercent}{0.0}
\definevar{MFAEmailNoAnswerPercent}{0.0}
\definevar{MFAHardwareTokenRegainedPercent}{100.0}
\definevar{MFAHardwareTokenLostPercent}{0.0}
\definevar{MFAHardwareTokenNoAnswerPercent}{0.0}
\definevar{MFAMobileAppRegainedPercent}{51.02}
\definevar{MFAMobileAppLostPercent}{42.86}
\definevar{MFAMobileAppNoAnswerPercent}{6.12}
\definevar{MFASMSRegainedPercent}{33.33}
\definevar{MFASMSLostPercent}{60.0}
\definevar{MFASMSNoAnswerPercent}{6.67}

\definevar{RecBackupCodeLostCount}{16}
\definevar{RecBackupCodeWaitingCount}{1}
\definevar{RecBackupCodeSuccessCount}{22}
\definevar{RecIndirectLostCount}{1}
\definevar{RecIndirectSuccessCount}{4}
\definevar{RecNoRecoveryLostCount}{8}
\definevar{RecNoRecoveryWaitingCount}{3}
\definevar{RecNoRecoverySuccessCount}{10}
\definevar{RecOtherLostCount}{3}
\definevar{RecOtherSuccessCount}{2}
\definevar{RecTOTPLostCount}{5}

\definevar{TimeHourMean}{32.86}
\definevar{TimeHourMedian}{3.28}
\definevar{TimeHourMin}{0.02}
\definevar{TimeHourMax}{646.27}
\definevar{TimeHourStd}{104.24}
\definevar{RecBackupCodeRegainedPercent}{56.41}
\definevar{RecBackupCodeLostPercent}{41.03}
\definevar{RecBackupCodeNoAnswerPercent}{2.56}
\definevar{RecIndirectRegainedPercent}{80.0}
\definevar{RecIndirectLostPercent}{20.0}
\definevar{RecIndirectNoAnswerPercent}{0.0}
\definevar{RecNoRecoveryRegainedPercent}{47.62}
\definevar{RecNoRecoveryLostPercent}{38.1}
\definevar{RecNoRecoveryNoAnswerPercent}{14.29}
\definevar{RecOtherRegainedPercent}{40.0}
\definevar{RecOtherLostPercent}{60.0}
\definevar{RecOtherNoAnswerPercent}{0.0}
\definevar{RecTOTPRegainedPercent}{0.0}
\definevar{RecTOTPLostPercent}{100.0}
\definevar{RecTOTPNoAnswerPercent}{0.0}

\definevar{TimeMinuteMean}{32.86}
\definevar{TimeMinuteMedian}{3.28}
\definevar{TimeMinuteMin}{0.02}
\definevar{TimeMinuteMax}{646.27}
\definevar{TimeMinuteStd}{104.24}

\definevar{RecDetailEmailAccessSufficientCount}{14}
\definevar{RecDetailEmailAccessSufficientPercent}{19.72}
\definevar{RecDetailRequiredIDUnavailableCount}{13}
\definevar{RecDetailRequiredIDUnavailablePercent}{18.31}
\definevar{RecDetailStrictPolicyCount}{10}
\definevar{RecDetailStrictPolicyPercent}{14.08}
\definevar{RecDetailKnowledgeAboutAcc.Count}{10}
\definevar{RecDetailKnowledgeAboutAcc.Percent}{14.08}
\definevar{RecDetailCouldNotContactCount}{6}
\definevar{RecDetailCouldNotContactPercent}{8.45}
\definevar{RecDetailEmailMFACount}{6}
\definevar{RecDetailEmailMFAPercent}{8.45}
\definevar{RecDetailNoRequirementsCount}{4}
\definevar{RecDetailNoRequirementsPercent}{5.63}
\definevar{RecDetailStillWaitingCount}{4}
\definevar{RecDetailStillWaitingPercent}{5.63}
\definevar{RecDetailTimeBasedCount}{3}
\definevar{RecDetailTimeBasedPercent}{4.23}
\definevar{RecDetailWebsiteMFABrokenCount}{1}
\definevar{RecDetailWebsiteMFABrokenPercent}{1.41}
\definevar{PagesTotal}{2,021}
\definevar{PagesMFA}{1,303}

\definevar{SkippedDuplicates}{44} 
\definevar{SkippedBroken}{7} 
\definevar{SkippedLang}{155} 
\definevar{PagesNoMFA}{512} 

\definevar{Excluded}{}

\definevar{RecoveryMethodsCodesTotal}{17}
\definevar{RecOptOut}{4}

\definevar{RecAccTotal}{71}
\definevar{RecAccSuccessCount}{37}
\definevar{RecAccSuccessPercent}{52.11}
\definevar{RecAccLostCount}{30}
\definevar{RecAccLostPercent}{42.25}
\definevar{RecAccWaitingCount}{4}
\definevar{RecAccWaitingPercent}{5.63}
\definevar{RecAccPick}{16}
\definevar{RecAccRand}{27}
\definevar{RecAccTop}{28}

\definevar{RecSkipTotal}{111}
\definevar{RecSkipData}{23}
\definevar{RecSkipReq}{25}
\definevar{RecSkipDone}{12}
\definevar{RecSkipBlock}{11}
\definevar{RecSkipOff}{3}
\definevar{RecSkipTrial}{17}
\definevar{RecSkipTOS}{20}

\definevar{MFAUsedSMSCount}{15}
\definevar{MFAUsedSMSPercent}{21.13}
\definevar{MFAUsedTOTPCount}{49}
\definevar{MFAUsedTOTPPercent}{69.01}
\definevar{MFAUsedEmailCount}{6}
\definevar{MFAUsedEmailPercent}{8.45}
\definevar{MFAUsedHWCount}{1}
\definevar{MFAUsedHWPercent}{1.41}

\definevar{RecoverySetCount}{50}
\definevar{RecoverySetPercent}{70.42}
\definevar{RecoveryNotSetCount}{21}
\definevar{RecoveryNotSetPercent}{29.58}

\definevar{RecUsedBackupCount}{39}
\definevar{RecUsedBackupPercent}{54.93}
\definevar{RecUsedNoneCount}{21}
\definevar{RecUsedNonePercent}{29.58}
\definevar{RecUsedIndirectCount}{5}
\definevar{RecUsedIndirectPercent}{7.04}
\definevar{RecUsedTOTPCount}{5}
\definevar{RecUsedTOTPPercent}{7.04}
\definevar{RecUsedOtherCount}{5}
\definevar{RecUsedOtherPercent}{7.04}

\definevar{RecSetBackupCount}{54}
\definevar{RecSetBackupPercent}{76.06}
\definevar{RecSetBackupSuccessCount}{27}
\definevar{RecSetBackupSuccessPercent}{50}
\definevar{RecSetBackupLostCount}{23}
\definevar{RecSetBackupLostPercent}{42.59}
\definevar{RecSetBackupWaitingCount}{4}
\definevar{RecSetBackupWaitingPercent}{7.41}
\definevar{RecNotSetBackupSuccessCount}{10}
\definevar{RecNotSetBackupSuccessPercent}{47.62}
\definevar{RecNotSetBackupLostCount}{11}
\definevar{RecNotSetBackupLostPercent}{52.38}
\definevar{RecStoryIndependentCount}{47}
\definevar{RecStoryIndependentPercent}{87.04}

\definevar{AccessDueToEmailCount}{26}
\definevar{AccessDueToEmailPercent}{70.27}
\definevar{AccessDueToKnowledgeCount}{7}
\definevar{AccessDueToKnowledgePercent}{18.92}
\definevar{AccessDueToNothingCount}{3}
\definevar{AccessDueToNothingPercent}{8.11}
\definevar{AccessDueToOtherCount}{1}
\definevar{AccessDueToOtherPercent}{2.7}

\definevar{RecSuggestedYesCount}{10}
\definevar{RecSuggestedYesPercent}{14.08}
\definevar{RecSuggestedNoCount}{35}
\definevar{RecSuggestedNoPercent}{49.3}

\definevar{RecContactEmailCount}{19}
\definevar{RecContactEmailPercent}{26.76}
\definevar{RecContactFormCount}{5}
\definevar{RecContactFormPercent}{7.04}
\definevar{RecContactChatCount}{19}
\definevar{RecContactChatPercent}{26.76}
\definevar{RecContactNoneCount}{28}
\definevar{RecContactNonePercent}{39.44}

\definevar{RecWarningsExpCount}{16}
\definevar{RecWarningsExpPercent}{22.54}
\definevar{RecWarningsDocCount}{13}
\definevar{RecWarningsDocPercent}{18.31}
\definevar{RecSoftWarningExpCount}{32}
\definevar{RecSoftWarningExpPercent}{45.07}
\definevar{RecSoftWarningDocCount}{35}
\definevar{RecSoftWarningDocPercent}{49.3}
\definevar{RecSetupExpCount}{42}
\definevar{RecSetupExpPercent}{59.15}
\definevar{RecSetupDocCount}{47}
\definevar{RecSetupDocPercent}{66.2}
\definevar{RecSMSWarningExpCount}{0}
\definevar{RecSMSWarningExpPercent}{0}
\definevar{RecSMSWarningDocCount}{7}
\definevar{RecSMSWarningDocPercent}{9.86}
\definevar{RecHelpNumDocCount}{18}
\definevar{RecHelpNumDocPercent}{25.35}
\definevar{RecDelayExpCount}{0}
\definevar{RecDelayExpPercent}{0}
\definevar{RecDelayDocCount}{6}
\definevar{RecDelayDocPercent}{8.45}
\definevar{RecRestrictedExpCount}{3}
\definevar{RecRestrictedExpPercent}{4.23}
\definevar{RecRestrictedDocCount}{22}
\definevar{RecRestrictedDocPercent}{30.99}
\definevar{RecForcedExpCount}{10}
\definevar{RecForcedExpPercent}{14.08}
\definevar{RecForcedDocCount}{5}
\definevar{RecForcedDocPercent}{7.04}
\definevar{RecTriggerExpCount}{50}
\definevar{RecTriggerExpPercent}{70.42}
\definevar{RecTriggerDocCount}{66}
\definevar{RecTriggerDocPercent}{92.96}

\definevar{CommMethodsSameCount}{9}
\definevar{CommMethodsSamePercent}{12.68}
\definevar{CommDetailsSameCount}{1}
\definevar{CommDetailsSamePercent}{1.41}
\definevar{CommMethodsSameAvg}{10.78}
\definevar{CommDetailsSameAvg}{29.88}
\definevar{CommWarnedRegainedExpCount}{9}
\definevar{CommWarnedRegainedExpPercent}{12.68}
\definevar{CommNotWarnedLostExpCount}{6}
\definevar{CommNotWarnedLostExpPercent}{8.45}
\definevar{CommWarnedNoSetupExpCount}{2}
\definevar{CommWarnedNoSetupExpPercent}{12.5}
\definevar{CommWarnedSetupExpCount}{14}
\definevar{CommWarnedSetupExpPercent}{19.72}
\definevar{CommBackupCodeForcedCount}{6}
\definevar{CommBackupCodeForcedPercent}{15}

\definevar{HandPickedInteresting}{6}
\definevar{HandPickedSecurity}{6}
\definevar{HandPickedUnclear}{2}
\definevar{HandPickedStrict}{2}

\definevar{RecFinancialReqCount}{42}
\definevar{RecFinancialSpecCount}{25}
\definevar{RecFinancialReqPercent}{50}
\definevar{RecFinancialSpecPercent}{56}

\definevar{RecAddressReqCount}{30}
\definevar{RecAddressSpecCount}{20}
\definevar{RecAddressReqPercent}{33.33}
\definevar{RecAddressSpecPercent}{60.00}

\definevar{RecIDReq}{13}

\usepackage[usetwoside=true]{mdframed}
\usepackage{tcolorbox}
\tcbuselibrary{breakable}
\tcbuselibrary{skins}
\usepackage{enumitem}
\usepackage{subcaption}
\definecolor{darkgray}{gray}{0.3}
\tcbset{}
\newtcolorbox{summaryBox}[2][]
{
    enhanced,
    breakable,
    frame hidden,
    borderline west = {2pt}{0pt}{lightgray},
    colback         = white,
    size            = fbox,
    left            = 0.3em,
    enlarge top by  = 0.5em,
    coltitle        = black,
    title           = {\color{darkgray} #2 },
    attach title to upper,
    #1,
}

\newlength\bubblesize
\newcommand\yes{\tikz[baseline=0.1ex] \fill[black]
(\bubblesize,\bubblesize) circle (\bubblesize);}

\newcommand\may{\tikz[baseline=0.1ex] \fill[gray]
(\bubblesize,\bubblesize) circle (\bubblesize);}

\copyrightyear{2023}
\acmYear{2023}
\setcopyright{acmlicensed}\acmConference[CCS '23]{Proceedings of the 2023
ACM SIGSAC Conference on Computer and Communications
Security}{November 26--30, 2023}{Copenhagen, Denmark}
\acmBooktitle{Proceedings of the 2023 ACM SIGSAC Conference on Computer
and Communications Security (CCS '23), November 26--30, 2023, Copenhagen,
Denmark}
\acmPrice{15.00}
\acmDOI{10.1145/3576915.3623180}
\acmISBN{979-8-4007-0050-7/23/11}

\begin{document}

\title[An Evaluation of the Security and Usability of Multi-Factor Authentication Recovery Deployments]{“We've Disabled MFA for You”: An Evaluation of the Security and Usability of Multi-Factor Authentication Recovery Deployments}

\author{Sabrina Amft}
\affiliation{%
\institution{CISPA Helmholtz Center for Information Security}
\city{Hanover}
\country{Germany}}

\author{Sandra H{\"o}ltervennhoff}
\affiliation{%
\institution{Leibniz University Hannover}
\city{Hanover}
\country{Germany}
}

\author{Nicolas Huaman}
\affiliation{%
\institution{Leibniz University Hannover}
\city{Hanover}
\country{Germany}
}

\author{Alexander Krause}
\affiliation{%
\institution{CISPA Helmholtz Center for Information Security}
\city{Hanover}
\country{Germany}
}

\author{Lucy Simko}
\affiliation{%
\institution{The George Washington University}
\city{Washington, DC}
\country{USA}
}

\author{Yasemin Acar}
\affiliation{%
\institution{Paderborn University \& \\ The George Washington University}
\city{Paderborn}
\country{Germany}
}

\author{Sascha Fahl}
\affiliation{%
\institution{CISPA Helmholtz Center for Information Security}
\city{Hanover}
\country{Germany}
}

\renewcommand{\shortauthors}{Amft et al.}

\acmConference[CCS '23]{The 30th ACM Conference on Computer and Communications Security}{November 26--30, 2023}{Copenhagen, Denmark}

\begin{abstract}
Multi-Factor Authentication is intended to strengthen the security of password-based authentication by adding another factor, such as hardware tokens or one-time passwords using mobile apps.

However, this increased authentication security comes with potential drawbacks that can lead to account and asset loss. If users lose access to their additional authentication factors for any reason, they will be locked out of their accounts. Consequently, services that provide Multi-Factor Authentication should deploy procedures to allow their users to recover from losing access to their additional factor that are both secure and easy-to-use.

In this work, we investigate the security and user experience of Multi-Factor Authentication recovery procedures, and compare their deployment to descriptions on help and support pages.

We first evaluate the official help and support pages of \var{PagesMFA} websites that provide Multi-Factor Authentication and collect documented information about their recovery procedures. Second, we select a subset of \var{RecAccTotal} websites, create accounts, set up Multi-Factor Authentication, and perform an in-depth investigation of their recovery procedure security and user experience.

We find that many websites deploy insecure Multi-Factor Authentication recovery procedures and allowed us to circumvent and disable Multi-Factor Authentication when having access to the accounts' associated email addresses. 
Furthermore, we commonly observed discrepancies between our in-depth analysis and the official help and support pages, implying that information meant to aid users is often either incorrect or outdated.

Based on our findings, we provide recommendations for best practices regarding Multi-Factor Authentication recovery.

\end{abstract}



\received{20 February 2007}
\received[revised]{12 March 2009}
\received[accepted]{5 June 2009}

\maketitle

\section{Introduction}
\label{sec_intro}

Username and password are the de facto standard for user authentication on the web and beyond. However, they come with many security problems: Users tend to choose easily guessable passwords and re-use them for multiple accounts~\cite{bonneau2012science,ur2015added,wang2019reuse,wash2016understanding, pearman2017reuse}, or suffer from insecure or hard-to-use password policies~\cite{tan2020practical,gerlitz2021please,lee2022password}. Additionally, service providers may implement insecure and inadequate password storage, leaving millions of passwords unprotected due to data breaches~\cite{selfkeydatabreaches, pw-hashes-stolen, plaintext-pw-storage, bauman2015half, ForbesCollection1DataBreach2019, raponi2020longitudinal}.
\textit{Multi-Factor Authentication} (\mfa{}) adds an extra factor and additional security to password-based authentication schemes, and has become important in many authentication deployments on the web.
With \mfa{}, users authenticate themselves using additional factors, \eg{}, biometric features, hardware tokens, smartphone applications, or secret information only they know~\cite{nist_mfa}. 

Prior work has found tensions between security and usability in many \mfa{} implementations, revealing that the complexity of \mfa{} setup is a barrier to widespread use~\cite{reese2019usability, reynolds2018tale, reynolds2020empirical}, and user acceptance of long-term \mfa{} use can suffer due to the increased log-in and setup time~\cite{farke2020you, abbott2020mandatory}. While most prior work focused on \mfa{} methods and deployment, recovery procedures for loss of \mfa{} are less well understood.
Hence, in this paper, we focus on this critical aspect of \mfa{} security and usability: We investigate the security and usability of \mfa{} recovery procedure deployments on the web. Recovery procedures are crucial to regaining access to accounts in case \mfa{} factors are lost, stolen, or broken. The deployment of a recovery procedure by service providers has a significant impact on authentication security and usability. While locking users out of an account after losing access to \mfa{} factors keeps up security assumptions of \mfa{}, it might contribute to users' frustration~\cite{mcdonald2021annoying}, users leaving a service, or even avoiding \mfa{} in the future. However, allowing users to access their accounts after losing their \mfa{} factors by returning to password-based authentication significantly decreases security, allowing attackers to more easily gain access to user accounts than with \mfa{}. 
While security and usability seem to be directly at odds in many current implementations of \mfa{} recovery, it is essential for service providers to relieve this tension through better communication, and policy and technical changes in order to provide \textit{both} security and usability to support users.

We aim to better understand the experience of users who have lost access to their \mfa{}, and systematically evaluate \mfa{} deployment and recovery, asking the following research questions:

\begin{enumerate}[nosep, label=\textbf{RQ\arabic*:}, left=2pt]
\item \textit{``What \mfa{} recovery procedures are commonly deployed on the web?''} Due to their crucial impact on authentication security and usability, the variety of deployed \mfa{} recovery procedures need to be better understood in the research community. We investigate the deployment and documentation of common \mfa{} recovery procedures on the web.
\item \textit{``How are \mfa{} recovery procedures on the web implemented and what is their impact on authentication security?''}
\mfa{} recovery procedures aim to support users getting back into their accounts in case of losing access to \mfa{} factors. We investigate how service providers balance the usability and security of \mfa{} recovery procedures. 
\item \textit{``How can the security and usability of \mfa{} recovery procedures be improved?''} Standards or established best practices for \mfa{} recovery procedures are missing. Based on our results, we make recommendations to balance the security and usability of \mfa{} recovery procedures for future deployments.
\end{enumerate}

First, we created a list of \var{PagesMFA} websites offering \mfa{}. 
We systematically evaluated \mfa{} recovery procedures deployed on these websites using their official help and support documentation. 
In a follow-up ERB-approved study, we performed an in-depth analysis of the security and usability of deployed \mfa{} recovery procedures for a subset of \var{RecAccTotal} websites. 
We created accounts, enabled and configured \mfa{}, and went through the entire recovery process.

In this paper, we make the following contributions:
\begin{itemize}
    \item \textbf{\mfa{} Recovery Procedure Deployment}:
    By analyzing official help and support pages, and FAQs, we find that most websites only offer one recovery procedure, which is most commonly a backup code or direct support contact. A quarter of pages do not mention recovery in their help.
    \item \textbf{User Experience of \mfa{} Recovery Procedures}: We perform an in-depth investigation of the user experience of \mfa{} recovery procedures on \var{RecAccTotal} websites. We could recover \var{RecAccSuccessPercent}\% of the accounts. In most cases, access to the associated email inbox was sufficient.
    \item \textbf{Recommendations}: Based on our results, we provide recommendations for future research on secure and usable \mfa{} recovery procedures. We also provide guidance for web developers to help deploy easier-to-use and secure \mfa{} recovery procedures, spanning all areas of recovery such as setup, communication, and recovery itself.
    \item \textbf{Replication}: We provide all material including communication templates, the study protocols and our codebooks to support methodological transparency, replicability, and validity, and to help future research on this topic as part of a replication package\footnote{\label{replication}We provide this material on our accompanying website~\cite{mfrwebsite} and in our replication package: \url{https://doi.org/10.25835/9v3k2sx0}.}. 
\end{itemize}

This work is structured as follows. 
In Section~\ref{sec_relwork}, we describe previous work on the subject. We present the methodology and results of our \studyone{} in Section~\ref{sec_meth_deployment}, followed by our practical \studytwo{} in Section~\ref{sec_meth_recoverytests}. In Section~\ref{sec_ethlimits}, we elaborate on the ethics and limitations of our work. We discuss our findings in Section~\ref{sec_discussion}, and conclude in Section~\ref{sec_conclusion}.

\section{Related Work}
\label{sec_relwork}

In this section, we present previous work on related areas and discuss our novelty and contributions compared to them. We first showcase related papers regarding \mfa{} security and usability, then discuss research on the recovery of account access regarding both single-factor authentication and recovery of additional factors.

\subsection{Multi-Factor Authentication Usability}
\label{rw_mfa}
Several previous works have investigated the usability of different popular \mfa{} methods. 

In 2018, Colnago et al. observed the enrollment of Duo two-factor authentication at their university. While adopters found it annoying, they also perceived it as easy to use and secure~\cite{colnago2018s}. 
In the same year, Reynolds et al. conducted two studies with Yubikeys, asking participants to set them up or to include them into their daily lives. Despite participants embracing hardware keys, the study uncovered severe usability problems, especially during setup~\cite{reynolds2018tale}.
In 2019, Reese et al. conducted a usability study with 72 participants comparing five different 2FA methods, confirming findings by Reynolds et al.~\cite{reese2019usability}.

Ciolino et al. also conducted both a lab and longitudinal study comparing SMS OTP and hardware keys, and similar to previous works reported that SMS is both faster to use and set up, and that hardware keys are perceived as less usable~\cite{ciolino2019two}.
In 2020, Reynolds et al. evaluated authentication data sets from two universities that introduced 2FA, finding that more than five percent of authentication attempts failed, mostly due to technical difficulties such as timeouts, or users canceling or mistyping the codes~\cite{reynolds2020empirical}. 
Similarly, Abbott and Patil examined authentication logs and surveyed university users after 2FA became mandatory for them. They found that user acceptance was not influenced when 2FA was required for some sensitive authentication purposes, but decreased when it was enforced for every login~\cite{abbott2020mandatory}. 
Also in 2020, Farke et al. accompanied the introduction of FIDO2 keys as a single-factor (\ie{}, a password-less login, not a \mfa{} method) in a small company, finding that while participants considered them usable, they stopped using them due to concerns such as login efficiency or missing browser support~\cite{farke2020you}. 

A more recent work by Lyastani et al. evaluated 85 websites regarding their \mfa{} usability by creating accounts and evaluating their \mfa{} communication and settings. They found that \mfa{} implementations are largely inconsistent between different websites, and that many designs have previously been identified as obstructive or problematic~\cite{lyastani2023systematic}.

Overall, previous work shows that the usability depends on the \mfa{} method used, \eg{}, SMS OTPs are often faster and more usable than hardware keys, although they are known to be less secure. Additionally, users commonly struggle with setup and continued usage, obstructing widespread voluntary adoption of \mfa{}. While our work does not investigate user sentiments, we discuss the user experience of \mfa{} methods and recovery procedures, and how website and service providers communicate \mfa{} and its security benefits for online security. 

\subsection{Account Recovery}
\label{rw_recovery}
While not many previous works have discussed the recovery of inaccessible \mfas{}, several have worked on account recovery in general, and its potential obstacles.

One of the currently still popular forms of account recovery, security questions, was criticized as early as 2008 when Rabkin addressed the issue of using security questions to regain account access. The work highlighted several issues, including how questions ask for easily obtainable knowledge, especially with the rise of social media~\cite{rabkin2008personal}.
A similar work by Schechter et al. in 2009 confirmed this, finding that in 17\% of attempts, user acquaintances were able to guess the answer to security questions, 13\% were easily guessable within the first five attempts in general. Further issues included usability problems, as 20\% of users forgot the answers themselves six months after setting them up~\cite{schechter2009nosecret}.
The usability of security questions was further researched by Bonneau et al. in 2015 in a survey. They found that 37\% of participants lied when answering security questions, and 40\% did not remember their answers when requiring them to recover their account~\cite{bonneau2015secrets}.

In 2006, Brainard et al. proposed a novel approach for account and \mfa{} recovery, in which a trusted person can vouch for account owners and help them regain access. They further discussed limitations and variants, like additional measures to prevent users from abusing this recovery method as its own \mfa{}~\cite{brainard2006fourth}.
A similar approach was investigated in 2009 through a lab study by Schechter et al.~\cite{schechter2009s}. In their study, 17 of 19 participants successfully used trusted individuals, however, similar to the answers to security questions, some users forgot who their trusted individual was.

In 2015, Hang et al. researched fallback authentication methods for smartphones, finding that current methods were largely sufficient, but that a minority of users struggled with it due to a lack of knowledge or alternative authentication options~\cite{hang2015locked}.
While most of these studies examined specific account recovery procedures, Neil et al. evaluated real-world account recovery advice on 57 popular websites. Their results indicated that help sections were often incomplete and that 39\% of websites were failing to address the topic at all~\cite{neil2021investigating}.

In 2021, Kunke et al. evaluated twelve account recovery procedures for passwordless authentication with FIDO2 regarding different criteria within the areas of usability, deployability, and security~\cite{kunke2021evaluation}. They found the recoveries to be lacking and, in some cases, such as security questions and backup passwords, jeopardize the idea of passwordless authentication.
In 2023, Gilsenan et al. performed a usability analysis of 22 popular TOTP 2FA mobile apps regarding their backup implementations, finding that most rely on less secure methods such as SMS or email for backups~\cite{gilsenan2023security}.

Furthermore, a recent study by Gerlitz et al. has also investigated how websites act when additional factors are lost. They were able to recover about half of their created accounts, and find that there are no best practices for recovery behavior or communication~\cite{gerlitz2023adventures}.

To summarize, previous work has evaluated the security, usability, and communication of fallback authentication methods. However, this typically happened in the context of account recovery, and with no regard for \mfa{} loss. In contrast, we evaluate existing \mfa{} recovery procedures by experiencing them first-hand, with special regard to its usability and (loss of) security benefits.

\section{\mfa{} Recovery Procedure Analysis}
\label{sec_meth}
This section presents the method and findings of our analysis of deployed \mfa{} recovery procedures. 
Our analysis spans two parts; we first investigated deployed \mfa{} recovery procedures for \var{PagesMFA} websites using their publicly available help and support pages. 
Second, we performed an in-depth analysis of \mfa{} recovery procedures for a subset of \var{RecAccTotal} websites. 
We simulated a user's experience after losing access to their \mfa{} factor. See Figure~\ref{fig_studyflow} for a summary of all steps in our methodology.

\usetikzlibrary{shapes,arrows,fit,matrix,positioning}

\begin{figure}[tb]
    \centering
    \footnotesize

    \begin{tikzpicture}[
        auto,
        block_main/.style ={rounded corners, draw={black}, fill=none, text width=0.9\columnwidth, text ragged, minimum height=3em, inner sep=3pt},
        block_dash/.style ={rounded corners, dashed, draw={black}, fill=none, text width=0.9\columnwidth, text ragged, minimum height=3em, inner sep=3pt},
        block_double/.style ={rounded corners, draw={black}, fill=none, text width=0.42\columnwidth, text ragged, minimum height=3em, inner sep=3pt},
        block_noborder/.style ={rounded corners, draw=none, fill=none, minimum height=1em, inner sep=0pt, text width=0.90\columnwidth, text centered},
        block_attach/.style ={rounded corners, draw=none, fill=none, inner sep=3pt, minimum width=0.9\columnwidth},
        line/.style ={draw, thick, -latex', shorten >=0pt, draw={black}}]
    ]
	\begin{scope}[node distance = 0.3cm]

        \node [block_main]
        (study1) {
            \textbf{1. MFA and Recovery Procedure Categorization}\\  
            \begin{enumerate}[leftmargin=0.5cm]
                \item Compile initial list of \var{PagesMFA} websites with \mfa{} support. 
                \item Identify \mfa{} and recovery procedures based on support and help pages.
                \item Select a sample of \var{RecAccTotal} websites for in-depth analysis.
            \end{enumerate} 
        };

       \node [block_double, below=2cm of study1.south west, anchor=west, text depth = 2.6cm,] (study2) {
            \parindent=0.35cm
            \textbf{2. In-Depth \mfa{} Recovery \\ 
            User Experience}\\
            \begin{enumerate}[leftmargin=0.5cm]
                \item Account creation and \mfa{} configuration on selected sample of \var{RecAccTotal} websites.
                \item Wait for one week.
                \item Go through \mfa{} recovery procedure.
                \item Debrief websites. 
            \end{enumerate} 
        };
        \node [block_double, right=of study2, text depth = 2.6cm,] (study3) {
            \parindent=0.35cm
            \textbf{3. Detailed Analysis of \\ Documented \mfa{} \\ Recovery Procedures}\\
            \begin{enumerate}[leftmargin=0.5cm]
                \item Collect \mfa{} documentation for selected sample of \var{RecAccTotal} websites.
                \item Perform in-depth analysis of documented \mfa{} recovery procedures and extend evaluation in step 1.
            \end{enumerate} 
        };
     
        \node [block_main, below=0.9cm of study2.south west, anchor=west] (compare) {
           Compare documented \mfa{} recovery procedures with our experiences in the in-depth recovery procedure evaluation in step 2.\\
        };
       \end{scope}

    \begin{scope}[every path/.style=line]

        \path ([xshift=2.9cm]study1.south) -- ([xshift=2.9cm]study1.south |- study2.north);
        \path ([xshift=-2.9cm]study1.south) -- ([xshift=-2.9cm]study1.south |- study3.north);
        \path ([xshift=0cm]study2.south) -- ([xshift=0cm]study2.south |- compare.north);
        \path ([xshift=0cm]study3.south) -- ([xshift=0cm]study3.south |- compare.north);
    
    \end{scope}
    
    \end{tikzpicture}
\caption{Summary of our methodology, consisting of the \mfa{} and recovery procedure categorization (cf. Section~\ref{sec_meth_deployment}) and a follow-up in-depth evaluation of deployed \mfa{} recovery procedures for \var{RecAccTotal} websites (cf. Section~\ref{sec_meth_recoverytests})}
   \label{fig_studyflow}
   
\end{figure}
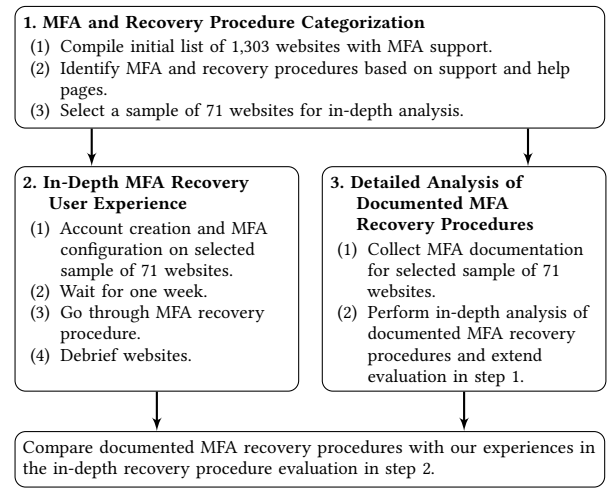

\subsection{Deployment of \mfa{} Recovery Procedures}
\label{sec_meth_deployment}
First, we conducted a systematic evaluation of \mfa{} recovery procedures deployed on \var{PagesMFA} websites\repref{} that use \mfa{}, based on the information provided on help and support pages.  
Overall, we investigate the impact of \mfa{} recovery procedures on authentication security and usability. 

\subsubsection{Methodology}
\label{sec_meth_deploymentmeth}
We aimed to simulate users who lost access to their \mfa{} factor and tried to find information on how to recover access to their accounts. While our overall focus was on the users' perspective on \mfa{} recovery procedures instead of a measurement of deployed \mfa{} methods on the web, we provide some context on the spread of \mfa{} methods and recovery procedures in the following.

Our dataset included all \var{PagesMFA} websites with \mfa{} deployments from \url{2fa.directory}~\cite{2fadirectory}, a high-quality crowdsourced collection of websites with information about \mfa{} deployment on the web. 
Most websites included on \url{2fa.directory} offer services to users, and typically collect user-connected information, \eg{}, addresses for online shopping, or financial information for online banking. 
The \url{2fa.directory} excludes self-hosted websites and websites that are not within the top 200,000 according to SimilarWeb~\cite{2faissue}. 
We verified the rank requirements by cross-checking our sample with the Tranco list~\cite{LePochat2019}\footnote{Based on \url{https://tranco-list.eu/list/82Q5V}.}, and found \var{PagesInTop200K} (\var{PercentInTop200K}\%) of the websites in our dataset within the top 200,000 websites, of which \var{TrancoCount_1} (\var{TrancoPercent_1}\%) were even within the top 20,000 \repref{}\extendedversion{(cf. Table~\ref{tab_tranco}in Appendix~\ref{sec_app_tabs})}.

Overall, we found \var{PagesTotal} websites on 2fa.directory. From these, we excluded \var{SkippedDuplicates} duplicates that were included in multiple categories, \var{SkippedBroken} unreachable websites. We further only examined websites that were either in English or German and excluded \var{SkippedLang} websites in other languages. Finally, \var{PagesNoMFA} pages did not offer \mfa{} and were therefore not relevant for our study. This gave us a total of \var{PagesMFA} websites. 
We focused on German and English websites, since all authors were either proficient in English or German. We excluded other languages since in some cases automated translations of help and support pages in other languages were ambiguous in the \mfa{} methods or recovery used, too vague, or inaccurate. While some websites provided official translations, they were not always well-maintained or even outdated. 
We further removed \var{PagesNoMFAExcluded} websites we could not reliably reach due to, \eg{}, geo-restrictions, or unavailable services.

To evaluate help and support pages, we used the support links provided in the \url{2fa.directory} database, and the websites' help and support sections that we could either find by manually going through the websites, or using Google with the following search term \texttt{"<website> two-factor authentication"}\footnote{\label{google-term} We found this search term to work well, as Google also detected various synonyms (such as multi-step verification) or alternative spellings (2-factor instead of two-factor). We complemented the Google search with our manual search.}. Within our Google search, we followed all links that were officially affiliated with the website in question (\ie{}, no third-party news websites) within the first page of Google results~\cite{googlectrstats, hembrooke2005google, hariri2011relevance}. We visited all of them and examined them for input regarding \mfa{} methods and recovery procedures.
We followed all links and relevant prompts, including "Trouble logging in?" or "Forgot your \mfa{}?" prompts. 
This did not include interactive processes to reset passwords, as we focused on lost \mfa{}.

Two researchers analyzed the help and support pages for documented \mfa{} recovery procedures using an inductive categorization approach~\cite{mayring2014qualitative}. 
We used a semi-open coding approach and an initial codebook including the five \mfa{} methods given on \url{2fa.directory} and several initial recovery procedures based on personal knowledge and previous work~\cite{kunke2021evaluation, rabkin2008personal}. 
We extended or updated the codebook accordingly whenever we identified a novel \mfa{} recovery procedure, and revisited previous codings. 
The full codebook and a more detailed description of \mfa{} methods and recovery procedures can be found in the replication package~\repref{}\extendedversion{in Table~\ref{tab_codebook_deployment}, and Section~\ref{sec_back} in the Appendix respectively}.

We resolved all coding conflicts as they came up, leading to a hypothetical agreement of 100\%. 
In line with common practice and previous high-quality qualitative research, we, therefore, chose to omit the calculation of an inter-rater reliability score~\cite{McDonald:2019:ReliabilityAndInterRaterReliability, mcdonald2021s, 255290, wermke2022committed}.

For the initial categorization, we coded for documented \mfa{} and recovery procedures. 
We used counts, means, and other aggregation measures to report our findings and generate descriptive statistics. 
During our evaluation, we made notes of every recovery procedure that stood out, either positively or negatively. This includes constraints on \mfa{}, \eg{}, if it was only provided for certain user groups or customers, interesting security-related statements, such as stressing that lost \mfa{} could not be recovered in any way, or interesting methods, such as receiving recovery codes via mail.
We used these notes as a basis for the detailed coding described in Section~\ref{sec_meth_recoverytests} and our best practice recommendations in Section~\ref{sec_discussion}.

\begin{table*}[t] 
    \caption{Summary of deployed \mfa{} methods and recovery procedures in our sample of \var{PagesMFA} websites. For example, we find that of \var{AllCountMethSMS} websites using SMS (\var{AllPercentMethSMS}\% of all websites),   \var{CountRecContactSupportIfMethSMS} (\var{PercentRecContactSupportIfMethSMS}\% of websites that offer SMS as main \mfa{}) offer support contact as a recovery procedure. Because some websites offered multiple \mfa{} and recovery procedures, not all percentages add to 100. We provide a more detailed description of these methods in our replication package\repref{}\extendedversion{and in Section~\ref{sec_back} in the Appendix}.}

    \label{tab_mfa-methods}
	\setlength{\tabcolsep}{2.7pt}
	\footnotesize
    \centering

    \aboverulesep=0ex 
    \belowrulesep=0ex 
    \definecolor{Gray}{gray}{0.93}
    
\begin{tabular}{lllp{5.8em}llllll}
\toprule

&  &  &  \cellcolor{white!} & & & & & & \\ 
\\[-2em]

\multicolumn{2}{c}{} &
\multicolumn{8}{c}{ \cellcolor{white!} \bfseries Deployed MFA Methods} \\[2pt]

 &  &  & \cellcolor{white!}  \textbf{All \mfa{}} & \textbf{SMS} & \textbf{Phone Call} &  \textbf{Email} &  \textbf{Hardware} & \textbf{Mobile App} & \textbf{Other \mfa{}}\\[1pt]

&  &  &  \cellcolor{white!} & & & & & & \\ 
\\[-2em]

& & 
\bfseries All Recovery & \var{PagesMFA} (100\%) &
\var{AllCountMethSMS} (\var{AllPercentMethSMS}\%) &
\var{AllCountMethCall} (\var{AllPercentMethCall}\%) &
\var{AllCountMethMail} (\var{AllPercentMethMail}\%) &
\var{AllCountMethHWToken} (\var{AllPercentMethHWToken}\%) &
\var{AllCountMethSWToken} (\var{AllPercentMethSWToken}\%) &
\var{AllCountMethOther} (\var{AllPercentMethOther}\%) \\

\midrule

\multirow{16}{*}{\rotatebox{90}{\bfseries  Deployed Recovery Procedures}} &
R01 &
Contact Website Support & 
\cellcolor{Gray} \var{AllCountRecContactSupport} (\var{AllPercentRecContactSupport}\%) &
\var{CountRecContactSupportIfMethSMS} (\var{PercentRecContactSupportIfMethSMS}\%) &
\var{CountRecContactSupportIfMethCall} (\var{PercentRecContactSupportIfMethCall}\%) &
\var{CountRecContactSupportIfMethMail} (\var{PercentRecContactSupportIfMethMail}\%) &
\var{CountRecContactSupportIfMethHWToken} (\var{PercentRecContactSupportIfMethHWToken}\%) &
\var{CountRecContactSupportIfMethSWToken} (\var{PercentRecContactSupportIfMethSWToken}\%)&
\var{CountRecContactSupportIfMethOther} (\var{PercentRecContactSupportIfMethOther}\%)\\

& R02 &
Backup Codes &  
\cellcolor{Gray} \var{AllCountRecCodes} (\var{AllPercentRecCodes}\%)  &
\var{CountRecCodesIfMethSMS} (\var{PercentRecCodesIfMethSMS}\%) &
\var{CountRecCodesIfMethCall} (\var{PercentRecCodesIfMethCall}\%) &
\var{CountRecCodesIfMethMail} (\var{PercentRecCodesIfMethMail}\%) &
\var{CountRecCodesIfMethHWToken} (\var{PercentRecCodesIfMethHWToken}\%) &
\var{CountRecCodesIfMethSWToken} (\var{PercentRecCodesIfMethSWToken}\%) &
\var{CountRecCodesIfMethOther} (\var{PercentRecCodesIfMethOther}\%) \\

& R03 &
Contact Local Admin & 
\cellcolor{Gray} \var{AllCountRecContactAdmin} (\var{AllPercentRecContactAdmin}\%) &
\var{CountRecContactAdminIfMethSMS} (\var{PercentRecContactAdminIfMethSMS}\%) &
\var{CountRecContactAdminIfMethCall} (\var{PercentRecContactAdminIfMethCall}\%) &
\var{CountRecContactAdminIfMethMail} (\var{PercentRecContactAdminIfMethMail}\%) &
\var{CountRecContactAdminIfMethHWToken} (\var{PercentRecContactAdminIfMethHWToken}\%) &
\var{CountRecContactAdminIfMethSWToken} (\var{PercentRecContactAdminIfMethSWToken}\%) &
\var{CountRecContactAdminIfMethOther} (\var{PercentRecContactAdminIfMethOther}\%) \\

& R04 &
Additional \mfa{} Method & 
\cellcolor{Gray} \var{AllCountRecAdd2FA} (\var{AllPercentRecAdd2FA}\%)  &
\var{CountRecAdd2FAIfMethSMS} (\var{PercentRecAdd2FAIfMethSMS}\%) &
\var{CountRecAdd2FAIfMethCall} (\var{PercentRecAdd2FAIfMethCall}\%) &
\var{CountRecAdd2FAIfMethMail} (\var{PercentRecAdd2FAIfMethMail}\%) &
\var{CountRecAdd2FAIfMethHWToken} (\var{PercentRecAdd2FAIfMethHWToken}\%) &
\var{CountRecAdd2FAIfMethSWToken} (\var{PercentRecAdd2FAIfMethSWToken}\%) &
\var{CountRecAdd2FAIfMethOther} (\var{PercentRecAdd2FAIfMethOther}\%) \\

& R05 &
Backup SMS/Phone Call & 
\cellcolor{Gray} \var{AllCountRecPhone} (\var{AllPercentRecPhone}\%)  &
\var{CountRecPhoneIfMethSMS} (\var{PercentRecPhoneIfMethSMS}\%) &
\var{CountRecPhoneIfMethCall} (\var{PercentRecPhoneIfMethCall}\%) &
\var{CountRecPhoneIfMethMail} (\var{PercentRecPhoneIfMethMail}\%) &
\var{CountRecPhoneIfMethHWToken} (\var{PercentRecPhoneIfMethHWToken}\%) &
\var{CountRecPhoneIfMethSWToken} (\var{PercentRecPhoneIfMethSWToken}\%) &
\var{CountRecPhoneIfMethOther} (\var{PercentRecPhoneIfMethOther}\%) \\

& R06 &
Backup Email & 
\cellcolor{Gray} \var{AllCountRecMail} (\var{AllPercentRecMail}\%)  &
\var{CountRecMailIfMethSMS} (\var{PercentRecMailIfMethSMS}\%) &
\var{CountRecMailIfMethCall} (\var{PercentRecMailIfMethCall}\%) &
\var{CountRecMailIfMethMail} (\var{PercentRecMailIfMethMail}\%) &
\var{CountRecMailIfMethHWToken} (\var{PercentRecMailIfMethHWToken}\%) &
\var{CountRecMailIfMethSWToken} (\var{PercentRecMailIfMethSWToken}\%) &
\var{CountRecMailIfMethOther} (\var{PercentRecMailIfMethOther}\%) \\

& R07 &
Dedicated Account Recovery System & 
\cellcolor{Gray} \var{AllCountRecRecoveryService} (\var{AllPercentRecRecoveryService}\%) &
\var{CountRecRecoveryServiceIfMethSMS} (\var{PercentRecRecoveryServiceIfMethSMS}\%) &
\var{CountRecRecoveryServiceIfMethCall} (\var{PercentRecRecoveryServiceIfMethCall}\%) &
\var{CountRecRecoveryServiceIfMethMail} (\var{PercentRecRecoveryServiceIfMethMail}\%) &
\var{CountRecRecoveryServiceIfMethHWToken} (\var{PercentRecRecoveryServiceIfMethHWToken}\%) &
\var{CountRecRecoveryServiceIfMethSWToken} (\var{PercentRecRecoveryServiceIfMethSWToken}\%) &
\var{CountRecRecoveryServiceIfMethOther} (\var{PercentRecRecoveryServiceIfMethOther}\%) \\

& R08 &
TOTP Seed & 
\cellcolor{Gray} \var{AllCountRecSecret} (\var{AllPercentRecSecret}\%)  &
\var{CountRecSecretIfMethSMS} (\var{PercentRecSecretIfMethSMS}\%) &
\var{CountRecSecretIfMethCall} (\var{PercentRecSecretIfMethCall}\%) &
\var{CountRecSecretIfMethMail} (\var{PercentRecSecretIfMethMail}\%) &
\var{CountRecSecretIfMethHWToken} (\var{PercentRecSecretIfMethHWToken}\%) &
\var{CountRecSecretIfMethSWToken} (\var{PercentRecSecretIfMethSWToken}\%) &
\var{CountRecSecretIfMethOther} (\var{PercentRecSecretIfMethOther}\%) \\

& R09 &
Trusted Device & 
\cellcolor{Gray} \var{AllCountRecTrustedDevice} (\var{AllPercentRecTrustedDevice}\%) &
\var{CountRecTrustedDeviceIfMethSMS} (\var{PercentRecTrustedDeviceIfMethSMS}\%) &
\var{CountRecTrustedDeviceIfMethCall} (\var{PercentRecTrustedDeviceIfMethCall}\%) &
\var{CountRecTrustedDeviceIfMethMail} (\var{PercentRecTrustedDeviceIfMethMail}\%) &
\var{CountRecTrustedDeviceIfMethHWToken} (\var{PercentRecTrustedDeviceIfMethHWToken}\%) &
\var{CountRecTrustedDeviceIfMethSWToken} (\var{PercentRecTrustedDeviceIfMethSWToken}\%) &
\var{CountRecTrustedDeviceIfMethOther} (\var{PercentRecTrustedDeviceIfMethOther}\%) \\

& R10 &
Photo/Official ID Proof & 
\cellcolor{Gray} \var{AllCountRecPhoto} (\var{AllPercentRecPhoto}\%) &
\var{CountRecPhotoIfMethSMS} (\var{PercentRecPhotoIfMethSMS}\%) &
\var{CountRecPhotoIfMethCall} (\var{PercentRecPhotoIfMethCall}\%)&
\var{CountRecPhotoIfMethMail} (\var{PercentRecPhotoIfMethMail}\%) &
\var{CountRecPhotoIfMethHWToken} (\var{PercentRecPhotoIfMethHWToken}\%) &
\var{CountRecPhotoIfMethSWToken} (\var{PercentRecPhotoIfMethSWToken}\%) &
\var{CountRecPhotoIfMethOther} (\var{PercentRecPhotoIfMethOther}\%) \\

& R11 &
MFA Not Needed for Login & 
\cellcolor{Gray} \var{AllCountRecAccessNotLost} (\var{AllPercentRecAccessNotLost}\%)  &
\var{CountRecAccessNotLostIfMethSMS} (\var{PercentRecAccessNotLostIfMethSMS}\%) &
\var{CountRecAccessNotLostIfMethCall} (\var{PercentRecAccessNotLostIfMethCall}\%) &
\var{CountRecAccessNotLostIfMethMail} (\var{PercentRecAccessNotLostIfMethMail}\%) &
\var{CountRecAccessNotLostIfMethHWToken} (\var{PercentRecAccessNotLostIfMethHWToken}\%) &
\var{CountRecAccessNotLostIfMethSWToken} (\var{PercentRecAccessNotLostIfMethSWToken}\%) &
\var{CountRecAccessNotLostIfMethOther} (\var{PercentRecAccessNotLostIfMethOther}\%) \\

& R12 &
Security Questions & 
\cellcolor{Gray} \var{AllCountRecSecQuestion} (\var{AllPercentRecSecQuestion}\%)  &
\var{CountRecSecQuestionIfMethSMS} (\var{PercentRecSecQuestionIfMethSMS}\%) &
\var{CountRecSecQuestionIfMethCall} (\var{PercentRecSecQuestionIfMethCall}\%) &
\var{CountRecSecQuestionIfMethMail} (\var{PercentRecSecQuestionIfMethMail}\%) &
\var{CountRecSecQuestionIfMethHWToken} (\var{PercentRecSecQuestionIfMethHWToken}\%) &
\var{CountRecSecQuestionIfMethSWToken} (\var{PercentRecSecQuestionIfMethSWToken}\%) &
\var{CountRecSecQuestionIfMethOther} (\var{PercentRecSecQuestionIfMethOther}\%) \\

& R13 &
Password Reset & 
\cellcolor{Gray} \var{AllCountRecPWReset} (\var{AllPercentRecPWReset}\%)  &
\var{CountRecPWResetIfMethSMS} (\var{PercentRecPWResetIfMethSMS}\%) &
\var{CountRecPWResetIfMethCall} (\var{PercentRecPWResetIfMethCall}\%) &
\var{CountRecPWResetIfMethMail} (\var{PercentRecPWResetIfMethMail}\%) & 
\var{CountRecPWResetIfMethHWToken} (\var{PercentRecPWResetIfMethHWToken}\%) &
\var{CountRecPWResetIfMethSWToken} (\var{PercentRecPWResetIfMethSWToken}\%) &
\var{CountRecPWResetIfMethOther} (\var{PercentRecPWResetIfMethOther}\%) \\

& R14 &
Other Recovery & 
\cellcolor{Gray} \var{AllCountRecOther} (\var{AllPercentRecOther}\%) &
\var{CountRecOtherIfMethSMS} (\var{PercentRecOtherIfMethSMS}\%) &
\var{CountRecOtherIfMethCall} (\var{PercentRecOtherIfMethCall}\%) &
\var{CountRecOtherIfMethMail} (\var{PercentRecOtherIfMethMail}\%) &
\var{CountRecOtherIfMethHWToken} (\var{PercentRecOtherIfMethHWToken}\%) &
\var{CountRecOtherIfMethSWToken} (\var{PercentRecOtherIfMethSWToken}\%) &
\var{CountRecOtherIfMethOther} (\var{PercentRecOtherIfMethOther}\%) \\

 \cmidrule{2-10}
& R15 &
Help Page Not Accessible & 
\cellcolor{Gray} \var{AllCountRecInaccesible} (\var{AllPercentRecInaccesible}\%) &
\var{CountRecInaccesibleIfMethSMS} (\var{PercentRecInaccesibleIfMethSMS}\%) &
\var{CountRecInaccesibleIfMethCall} (\var{PercentRecInaccesibleIfMethCall}\%) &
\var{CountRecInaccesibleIfMethMail} (\var{PercentRecInaccesibleIfMethMail}\%) &
\var{CountRecInaccesibleIfMethHWToken} (\var{PercentRecInaccesibleIfMethHWToken}\%) &
\var{CountRecInaccesibleIfMethSWToken} (\var{PercentRecInaccesibleIfMethSWToken}\%) &
\var{CountRecInaccesibleIfMethOther} (\var{PercentRecInaccesibleIfMethOther}\%)  \\

& R16 &
No \mfa{} Recovery Available & 
\cellcolor{Gray} \var{AllCountRecNone} (\var{AllPercentRecNone}\%)  &
\var{CountRecNoneIfMethSMS} (\var{PercentRecNoneIfMethSMS}\%) &
\var{CountRecNoneIfMethCall} (\var{PercentRecNoneIfMethCall}\%) &
\var{CountRecNoneIfMethMail} (\var{PercentRecNoneIfMethMail}\%) &
\var{CountRecNoneIfMethHWToken} (\var{PercentRecNoneIfMethHWToken}\%) &
\var{CountRecNoneIfMethSWToken} (\var{PercentRecNoneIfMethSWToken}\%) &
\var{CountRecNoneIfMethOther} (\var{PercentRecNoneIfMethOther}\%) \\

& R17 &
No \mfa{} Recovery Documented & 
\cellcolor{Gray} \var{AllCountRecNoMention} (\var{AllPercentRecNoMention}\%) &
\var{CountRecNoMentionIfMethSMS} (\var{PercentRecNoMentionIfMethSMS}\%) &
\var{CountRecNoMentionIfMethCall} (\var{PercentRecNoMentionIfMethCall}\%) &
\var{CountRecNoMentionIfMethMail} (\var{PercentRecNoMentionIfMethMail}\%) &
\var{CountRecNoMentionIfMethHWToken} (\var{PercentRecNoMentionIfMethHWToken}\%) &
\var{CountRecNoMentionIfMethSWToken} (\var{PercentRecNoMentionIfMethSWToken}\%) &
\var{CountRecNoMentionIfMethOther} (\var{PercentRecNoMentionIfMethOther}\%) \\

\bottomrule
\end{tabular}
\end{table*}

\subsubsection{Results}
\label{sec_meth_deploymentresults}
Overall, mobile applications were the most popular \mfa{} method, deployed on \var{AllCountMethSWToken} pages (\var{AllPercentMethSWToken}\%), mostly as TOTP generator apps. Second most popular were SMS (\var{AllCountMethSMS}, \var{AllPercentMethSMS}\%).  
All other methods, \ie{}, phone calls, email, hardware tokens, and various other methods were noticeably less widespread and present in between \var{AllPercentMethOther}\%--\var{AllPercentMethMail}\% (\var{AllCountMethOther}--\var{AllCountMethMail}) of our sample.
On average, websites offered \var{AllMethMean} (median: \var{AllMethMedian}) \mfa{} methods. We provide details for \mfa{} methods and their recovery procedures in Table~\ref{tab_mfa-methods}. 

Websites offered an average of \var{AllRecMean} (median: \var{AllRecMedian}) different recovery procedures. Most websites used direct support contacts (R01) (\var{AllCountRecContactSupport}; \var{AllPercentRecContactSupport}\%) and backup codes (R02) (\var{AllCountRecCodes}; \var{AllPercentRecCodes}\%). However, \var{AllCountRecNoMention} (\var{AllPercentRecNoMention}\%) websites that deployed \mfa{} did not provide publicly accessible information about \mfa{} recovery procedures at all (R17). While some websites might provide \mfa{} recovery information behind a login, we argue that users locked out due to \mfa{} loss would benefit greatly from public recovery information. 

While \var{AllCountMethHWToken} (\var{AllPercentMethHWToken}\%) websites offered \mfa{} using hardware tokens, no website in our sample offered them as a recovery procedure. 
In most cases, users could only use hardware tokens with another \mfa{} method -- a requirement we did not find for other \mfa{} options. 
While websites did not justify this decision, we suspect known usability issues of hardware tokens and a perceived higher likelihood of becoming inaccessible~\cite{reynolds2018tale, ciolino2019two, farke2020you, marky2022nah}.

We compared \mfa{} recovery procedures between website categories provided by the \url{2fa.directory} categories, \eg{}, \textit{Banking}, or \textit{University} websites.
While most recovery procedures were uniformly distributed across website categories, some recovery procedures were more common in certain categories. Providing selfies with handwritten notes and governmental IDs (R10) almost exclusively appeared on \textit{Cryptocurrency} websites. \textit{Universities} were more likely to rely on local administrators (R03) for \mfa{} recovery. \extendedversion{Figure~\ref{fig_recPresenceCategories} in the Appendix illustrates the distribution of \mfa{} recovery procedures across website categories in our dataset.} 

\begin{summaryBox}{Key Points | \mfa{} Recovery Procedure Analysis}{}
More than 80\% of the websites we analyzed deployed mobile apps and SMS for \mfa{}.
Most offered one \mfa{} option, and one recovery procedure; 
Most websites used backup codes or direct support contact as their recovery procedure. 
A quarter of websites did not provide public \mfa{} recovery information. 
\end{summaryBox}

\subsection{Recovery User Experience}
\label{sec_meth_recoverytests}
The previous experiment sheds light on the deployment frequencies of different \mfa{} recovery procedures, including recovery instructions for \var{PagesMFA} websites supporting \mfa{}.

In this section, we investigate the user experience of deployed \mfa{} recovery procedures.

Therefore, we selected a subset of \var{RecAccTotal} websites, created accounts, and performed the entire \mfa{} recovery procedure process. 
We focused on authentication security and usability and how well the implemented \mfa{} recovery procedures were documented.

\subsubsection{Methodology}
\label{sec_meth_recoverytestsmeth}
Since we were further interested in detailed but also diverse insights into \mfa{} recovery procedures that cannot be achieved from collecting help and support pages alone, we selected a subset of \var{RecAccTotal} websites for an in-depth analysis of the user experience. Similar to previous work~\cite{Oltrogge:2021, lee2022password, snyder2017most, neil2021investigating, gilsenan2023security}, we chose to comprise our sample from different sources, including \var{RecAccTop} of the highest ranking, \var{RecAccRand} random, and \var{RecAccPick} hand-picked websites. 
We chose the \var{RecAccPick} handpicked websites based on interesting edge cases in our larger dataset. This included \var{HandPickedSecurity} websites that offered security products such as SSL certificates or antivirus programs, and \var{HandPickedInteresting} websites that mentioned unusual recovery procedures. 
Examples of unusual recovery procedures were preventive measures such as enforcing downloads of recovery codes, offering users to store various types of personal data which would be later used to recover the account, or multiple tests of the availability of their chosen recovery. 
Similarly, we chose some websites based on their recovery procedures, including a website that involved team members as witnesses to vouch for or against a recovery decision or enforcing fixed waiting times before \mfa{} would be reset. 
Finally, we chose \var{HandPickedStrict} websites that explicitly mentioned that the account would be lost if users lose their \mfa{}, including advice to buy multiple YubiKeys, and \var{HandPickedUnclear} websites with unclear or no instructions to see what would happen in these cases.

From an initial qualitative sample of several websites, sampled after our three criteria above, we excluded websites that required pre-existing contract or group memberships (\var{RecSkipReq}), or data we could not or did not want to provide to protect the privacy of the researchers conducting this study, \eg{}, credit or ID card data (\var{RecSkipData}). We further excluded websites with terms of service that prohibited our use case (\var{RecSkipTOS}).
Other reasons included that the website shared its account with another website already in our sample (\var{RecSkipDone}), that we were blocked on due to unspecified reasons (\var{RecSkipBlock}), or websites whose free versions were only time-limited trials (\var{RecSkipTrial}). 
Finally, \var{RecSkipOff} domains forwarded us to other websites or were offline since the \studyone{}. An anonymized overview of excluded websites is given in our replication package \repref{}\extendedversion{and in Table~\ref{tab_skippedwebsites} in the Appendix~\ref{sec_app_tabs}}.
Whenever we could not create accounts, we replaced a website with the next best alternative where possible-- \eg{}, if a top-ranked website required us to have a contract with them to acquire an account such as a bank, we replaced it with the next most popular website not yet included. 
We ended up with a sample of \var{RecAccTotal} websites, on which we tested \mfa{} recovery.
Despite the exclusion of websites, we feel confident that our in-depth analysis of the actual \mfa{} recovery process on \var{RecAccTotal} websites provides deep and diverse insights valuable to the research community. Additionally, the distribution of offered \mfa{} methods in our final sample is comparable to the overall distribution on all \var{PagesMFA} websites.

We created accounts on all \var{RecAccTotal} websites using free account plans, enabled \mfa{}, and triggered the respective \mfa{} recovery procedure. 
Using screencasts, we made video recordings of the entire procedure to compare it with the official documentation during later evaluation. 
To receive email and text messages for account creation, \mfa{} setup, or \mfa{} recovery, we used a Google email address created for the study and an eSIM with a US-based phone number. 

During the \mfa{} setup, we simulated a user who follows instructions but only fulfilled the minimum requirements, following findings from previous work~\cite{xiao2014social,wash2016understanding, herley2009so}. During the process, we noted all steps, whether we regained access, and how long recovery took. We also captured security and usability details. 
For each website, we followed the protocol below: 
\begin{enumerate}
    \item We enabled the first \mfa{} method provided by a website. 
    We set specific or additional \mfa{} and recovery procedures if the website gave explicit instructions to do so. 
    This decision was further driven by the assumption that the first listed \mfa{} method and recovery procedure is preferred by a website and likely best supported.
    \item After one week had passed, we revisited the websites. 
    This decreased suspicion because we did not immediately lose access after creating the accounts.
    \item We tried to log in without our supposedly lost \mfa{} and recovery procedures. 
    If the respective website allowed account recovery for lost or inaccessible \mfa{} devices, we followed it without using our configured recovery procedure (\ie{}, we did not use our backup codes to regain access), to simulate the experience of a user who has lost \mfa{} completely. 
    \item If a website did not provide an automated emergency solution for login without \mfa{} or the recovery procedure we configured, we contacted the website's support directly. In our initial request\repref{}\extendedversion{(cf. Appendix~\ref{sec_app_texts})}, we said that we recently lost a backpack containing our phone and wallet, and found ourselves without access to the configured \mfa{} factor.
    \item If there was no way to contact the website (\eg{}, because support channels were only available for paying account plans), we stopped, marked the account as lost, and continued with the next website. 
\end{enumerate}

For an in-depth analysis of \mfa{} recovery procedure experience and comparison with their documentation, we qualitatively assessed the help and support pages of the \var{RecAccTotal} websites in our data set. We collected all relevant help and support pages for all websites: 
\begin{itemize}
    \item We performed a Google search for \texttt{"<website> two-factor authentication"}\textsuperscript{\ref{google-term}} similar to the one for our \studyone{}.
    \item We collected the help and support sections linked in the \url{2fa.directory} database.
    \item We manually navigated websites, starting with their main pages and using their search functions if available.
\end{itemize}

Following the steps above, we saved screenshots of help pages and expanded our codebook from the initial categorization \repref{}\extendedversion{(see Table~\ref{tab_codebook_deployment})}. We added details such as whether a website explicitly warned their users of account loss in case of losing access to their \mfa{}, if \mfa{} was required in general, and how often it needed to be entered, or whether users were required to set a recovery procedure \repref{}\extendedversion{(see Table~\ref{tab_codebook_detail} for the codebook expansion)}. We coded both the \mfa{} account setup processes and the information provided on the help pages. We compared them to identify potential shortcomings or contradictions in how providers communicate \mfa{} to their users.
Our institution's ethical review board (ERB) approved the study, and we took measures to protect the websites' identities, their resources, and the time of their staff (cf. Section~\ref{sec_ethics}).

\subsubsection{Results}
\label{sec_meth_recoverytestsresults}

Below, we provide details for our \mfa{} recovery procedures evaluation. Furthermore, we compare our user experience of \mfa{} recovery procedures with their official \mfa{} recovery documentation.

\boldparagraph{Account Recovery Success Rates}
Overall, we created \var{RecAccTotal} accounts in August and September 2022. We regained access to \var{RecAccSuccessCount} (\var{RecAccSuccessPercent}\%), and lost access to \var{RecAccLostCount} (\var{RecAccLostPercent}\%). 
As of the time of submission, we did not receive an answer from \var{RecAccWaitingCount} (\var{RecAccWaitingPercent}\%) service providers. 
Figure~\ref{fig_sankeyMFARecovery} in the Appendix connects our website sample to \mfa{} methods and recovery success.
We configured \textit{mobile apps} (\var{MFAUsedTOTPCount}, \var{MFAUsedTOTPPercent}\%), \textit{SMS} (\var{MFAUsedSMSCount}, \var{MFAUsedSMSPercent}\%), \textit{email} (\var{MFAUsedEmailCount}, \var{MFAUsedEmailPercent}\%) and finally \textit{hardware token} (\var{MFAUsedHWCount}, \var{MFAUsedHWPercent}\%) as primary \mfa{} methods. 
We found that websites that deployed SMS-based \mfa{} had slightly worse recovery success rates than TOTP or other app-based approaches.
As we retained access to our email inbox during our study, \ie{}, we never lost access to the main \mfa{} method on the respective website, email had a perfect success rate. We decided not to further bias our results due to a change in methodology by treating websites with email as the main \mfa{} differently from those that deployed other \mfa{} methods. 

\begin{figure}[tb]
     \centering
         \centering
         \includegraphics[width=0.9\columnwidth]{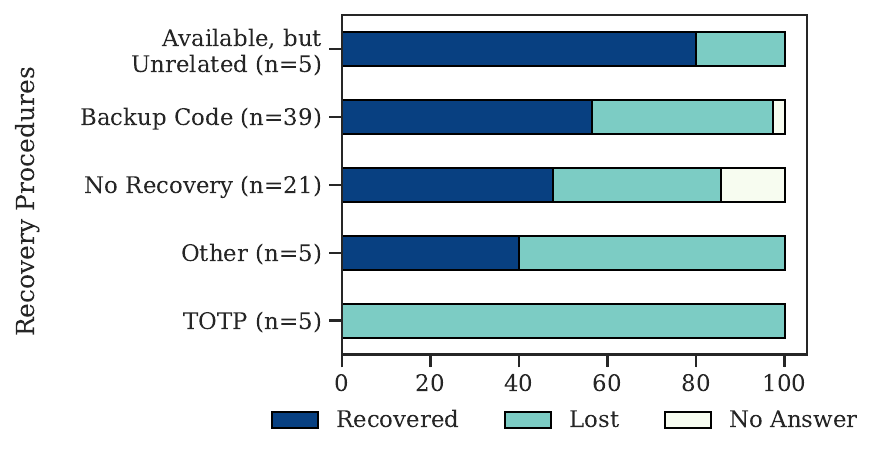}
    \caption{Overview of the recovery success rates per recovery procedure in percent per method. \textit{Other} encompasses two phone-based recoveries and contact support. \textit{TOTP} also includes the TOTP secrets. Numbers do not add up to \var{RecAccTotal} due to websites offering multiple recovery procedures.}
        \label{fig_recoveryrates}
\end{figure}

Overall, we set up at least one \mfa{} recovery procedure on \var{RecoverySetCount} (\var{RecoverySetPercent}\%) websites. 
However, \var{RecoveryNotSetCount} (\var{RecoveryNotSetPercent}\%) either did not offer any \mfa{} recovery option or did not prompt us to configure a recovery option during \mfa{} setup. 
The recovery rates for all \mfa{} recovery procedures are illustrated in Figure~\ref{fig_recoveryrates}. 

We configured and stored \textit{backup codes (R02)} for \var{RecUsedBackupCount} websites.
Only \var{CommBackupCodeForcedCount} (\var{CommBackupCodeForcedPercent}\%) of those websites tried to verify backup code availability during setup, \eg{}, by asking for the code or confirming that we stored the code.
We simulated having lost access to our backup codes for the recovery requests.
We could regain access to only slightly more than half (\var{RecBackupCodeSuccessCount}, \var{RecBackupCodeRegainedPercent}\%), and lost \var{RecBackupCodeLostCount} accounts (\var{RecBackupCodeLostPercent}\%). The remaining \var{RecBackupCodeWaitingCount} website did not reply to our recovery request.

We were unable to regain access to any account that used \textit{TOTP apps or their seeds} for \mfa{} recovery (\var{RecTOTPLostCount}, \var{RecTOTPLostPercent}\%). 
On \var{RecUsedIndirectCount} websites, we tried to recover our account using \textit{available, but unrelated} procedures, \ie{}, alternative contact methods such as backup email (R06) or phone numbers (R05) we set up during account creation that were never officially dedicated as recovery or mentioned during \mfa{} setup. 
Despite the vague communication, this led to indeed \var{RecIndirectSuccessCount} (\var{RecIndirectRegainedPercent}\%) accounts being recovered from \mfa{} loss. 
For \textit{other} recovery procedures, mostly consisting of phone-based (R05) recovery or the direct mention of support channels (R01) as a backup solution, we regained access to \var{RecOtherSuccessCount} (\var{RecOtherRegainedPercent}\%) accounts. 
Finally, we did \textit{not set up} (R16) \mfa{} recovery procedures on \var{RecUsedNoneCount} websites, as this was not part of the \mfa{} setup process. Despite the lack of a recovery procedure, we could regain \var{RecNoRecoverySuccessCount} (\var{RecNoRecoveryRegainedPercent}\%) accounts. 
Overall, we configured \mfa{} recovery for \var{RecSetBackupCount} accounts. We successfully regained access to \var{RecSetBackupSuccessCount} (\var{RecSetBackupSuccessPercent}\%), and lost access to \var{RecSetBackupLostCount} (\var{RecSetBackupLostPercent}\%) accounts. We received no answer from \var{RecSetBackupWaitingCount} (\var{RecSetBackupWaitingPercent}\%).

\begin{figure}[tb]
    \centering
    \includegraphics[width=0.8\columnwidth]{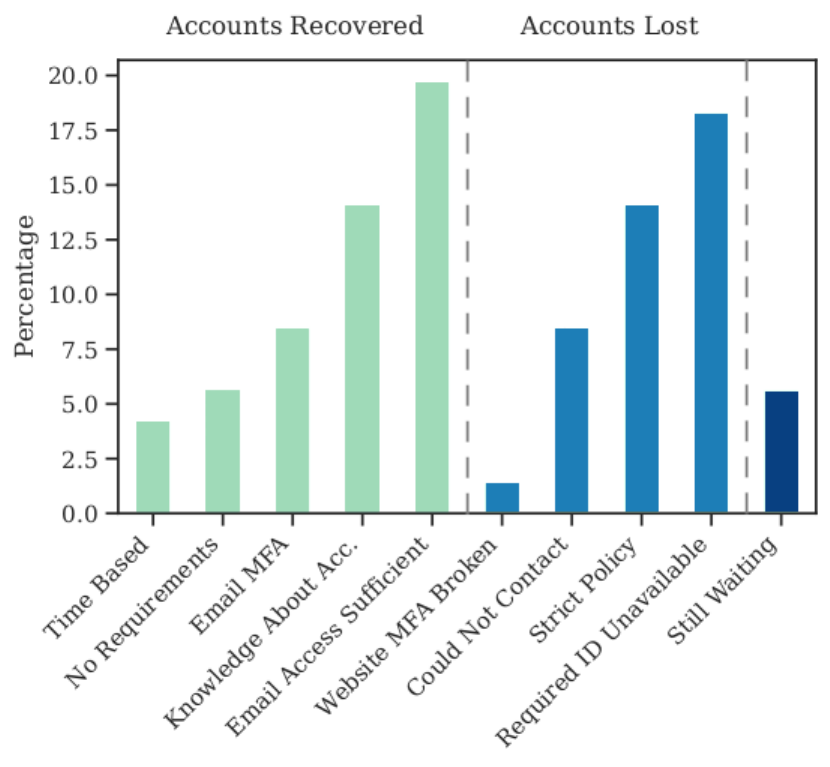}
    \caption{Summary of account recovery and loss split by root cause. We distinguished between recovery due to email as the main \mfa{} method, and email as the main recovery communication channel due to which access is regained.}
    \label{fig_detailedreason}
\end{figure}

\boldparagraph{Account Recovery User Experience}
During our user experience, we found \mfa{} recovery implementations to differ vastly, \eg{}, websites offering different \mfa{} or recovery procedures, or some including the recovery procedure configuration in \mfa{} setup while others do not. We provide details on the reasons for accepting or denying manual account recovery, finding similarly inconsistent and contradictory recovery philosophies, \eg{}, some websites argued that because our account was barely used, there was no harm in allowing us back in, while others argued that this meant there was no harm in us losing access and creating a new account. Our findings hereof are described in the following paragraphs and illustrated in Figure~\ref{fig_detailedreason}.

Among the \var{RecAccLostCount} accounts to which we did not regain access,  \var{RecDetailCouldNotContactCount} (\var{RecDetailCouldNotContactPercent}\%) were because we \textit{could not contact} the website's administrator or customer support team. Reasons included that on some websites, our free account plan was not eligible to receive any support, leaving us, and any user with a free account, no way to reach out and ask for help. Additionally, some websites required us to use contact forms for which we did not have all the necessary information, \eg{}, credit card information or banking transactions, IMEI phone identifiers, or information about the specific program from the vendor's offer we were using, the information we were not asked to provide or set up during account creation. Another website required us to provide our location and an order number, as the contact form assumed that only paying customers with questions or problems regarding their orders would reach out. Finally, one website had removed their contact information from their website, arguing that due to a high request volume, only specific user groups were allowed to contact them. We refrained from reaching out with a dishonest request.

Of the websites we were able to contact, \var{RecDetailStrictPolicyCount} (\var{RecDetailStrictPolicyPercent}\%) had \textit{strict policies} regarding recovery and would not allow access without the configured \mfa{} or recovery procedure. Finally, \var{RecDetailRequiredIDUnavailableCount} (\var{RecDetailRequiredIDUnavailablePercent}\%) websites required us to provide various sensitive and \textit{identifying documents}, such as governmental ID cards or bank transaction receipts; when we refused, we lost access to these accounts. 
However, we successfully recovered \var{RecDetailEmailMFACount} (\var{RecDetailEmailMFAPercent}\%) accounts due to being able to access our primary \textit{email \mfa{}} method and \var{RecDetailKnowledgeAboutAcc.Count} (\var{RecDetailKnowledgeAboutAcc.Percent}\%) due to being able to provide specific \textit{knowledge about the account} during the recovery procedure.
This encompassed metadata or contextual information, such as our organization name, our address, the account creation date, typical login locations as well as our ISP or IP. One website asked us a series of questions about our typical behavior and settings, many of which were potentially public knowledge, such as the date of our last broadcast or other connected platforms. We recovered one account by answering security questions that had been part of the account creation process. Three websites asked about potentially sensitive, personal, or identifying information such as, \eg{}, recent orders or payment information---which we could not provide as we never completed a transaction on the website---or use of an SSH shell or GitHub recovery---which we had never connected---but several websites allowed us to regain access despite failing these requirements.

For \var{RecDetailNoRequirementsCount} (\var{RecDetailNoRequirementsPercent}\%) accounts, we regained access immediately after our initial request with no further questions asked and \textit{no requirements} or needing to go through any recovery process. In \var{RecDetailTimeBasedCount} (\var{RecDetailTimeBasedPercent}\%) cases, the recovery was \textit{time based}. Here, \mfa{} was disabled after a waiting period of 3--14 days. In these cases, our primary email account received multiple reminders and warnings, designed to give the legitimate account owners time to interrupt the recovery in case a malicious or unknown third party made the recovery request.

Overall, we found that most of our requests to regain access were fulfilled after just a few emails from our primary email account and that we rarely required more than contextual knowledge about either the account or us as the main user. 
Our findings suggest that the security of many deployed \mfa{} recovery procedures is similar to that of the email account connected to the \mfa{} we wished to recover, or that of security questions, as many requested data points were often easily guessable or public knowledge such as connected platforms or our companies address, or accessible via our email inbox such as account creation dates or ISPs.

\boldparagraph{Correspondence with Website Support Teams}

The majority of accounts, to which we regained access, it was \textit{sufficient to have email access} (\var{AccessDueToEmailCount}, \var{AccessDueToEmailPercent}\%); this encompasses accounts that used email as their main \mfa{} or recovery procedure, as well as accounts that asked us to confirm our request by email to disable \mfa{} and regain access. While only confirmed by one website explicitly, we assume this was required as a form of phishing protection. 

In two cases, websites typically used for commercial purposes tried to ask our account owner to allow them to disable \mfa{} for us. As they assumed a hierarchy within the account -- \eg{}, a dedicated main user, admin or owner, and several team members--but found us to be the only user, we received messages to the same email both as a user who applied for a \mfa{} reset, and as account owner that can allow or deny this request. Hence, we were able to allow ourselves back into the account. While this might be an edge case of their recovery procedure, it might lead to a decreased security of important administrator roles in comparison to normal users, as this cancels the extra obstacle of requiring admin approval. 

Overall, we set up a recovery procedure during the account and \mfa{} setup for \var{RecoverySetCount} (\var{RecoverySetPercent}\%) websites. When we initially applied for account recovery\repref{} \extendedversion{(cf. Appendix~\ref{sec_app_texts} for the message)}, we did not mention these recovery procedures, but only stressed that we lost our wallet and phone. While this would render most pre-set recoveries, such as backup codes, still usable (\var{RecStoryIndependentCount}, \var{RecStoryIndependentPercent}\%), only a minority of \var{RecSuggestedYesCount} (\var{RecSuggestedYesPercent}\%) websites mentioned using the pre-set recovery procedure, with one of them disabling our \mfa{} without further requirements after we responded that we lost the backup codes as well. \var{RecDetailStrictPolicyCount} (\var{RecDetailStrictPolicyPercent}\%) websites insisted on only allowing us access if we used either the original \mfa{} or the recovery procedure we set up.
Only a few websites enforced and verified the configuration of a \mfa{} recovery procedure. During our experiment, \var{RecForcedExpCount} (\var{RecForcedExpPercent}\%) websites required us to set up and test the recovery procedure before being allowed to complete the \mfa{} setup or required a manual confirmation that backup codes were stored before finalizing the \mfa{} configuration. 
\boldparagraph{Documented \mfa{} (Recovery) Procedures}
In this paragraph, we discuss our evaluation of how websites communicated the details of \mfa{} and recovery procedures with users, encompassing both our user experience during the account and \mfa{} setup process, as well as the publicly accessible help and support pages.

\begin{figure}[tb]
    \centering
    \includegraphics[width=0.9\columnwidth]{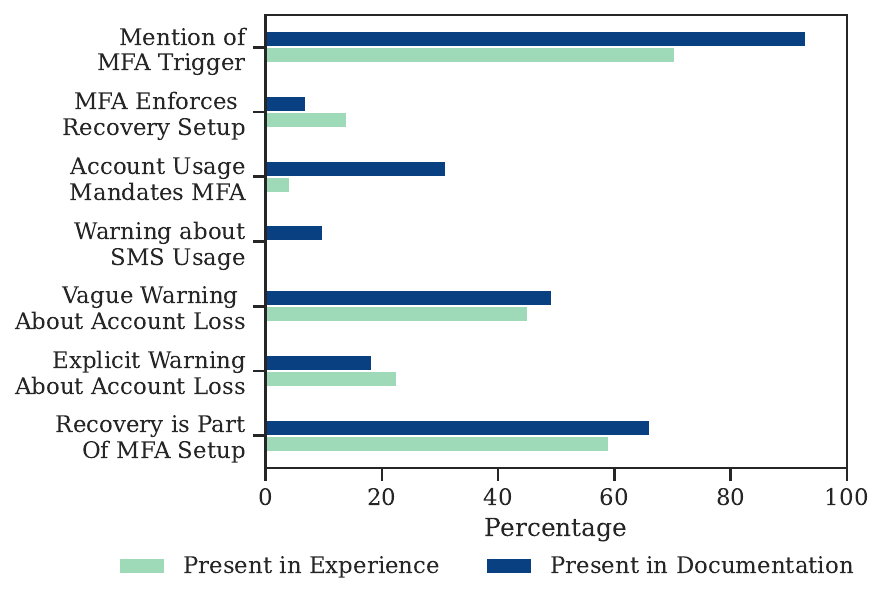}
    \caption{Percentage of occurrence of certain topics or hints within either our own experience when setting up accounts and \mfa{}, or the respective help and support pages. Note that not all services offer SMS, \ie{}, related topics are not necessarily applicable for all tested websites.}
    \label{fig_diff_expVSdoc}
\end{figure}

Building on our evaluation of help pages in our initial categorization (Section~\ref{sec_meth_deploymentmeth}), for our user experience, we revisited each of the \var{RecAccTotal} websites' help and support pages in a more in-depth investigation of their \mfa{} and recovery procedures and formed a basis for comparison with our user experience. 
We collected additional data about, \eg{}, whether the recovery setup was part of enabling \mfa{}, if users were warned about account loss in case of \mfa{} loss, or whether account restrictions or forced setup of recoveries were mentioned. We next discuss interesting findings and differences between the documentation and our own experience. Figure~\ref{fig_diff_expVSdoc} displays the frequency with which the topics in this section are present in either our experience or the official help and documentation pages.
Overall, our recovery experience never matched the documented recovery procedure perfectly. Even if only the offered \mfa{} methods are considered, we still only find \var{CommMethodsSameCount} (\var{CommMethodsSamePercent}\%) of all help and support pages to match our experience regarding available \mfa{} methods and recovery procedures. This difference suggests that recovery procedures are commonly not communicated properly during setup, or that documentations are outdated or incomplete.

Our detailed analysis also reveals that at most two-thirds of websites discussed recovery procedures during \mfa{} setup, slightly more frequently present in documentations (\var{RecSetupDocPercent}\%) than in (our) user experience (\var{RecSetupExpPercent}\%).

Although a perfectly secure \mfa{} implementation would include account loss for users unable to prove their identity, we found only \var{RecWarningsExpCount} (\var{RecWarningsExpPercent}\%) websites to \textit{explicitly warn us} about the potential danger of irrevocably losing access. This was even rarer within official documentation (\var{RecWarningsDocCount}, \var{RecWarningsDocPercent}\%). 
However, \var{RecSoftWarningDocCount} websites (\var{RecSoftWarningDocPercent}\%) included more \textit{vague warning} phrasings that mentioned account loss but refrained from portraying it as a guaranteed consequence of losing \mfa{} and its recovery. Overall, we find that most of these do not convey the urgency of keeping the \mfa{} secured, or do not even clearly state the connection between losing one's \mfa{} and therefore losing access to the account. 
In the following, we examine aspects from Figure~\ref{fig_diff_expVSdoc} in tandem with our recovery rates.

We further find that the warnings did not always portray the truth: \var{CommWarnedRegainedExpCount} (\var{CommWarnedRegainedExpPercent}\%) websites allowed us back into our accounts despite having warned us explicitly that they would not. On the other hand, we found that we lost \var{CommNotWarnedLostExpCount} (\var{CommNotWarnedLostExpPercent}\%) accounts without receiving any prior warning. As such, we emphasize that when warning users about account loss in case of \mfa{} loss, websites should also introduce users to the recovery process, particularly during \mfa{} setup.
We find that of the websites that do warn about account loss, only \var{CommWarnedSetupExpCount} (\var{CommWarnedSetupExpPercent}\%) adhere to this, suggesting that users are often left alone with the task of researching and enabling \mfa{} recovery procedures, despite the danger of account loss.

In our initial evaluation of deployed \mfa{} and recovery procedures, we found SMS to be a popular \mfa{} solution despite its known insecurities~\cite{siadati2017mind, WiredTwitterCEOHacked2019, lee2020simswap, jover2020security}. We were therefore interested in the number of websites that acknowledged these insecurities and \textit{warned about SMS} and found that while SMS was available for roughly half of all pages, websites rarely (\var{RecSMSWarningExpPercent}\% during user experience;\var{RecSMSWarningDocPercent}\% of documentations) admitted its insecurity.

Further, our analysis indicates that websites were often inconsistent regarding their communication of \textit{mandatory \mfa{} usage} or the \textit{enforcement of recovery procedures}. We find usage restrictions mentioned more commonly within official documentation (\var{RecRestrictedDocPercent}\%) than we encountered (\var{RecRestrictedExpPercent}\%). This is likely because websites most commonly mentioned requiring account administrators to enable \mfa{}, and the accounts we created might not have had this status. 
Regarding enforcing recovery setup, only a small portion of websites includes this (\var{RecForcedExpPercent}\% during user experience; \var{RecForcedDocPercent}\% of documentations). As websites cannot verify that users did, \eg{}, store their backup codes, we considered every mechanism that tried to only allow users to continue the process after the recovery was set, \eg{}, by using checkboxes in which users confirm that they stored the backup, asking users to use the backup as a proof, or setting up recovery before \mfa{} itself is set.

Websites often mention in which situations \mfa{} is triggered, \eg{}, any login, only logins from new locations or devices, or specific functionalities such as accessing security-relevant settings or withdrawing funds. However, we find the \textit{\mfa{} trigger} more commonly mentioned on help and support pages (\var{RecTriggerDocPercent}\%) than as a part of the \mfa{} setup (\var{RecTriggerExpPercent}\%), although, \eg{}, how commonly users are required to provide the information can be an important reason for or against enabling \mfa{}.

Finally, we examined the number of pages within the documentation, help sites, and FAQs that discussed \mfa{} and its recovery. In cases where a website offered multiple services, we included only those related to the specific account we set up, and argue that ideally, all information should be present on a few pages to help users find information quickly without getting lost within the documentation and to help website providers keeping all information current as there are fewer pages requiring updates. We find that \var{RecHelpNumDocCount} (\var{RecHelpNumDocPercent}\%) websites within our sample used more than three different pages to discuss all details of \mfa{} and its recovery, therefore potentially obstructing a user's search for help, and involuntarily burying information within too many pages. 

\boldparagraph{Sample Characteristics}
This study explored different \mfa{} recovery implementations that services deployed in an in-depth study of recovery procedures. 
While our sample is not generalizable, we provide an overview of the websites within our sample and the respective recovery results to contextualize our findings. An anonymized overview of websites in our sample can be found in our replication package\extendedversion{and in Table~\ref{tab_examinedwebsites} in the Appendix}\repref{}.
In general, it is difficult to quantify the value of accounts, as this is not only dependent on financial information linked to them, but also on social, physical, or sentimental value, and therefore subjective for each user. For example, while losing a social media account can be meaningless for some, it might be the base of income for others. Likewise, the precise threat model is dependent on how valuable users estimate their accounts to be, which results in a similar subjective outcome. However, we deem the value of an account to be highly relevant to users, as this has an impact on their security decisions and how they protect accounts.

To estimate the risk and damage for users when they lose access to an account, or when a malicious third party is wrongfully given access, we took notes on which types of data a website requires to function, including financial data (\eg{}, credit card or banking information), addresses (typically in the form of delivery or invoice addresses), or others such as telephone numbers. We find the majority (\var{RecFinancialReqCount}) of websites to require financial data to utilize the website, typically because its main purpose is related to investments, shopping, or money pooling, and while accounts can be created without providing payment data, they are unusable with regard to the website's purpose. Additionally, \var{RecFinancialSpecCount} websites do not require storing payment data, but offer \eg{}, voluntary subscriptions. In our recovery experiment, we were able to recover half of these websites (\var{RecFinancialReqPercent}\% for required data; \var{RecFinancialSpecPercent}\% for voluntary data), showing that the improper \mfa{} implementations we encountered indeed endangered valuable user data.
Similarly, a large portion requires (\var{RecAddressReqCount}) or optionally collects (\var{RecAddressSpecCount}) an address, and again was able to regain access to a significant number of websites (\var{RecAddressReqPercent}\% for required data; \var{RecAddressSpecPercent}\% for voluntary data).

Overall, most websites in our sample offer paid services, therefore collecting personal data. While financial data is typically not displayed in plain text (\ie{}, attackers cannot gain the credit card number as only the last digits are shown), it might still be abused by an attacker if malicious orders are placed using stored payment information. Other information, such as personal addresses or phone numbers, could be leaked and abused. In cases where the account is lost without a malicious party gaining access, users might lose access to products they legitimately bought, or lose content with sentimental value, such as social media they maintained for years, or photos and memories they shared. However, we found some websites to re-allow access in case an invoice or bank statement was provided, as this was accepted to identify the user. 

When regarding the different sample sources, \ie{}, whether the account was added to our sample because it was interesting, top-ranked, or randomly selected, we see almost no differences regarding the success of our recovery study. Overall, we successfully regained half of all samples, with the largest difference being that we received any kind of answer or reaction from all top-ranked accounts, therefore losing more accounts than in other sample types. However, the non-answer of \var{RecAccWaitingCount} handpicked and random accounts can also be considered as a form of account loss. Similarly, we found top-ranked accounts to more commonly offer contact forms rather than email addresses, which might be due to them being more professionally structured due to their popularity and large user bases.
While the recovery results were similar for all samples, we found handpicked accounts to diverge from the other two samples in some cases, \eg{}, by less commonly offering SMS or software tokens and overall having less variety in their offered \mfa{} methods, as well as offering no contact forms, but usually, email addresses to reach out to support staff. 

While we excluded websites that required the provision of, \eg{}, governmental IDs at start-up to protect the sensitive documents of the researchers conducting this study, we found that \var{RecIDReq} websites later required them for recovery purposes, leading to account loss on our side. This includes websites that mainly deal with, \eg{}, cryptocurrencies, payments, or IT-focused websites that offered, \eg{}, domain registrations, and website hosting. However, we also encountered similar websites within the sample that allowed us access without requiring a governmental ID, \ie{}, we regained access to potentially valuable accounts with typically only access to the respective email inbox.

\begin{summaryBox}{Key Points | Recovery User Experience}{}
We created \var{RecAccTotal} accounts, lost access to our \mfa{}, and \textbf{recovered} access to \textbf{half} of them. \textbf{Three quarters} of the websites prompted us to set up \textbf{any kind of recovery} during \mfa{} setup, typically \textbf{backup codes}. Reasons for success included \textbf{email access} and \textbf{contextual knowledge}; reasons for account loss were \textbf{strict policies}, \textbf{lack of ID}, and \textbf{lack of contact options}. 
User experience and documentations \textbf{rarely matched}, with documentations usually providing \textbf{additional information}. 
\end{summaryBox}
\section{Ethics and Limitations}
\label{sec_ethlimits}
Below, we discuss ethical concerns and the limitations of our work. 

\subsection{Ethics}
\label{sec_ethics}

This work was approved by our institution's ethical review board.
In the first part of our evaluation, we manually examined non-personal, publicly available data. We further took care to not cause unusual resource burdens, and refrain from naming any website in the paper.
In the second part of this work we contacted human support staff who were not initially aware that our support requests were part of an academic research project.
However, we carefully designed our study according to the Menlo Report~\cite{2012-dittrich-mraf} and its recommendations for a deception study. Overall, we deem the harm caused by our study to be justified by the benefits to society offered by the evaluation of current \mfa{} setups and recovery procedures and the resulting recommendations for developers and website operators. During our work, we caused an additional workload with our dishonest support requests. To minimize the burden on individuals, we took care to exclude websites working solely with volunteers. Furthermore, we limited ourselves to creating only one account per service and sent only one support request. 
We kept all communication as concise as possible and did not send out reminders. 
We also refrained from collecting sensitive data, and did not evaluate the support staff members themselves.

For each website, we studied the terms of service and refrained from registering with \var{RecSkipTOS} services for which we would have violated them.
As we were unable to ask for prior consent due to the nature of the study, we followed best practices and sent a post hoc notification and debriefing email to all services we evaluated\repref{} \extendedversion{(see Appendix~\ref{sec_app_texts})}. We informed them about the fully anonymous use of their data and allowed them to drop out of the study, in which case we deleted all collected data for the respective website, and did not use it in our work. A total of \var{RecOptOut} websites declined participation and were subsequently removed from our data set. Overall, we deem the deception in this work to be necessary for our goal of evaluating unbiased \mfa{} recovery procedures, as otherwise support staff members might have adapted their behavior. All details of our methodology, including all texts we used to communicate with human support workers, were part of our approved ERB application.

\subsection{Limitations}
\label{limits}
In observational studies like ours, there are multiple potential sources of bias or error.
First, regarding our data source, we obtained a list of websites offering \mfa{} using the 2fa.directory database. As we illustrate (c.f. Section~\ref{sec_meth_deployment}) this database offers a public repository, where volunteers can contribute data according to certain quality criteria~\cite{2fadcontribution, 2faexclusion, 2faissue}. The quality criteria nicely align with our research goals, since they require high popularity ranks. We, therefore, decided to use 2fa.directory instead of manually going through top websites.
During our evaluation, we noticed that the list is slightly biased towards including more technical websites (\eg{}, \var{percentTechCategories}\% of all \var{PagesMFA} websites belong to technical or related categories). However, these websites are likely more tech-affine and therefore may be more likely to offer \mfa{}. This suggests that our measurement constitutes an upper bound to \mfa{} recovery procedure quantity.
To categorize the websites, we restricted ourselves to publicly available information, and might therefore have missed information only accessible to authorized users. However, this reflects the perspective of users who lost access, search information on \mfa{} recovery procedures, and can also only rely on publicly available data.

The user experience study had additional limitations. We could not create accounts on \var{RecSkipTotal} websites, including websites that required us to provide a government-issued ID, sign a (paid) contract for, \eg{}, a bank or investment service and utility providers for, \eg{}, gas or electricity, be part of a certain group such as enrolled students, or own certain devices such as IoT devices. This decision was made to protect the privacy of the Ph.D. students conducting this research. Additionally, our sample of service providers is globally distributed, and some services abroad did not allow us to create accounts located in Germany. We further focused on free account plans and did not pay for any service. While premium account tiers could have led to different results (\eg{}, because we would have received premium support), we think that our results still provide valuable insights for the community, and reflect the experience of users who are unable to pay for such services. Furthermore, we excluded \var{SkippedLang} websites that were not primarily in German or English.
While we aimed to sample high-ranked, handpicked, and random websites to increase diversity, and resampled whenever we needed to skip a website (cf. Section~\ref{sec_meth_deploymentmeth}), the number of handpicked websites was limited and could therefore not be arbitrarily extended, resulting in uneven group sizes. 
Our study is further limited by our choice to not utilize the recovery codes we received on some websites and to retain our email inbox. However, our goal was to treat all websites equally, and we argue that losing our recovery procedure along with the main \mfa{} sufficiently reflects reality, especially since the configuration is rarely enforced.

Finally, during our detailed analysis of our experience and official help and documentation pages, we might have missed information due to not following every path during the account and \mfa{} setup. We especially did not follow links to the documentation given during \mfa{} setup, but only the texts and prompts were shown during the process.
\section{Discussion}
\label{sec_discussion}

Below, we discuss our results and address our research questions.

\subsection{RQ1: Diverse Landscape of Recovery Procedures}
\label{sec_discussion_rq1}
We identified \var{RecoveryMethodsCodesTotal} different recovery procedures based on the public help and support pages of \var{PagesMFA} websites (cf. Section~\ref{sec_meth_deploymentresults}). 
Most providers offered support telephone hotlines, email addresses, or local IT staff, followed by backup codes. All procedures offer limited usability and security: Users may reach out to support channels and IT staff over insecure channels, such as unencrypted email~\cite{li2018email, conf/oakland/stransky22}. While backup codes are easy to distribute, they have similar issues as passwords: users can lose them or store them insecurely. 

The high diversity of \mfa{} recovery procedures we found is in line with work by \citeauthor{lyastani2023systematic} who investigated \mfa{} deployments without evaluating recovery. 
Implementations between websites and service types differed vastly, and we could not identify best practices regarding \mfa{} recovery procedures. Due to the lack of standards or best practices, most website providers deployed custom procedures, often leading to contradictions between websites and making the lives of their users unnecessarily complicated. 
For example, we both encountered websites suggesting setting up multiple backup methods, while others encouraged their users to only configure one recovery procedure to reduce their attack surface. 
While we agree that too many backup options decrease security, it is sensible to not rely on a single method and to clearly communicate the offered methods and the significance of a working recovery.
Similarly, some websites warned us about using SMS-based services due to various possible attacks on mobile networks, while others praised SMS for its availability and usability. 
In general, SMS should be avoided due to several vulnerabilities~\cite{siadati2017mind, WiredTwitterCEOHacked2019, lee2020simswap, jover2020security} -- however, due to its advantages, it can be a valid method for low-risk services if there is an alternative for users that are uncomfortable with sharing their private phone numbers.
After reaching out to their support, some providers recommended we create new accounts instead of recovering the existing ones. They made this recommendation after manually checking our accounts and seeing that they were not frequently used. 

Another provider re-allowed us access without identification \textit{because} the account was empty and not much used. 
In this specific case, we deem both methods valid for empty accounts but argue that account deletion is the safer option to give users full control over the stored data in case recovery procedures are abused.
In all of these cases, the deviating arguments have a well-meaning, true core, and in many cases, compromises based on the nature of the service are necessary.
We discuss shortcomings and recommendations in Section~\ref{sec_discussion_rq3}.

\subsection{RQ2: Inconsistencies Compromise Security \& Usability}
\label{sec_discussion_rq2}

Overall, we found examples of both websites that prioritized security, and did not allow us access without the proper \mfa{} or recovery procedure we configured, and websites that helped us to regain access without having access to our second factor, lowering authentication security (cf. Section~\ref{sec_meth_recoverytestsresults}). However, from a user's point of view, this means either sacrificing security or usability, which can be especially critical for accounts that manage important resources such as monetary funds or medical data.

We identified inconsistencies between the authentication security associated with the use of \mfa{}, and the recovery procedures implemented on many websites. 
Using \mfa{} is expected to strengthen the security of authentication by limiting account access to users who know the account password and are in possession of the configured additional authentication factor. 
However, having access to the email address we used for account creation was sufficient for most of the accounts we regained access to. 
This effectively reduces the security of \mfa{} to that of email security, which has been criticized in prior work about account recovery~\cite{li2018email, al2018email}, and would allow an attacker with email access to circumvent \mfa{}. 
Whenever we were required to provide contextual evidence such as our address or the nature of connected platforms, the security decreased to the level of security questions, which are also known to be insecure and guessable~\cite{rabkin2008personal, bonneau2015secrets}. 

Although we never said anything about having lost backup codes as well, services typically did not request them to recover our \mfa{}. 
This might be based on experiences with prior users that typically lose both \mfa{} and recovery, or that many users did not configure a recovery procedure. However, it also means that services refrained from utilizing their own suggested recovery. %
Websites that distribute backup codes should adhere to these security decisions and not accept other, even less secure alternative identifications. 

We ascribed many of the rejected recovery requests to usability issues. This includes not being able to contact the support, \eg{}, because contact is only available to logged-in or paying users. This can be especially frustrating when users are left without any other recovery option. 

In other cases, providers requested any form of ID or data without previously informing us that they would be part of the \mfa{} recovery procedure. 
While governmental IDs are a valid way to verify identities in some situations, they are not suitable to do so for accounts that often know little more about their users than full names, birthdates, or email addresses. 
Furthermore, even video-based identification processes have been successfully circumvented, allowing attackers to falsify documents and successfully impersonate their victims~\cite{CCChacksvideoident}. 

We find that websites rarely prepare users to provide ID identification, as they are usually not prompted to provide it during account setup, or informed that it will become relevant during recovery. However, the availability of these documents can otherwise be an issue, and lead to higher obstacles during recovery later on.
For example, not all countries, including the USA and Switzerland, require their citizens to own governmental IDs. Further, the causes of \mfa{} loss might also affect the availability of recovery: a lost or stolen purse might mean that not only the TOTP app on a user's phone, but also their wallet and ID might be gone, or an emergency, such as a house fire, could destroy both mobile phones and identifying documents. Finally, users who are not aware of the relevance of identifying information for \mfa{} recovery might provide false information to protect their privacy.

Some websites had exceptional expectations. 
For example, some help and support pages suggested setting up multiple hardware keys in case one was lost or broken. 
While having a backup configured is, in general, sensible, hardware keys are expensive, and having to set up multiple keys can impose a financial burden excluding users from setting up \mfa{} or its recovery. However, this is also encouraged by official FIDO guidelines regarding security keys~\cite{fidouxguide}. 

Other websites downplayed the process of regaining access to their users' phone numbers, by suggesting users get a replacement SIM card with which they can access their original \mfa{} again. However, this replacement can require significant time to arrive, especially if the user needs replacement during, \eg{}, a vacation.

We find the documentation to largely match our experience, but none of them matches perfectly. 
While we expect help and support pages to be more detailed, we argue that some details, such as warnings about account loss, or the intended recovery procedure should also be communicated during \mfa{} setup.  
However, they are often not found at all in either, therefore not sufficiently preparing users for a potential loss of \mfa{}.
We further find that websites often do not match their documentation regarding the most basic elements, such as the offered \mfa{} and setup methods, which suggests that help and support pages might be outdated and not reflect the current recovery procedures. 
This could lead to unpleasant surprises as users are not properly prepared for \mfa{} recovery. 

\subsection{RQ3: Improving \mfa{} Recovery Procedures}
\label{sec_discussion_rq3}

The three most pressing problems we identified related to unreachable support teams, insufficient \mfa{} recovery procedure documentation, and information that was needed for recovery but not configured during setup.
Based on our findings, we make the following recommendations: 

\boldparagraph{Prepare} Our findings illustrate that recovery procedures require a sufficient setup that should be part of the \mfa{} setup. 
We suggest that all information service providers requests during recovery are collected at account setup. Users need to be warned about everything that is strictly necessary, \ie{}, they should be told about required documents and how to avoid recovery failures due to security reasons. 
However, asking for too much or too sensitive information can be risky in regard to, \eg{}, information abuse and data breaches~\cite{karunakaran2018data, mayer2021now} and might deter users. 
Future research should work on identifying the right balance between the interests of both user privacy and \mfa{} recovery. 
While we agree that limiting the number of \mfa{} recovery procedures is useful to not open up multiple attack vectors, having fallbacks available is in line with general security recommendations depending on the respective use case and threat model~\cite{attacksurfacemetric, dingledine2004tor}.
As some procedures, such as SMS or security questions, are known to be insecure or impractical, we suggest offering backup codes, additional unique communication channels, \eg{}, bank statements where possible to identify the user as account owners, or additional (desktop) devices with TOTP.

\boldparagraph{Communicate} The process of setting up, disabling, or recovering \mfa{} and what to expect in the worst case should be clearly and directly communicated during the setup process and be part of publicly available help and support pages. 
We argue that having too many help pages can be confusing and lead to both users not finding the information they need~\cite{redmiles2020comprehensive}, and website maintainers struggling to keep them all up-to-date. 
We, therefore, suggest keeping the number of different help and support pages as low as possible and using structural elements like HTML \texttt{tabs} to help users more easily find the required information for their platform or \mfa{} method.

\boldparagraph{Maintain} Recovery procedures should be tested regularly. Websites could prompt users to, \eg{}, enter one of their backup codes or a TOTP to make sure that the recovery procedure is still available and working. 
Previous work has shown that pop-ups and security warnings can be perceived as obtrusive or might be ignored by users~\cite{Akhawe:2013wx, Sunshine:2009, weinberger:2016, bailey2021have}. Hence, future work is required to find a good trade-off between verifying that recovery procedures are still available, and obstructing user workflows too much.

\boldparagraph{Recover} For \mfa{} recovery, we urge websites to adhere to their documented recovery procedures. 
When websites deviate from their documented recovery procedures, \eg{}, by restoring access despite warning about definite account loss, or by allowing additional recovery measures, they might create opportunities for attackers that users are not aware of, or create (false) expectations for other websites. 
Additionally, honest and clear communication about the consequences of \mfa{} loss can help users make informed and empowered decisions about their authentication security.
We advise against using sensitive documents, such as governmental IDs or utility bills, as identity proof, as they often cannot reliably identify authorized users. However, if this kind of identification is sensible and necessary, websites should take precautions to collect it early on and make sure that users are informed of this part of their recovery procedure. 

\subsection{Putting our Work into Context}
Previous work often focused on investigating the theoretical usability of recovery procedures in different areas~\cite{reynolds2020empirical, lyastani2023systematic, rabkin2008personal, hang2015locked, neil2021investigating, kunke2021evaluation, gilsenan2023security}. 
In contrast, our work studies the first-hand user experience of \mfa{} recovery procedures on websites: We went through the actual \mfa{} recovery procedure processes implemented by \var{RecAccTotal} websites and reported lessons learned. 
We could regain access to almost half of the websites we tested based on having access to our email inbox. 
Hence, our study illustrates that \mfa{} recovery procedures' security on websites is affected by similar security risks as TOTP apps or smartphone security implementing insecure methods as fallback~\cite{reynolds2020empirical, gilsenan2023security, kunke2021evaluation, schechter2009s}.

Previous work evaluated security documentation for setting up \mfa{}~\cite{reynolds2018tale, lyastani2023systematic}, and encountered issues around unclear instructions. 
Our work shows many websites did not sufficiently document their \mfa{} recovery processes, \eg{} instructions described recovery processes that did not match the implemented deployments. Some websites did not provide documentation at all. 
\section{Conclusion}
\label{sec_conclusion}

\mfa{} recovery procedures need to balance security and usability by allowing authorized users to access accounts without locking them out due to \mfa{} loss. We are the first to analyze deployed \mfa{} recovery procedures and related documentations on the web.
In this work, we first categorized the recovery procedures deployed on \var{PagesMFA} websites, then conducted an in-depth analysis in which we created \var{RecAccTotal} accounts, configured \mfa{} and recovery procedures, and finally tried to recover them from supposed \mfa{} loss.
We found that most websites only offered a limited selection of recovery procedures and that these were not necessarily part of the \mfa{} setup process. 
In the recovery process, websites often did not adhere to their documentation and allowed us access without requiring the configured \mfa{} or recovery. 
In most cases, having access to the account's associated email address was sufficient for account recovery.
Overall, we recommend websites clearly and correctly document their \mfa{} recovery procedures. Measures to detect loss of \mfa{} or recovery procedures before users are locked out should be in place.

\begin{acks}
We thank the included websites for allowing us to use their data and our insights into their processes for this work. Furthermore, we thank our reviewers for their valuable feedback. Finally, we thank Philip Klostermeyer and Juliane Schmüser for proofreading our work. 
Funded by the and by the VolkswagenStiftung Niedersächsisches Vorab – ZN3695 and by the Deutsche Forschungsgemeinschaft (DFG, German Research Foundation) under Germany´s Excellence Strategy - EXC 2092 CASA – 390781972.
\end{acks}



\printbibliography

\appendix
\extendedversion{

\section{\mfa{} Methods and Known Recovery Procedures}
\label{sec_back}

Below, we provide background information for \mfa{} approaches and recovery procedures we encountered. We also discuss their impact on authentication security and usability.

\subsection{Email}
\label{sec_back_email}
\mfa{} based on \textbf{email}, commonly uses pre-generated one-time passcodes (OTP) or a verification link sent to the user via email. 
Consequently, users retain account access as long as they can access their email account. However, email security is lacking when compared to other mechanisms, as emails are usually sent in plaintext~\cite{conf/oakland/stransky22}, and email account security often depends on password-based authentication~\cite{wang2018next, hanamsagar2016users}. Password resets also often occur via email~\cite{google1, netflix, twitter, wikipedia, facebook1}, effectively downgrading security to the email accounts' password. It is further possible to use email OTPs as a recovery procedure for \mfa{}\footnote{\label{addmeth} In our evaluation, we only considered this as recovery if it was explicitly a secondary backup, and otherwise considered it as an \textit{additional \mfa{} method}.}.

\subsection{SMS \& Phone Call OTP}
\label{sec_back_phone}
A common alternative to email are OTPs via \textbf{SMS}. Apart from that, users can receive automated \textbf{phone calls} to provide OTPs. Most people own a phone~\cite{mobilefactsheet}, making SMS or phone calls popular additional authentication factors. The recovery of this method is potentially expensive and time-consuming because it requires users to obtain a new device and regain access to their phone number. While popular, telephone-based \mfa{} has proven to be insecure due to its susceptibility to various severe attacks~\cite{siadati2017mind, WiredTwitterCEOHacked2019, lee2020simswap, jover2020security}, that enable third parties to intercept traffic or change the user's phone number to the attacker's device via social engineering. Similar to email, a secondary phone number can be used to offer SMS or phone calls as \mfa{} recovery\textsuperscript{\ref{addmeth}}.

\subsection{Mobile Applications}
\label{sec_back_apps}
\textbf{Mobile applications}, such as time-based one-time password (TOTP) apps like Google Authenticator~\cite{googleauth}, are a common \mfa{} method. They generate one-time passwords~\cite{RFC6238} (OTPs) that users can provide to prove their identity\textsuperscript{\ref{addmeth}}. The app is set up by scanning a confidential QR code or entering a secret key provided by the respective website, which serves as a seed for the TOTP algorithm. As long as both server and client have the same current time and share the same secret, they can independently compute the same TOTP values. 
Besides this, some apps like Authy~\cite{authy}, or password manager apps such as LastPass~\cite{lastpass} or 1Password~\cite{1password} require users to register an account and offer a cloud backup of the supported \mfa{} as a trade-off. While this eases recovery, it also means that users are not fully in control over their data, as service providers also have access to the backups. Related to this, LastPass leaked user data during two hacks in 2022~\cite{lastpassHack}. Lastly, some websites provide custom apps for their respective services, and include \mfa{} in the forms of, \eg{}, TOTP, usage of biometric sensors embedded in smartphones, or push notifications. The provided security and usability in these cases is limited by the app vendor and its implementation~\cite{mutchler15:mobilewebapps}.

\subsection{Hardware Tokens}
\label{sec_back_hw}
\textbf{Hardware tokens} (\eg{}, smart-cards), code generators (\eg{}, Transaction Authentication Number (TAN) generators), or Universal Second Factor (U2F)~\cite{specfido1} hardware keys, are established \mfa{} methods. WebAuthn supports the use of trusted platform modules, including biometric identity checks~\cite{rfcfido2}.
The adoption of U2F and WebAuthn has been limited~\cite{das2018johnny}, often due to their complex initial setups and device requirements. However, recent improvements provide better usability~\cite{reese2019usability}. The protocol offers verified security guarantees~\cite{basin2010degrees} and the main assumption is the secrecy of the chosen hardware token, resulting in the credential proof being \textit{something you have}.

\subsection{User-Dependent Methods}
\label{sec_back_user}
Below, we illustrate approaches that are dependent on the user, their input, or inherent features.
First, \textbf{biometry-based} \mfa{} commonly relies on apps that use the internal biometry security options of, \eg{}, mobile phones to authenticate. The main advantage compared to mobile applications is that authentication uses biometric features, requiring no additional device or memorization. 
While technically secure, there have been reports of malfunctioning biometrics~\cite{SiblingUnlocksFaceID,FaceUnlockAsleep, WiredChildUnlocksiPhone2017, ForbesFaceIDHacked2019, ForbesTouchIDHacked2013}, and they further require very sensitive personal information for authentication.
Another approach are \textbf{secret questions} that supposedly only the authorized user can answer, and \textbf{backup codes} that users are given and supposed to store securely or print, or the storage of the \textbf{TOTP secret}, \ie{}, the secret key or QR code shown during \mfa{} setup. 
While these methods are commonly deployed as recovery or alternative to \mfa{}, the security is often lacking: security questions can easily be guessed, especially since they tend to require information that is often known to acquaintances or social media contacts~\cite{schechter2009nosecret,rabkin2008personal,bonneau2015secrets}, and user-set passwords are often weak or re-used~\cite{bonneau2012science,ur2015added,wang2019reuse,wash2016understanding, pearman2017reuse}. Methods such as backup codes or stored secrets are easily lost or allow attackers to bypass \mfa{} completely if they gain access. 

\subsection{Website-Dependent Methods}
\label{sec_back_website}
We also distinguish methods that are dependent on the websites. 
In case access to \mfa{} is lost, users can often \textbf{contact the website support}. In other cases, especially for team-accounts or self-hosted tools, the account is managed by \textbf{local administrators} that can be contacted. Both cases can include fixed waiting times or identity verification for account recovery. In some cases, this process is more straightforward, as users are asked to provide a \textbf{photo proof} of them holding, \eg{}, handwritten notes or governmental IDs. Other websites utilize automated \textbf{dedicated recovery systems}, or disable \mfa{} automatically when the \textbf{password is reset}. 
The security of these methods depends on the respective communication channel, such as email or HTTPS. While easy to use, the usability can suffer from, \eg{}, long response times or queries for additional information, as methods involving any form of communication are often dynamic and resolved on a case-by-case basis.

\subsection{Additional Methods}
\label{sec_back_other}
In some cases, \textbf{\mfa{} is not needed for login}, but required for certain service features such as withdrawing funds. 
Some websites allow to omit \mfa{} for \textbf{trusted devices}. While used as recovery, those methods are often bound to a time-limit of, \eg{}, 30--90 days after which \mfa{} is required again, leading to a decreased backup usability if \mfa{} is lost outside this time frame.

\textbf{Other recovery} advice might combine some of the previously mentioned factors, or give additional advice to, \eg{}, contact mobile providers to replace inaccessible SIM cards.
Finally, the recovery can simply consist of setting up \textbf{additional \mfa{} methods}, \ie{}, extra methods that can all act as a backup for whichever method was lost by the user. While this has been an official FIDO recommendation~\cite{fidouxguide}, it also increases the attack vectors for malicious third parties.

While neither a \mfa{} nor recovery procedure, we encountered additional edge cases in which we were unable to determine the methods. This includes \textbf{inaccessible help} pages we were unable to visit due to, \eg{}, geo-blocking or deleted pages, and websites either mentioning explicitly that they had \textbf{no recovery} methods, or websites that did \textbf{not mention} recovery at all.

\section{General Appendix}}

\label{sec_appendix}

\extendedversion{

\subsection{Recovery Test Templates}
\label{sec_app_texts}

\boldparagraph{Recovery Request Message}
\label{app:rec_request}
\noindent Subject: Lost Second Factor \\
Hello,\\
I recently registered an account on your website and enabled 2FA. Yesterday, my backpack with my wallet and phone was stolen, so now I cannot access anything, and am unable to get past the 2FA. 
What do I need to do to regain access? My \textit{username/email/account number} is \textit{data}.\\
Best regards,\\
\textit{First Author}
\boldparagraph{Debriefing Message}
\label{app:debriefing}
Dear \textit{website} support,\\
We are a research team at the \textit{Anonymized For Review Institution, Country,} and  study the usability of multi factor authentication (MFA). We are currently investigating the trade-off between usability and security of multi factor authentication recovery. As a part of this research, we created accounts on several websites, enabled MFA, and tried to recover our accounts after a while due to supposedly inaccessible MFA. These websites were chosen because they were listed on 2fa.directory {[1]}, ranked highly on Tranco {[2]}, or because their documentation implied interesting MFA methods.\\
As you might already guess, we write to you today because your website was among the ones we chose. We wanted to inform you about your involuntary participation, as well as give you the time to decline and have your data removed.
On \textit{date}, we send you an \textit{email/chat request} \textit{(request ID if possible)}, which was part of our research, i.e., it was not a real users’ request to recover their MFA. While we are not content about deceiving you and sending a false request, we wanted to assure you that it was necessary for the goal of our research. We wanted to truly gauge the experience of a typical user, and to assess the usability and security of MFA procedures on the web. Due to this goal, we were sadly unable to communicate the true purpose of our request upfront. Please be assured that we only sent one line of request, and that we kept it as short as possible to decrease the load on your staff caused by our inquiry.\\
As mentioned, we would like to give you the opportunity to have your data removed and drop out of our data set before we de-identify it and use it as part of a scientific article. If you let us know until October 10th\footnote{This initially encompassed two weeks, but if any further opt-outs would have reached us until submission, we would have removed them as well.}, we will irrevocably delete all communication with or data about your website. Please be assured that we currently only store the data on encrypted, self-hosted servers, and that only involved researchers have access to our results. Furthermore, we will never publish the name of your website or any involved staff member. We will de-identify your website in our work, and we will never mention your service’s or website name. 
We hope that our study did not cause too much harm, and that you do not have objections against us using the data. In any case, we thank you for being able to create an account and collect valuable experiences, for the helpful support you gave us, and want to apologize for any inconvenience caused by our request.\\
If you are interested in more details of our research, please visit our project website: \textit{https://anonymizedforreview.tld/projects/multifactor-recovery/}. \\
In case you are interested in the results of our research, we would be happy to share the paper once it is published. Finally, we are also interested in deepening our understanding of MFA and its recovery by also including your perspective in future research (e.g. an interview). If you would be available for a follow-up study, we would be delighted to hear from you!\\
Best regards,\\
\textit{First Author} (PhD Candidate)\\
\textit{Anonymized For Review Institution}\\
{[1]} \url{https://2fa.directory/int/}\\
{[2]} \url{https://tranco-list.eu/}}

\extendedversion{
\subsection{Additional Tables}
\label{sec_app_tabs}

\begin{table}[htb] 
    \caption{Distribution of the initial \var{PagesMFA} websites over various TLDs.}
    \label{tab_tld}
	\setlength{\tabcolsep}{7pt}
	\footnotesize
    \centering
\begin{tabularx}{0.7\columnwidth}{lrr}
\toprule
\textbf{Top-Level Domain}  & \textbf{Counts} & \textbf{Relative}\\ 
\midrule

\hspace{5pt}  ca  & \var{TLDCount_ca} & \var{TLDPercent_ca}\% \\
\hspace{5pt}  ch  & \var{TLDCount_ch} & \var{TLDPercent_ch}\%  \\
\hspace{5pt}  co.uk  & \var{TLDCount_co.uk} & \var{TLDPercent_co.uk}\% \\
\hspace{5pt}  com & \var{TLDCount_com} & \var{TLDPercent_com}\% \\
\hspace{5pt}  com.au & \var{TLDCount_com.au} & \var{TLDPercent_com.au}\%  \\
\hspace{5pt}  de  & \var{TLDCount_de} & \var{TLDPercent_de}\% \\
\hspace{5pt}  edu  & \var{TLDCount_edu} & \var{TLDPercent_edu}\% \\
\hspace{5pt}  gov & \var{TLDCount_gov} & \var{TLDPercent_gov}\%  \\
\hspace{5pt}  io  & \var{TLDCount_io} & \var{TLDPercent_io}\% \\
\hspace{5pt}  net  & \var{TLDCount_net} & \var{TLDPercent_net}\% \\
\hspace{5pt}  org  & \var{TLDCount_org} & \var{TLDPercent_org}\%  \\
\hspace{5pt}  other  & \var{TLDCount_other} & \var{TLDPercent_other}\%  \\
\bottomrule
\end{tabularx}
\end{table}

\begin{table}[htb] 
    \caption{Distribution of the initial \var{PagesMFA} websites over the listed bins of Tranco ranks.}
    \label{tab_tranco}
	\setlength{\tabcolsep}{7pt}
	\footnotesize
    \centering
\begin{tabular}{lrr}
\toprule
\textbf{Tranco Ranking}  & \textbf{Counts} & \textbf{Relative}\\ 
\midrule

1 - 20000 & \var{TrancoCount_1} & \var{TrancoPercent_1}\% \\
20001 - 40000 & \var{TrancoCount_20001} & \var{TrancoPercent_20001}\%  \\
40001 - 60000 & \var{TrancoCount_40001} & \var{TrancoPercent_40001}\% \\
60001 - 80000 & \var{TrancoCount_60001} & \var{TrancoPercent_60001}\% \\
80001 - 100000 & \var{TrancoCount_80001} & \var{TrancoPercent_80001}\%  \\
100001 - 120000 & \var{TrancoCount_100001} & \var{TrancoPercent_100001}\% \\
120001 - 140000 & \var{TrancoCount_120001} & \var{TrancoPercent_120001}\% \\
140001 - 160000 & \var{TrancoCount_140001} & \var{TrancoPercent_140001}\%  \\
160001 - 180000 & \var{TrancoCount_160001} & \var{TrancoPercent_160001}\% \\
180001 - 200000 & \var{TrancoCount_180001} & \var{TrancoPercent_180001}\% \\
$\geq$ 200001 & \var{TrancoCount_200001} & \var{TrancoPercent_200001}\%  \\

\bottomrule
\end{tabular}
\end{table}

We were further interested in the top-level domains (TLDs) of websites in our sample. Due to our removal of \var{SkippedLang} websites in languages we were not fluent in, we found the vast majority of remaining websites (\var{TLDCount_com}, \var{TLDPercent_com}\%) to be .com websites, with the remaining being other English (\eg{}, .ca, .com.au, .co.uk) or German (.de, .ch) TLDs. Additionally, we find some neutral ones (.io, .net, .org), and .edu and .gov, which we almost exclusively found in the respective categories of Education, Universities and Government (see Table~\ref{tab_tld}). We find no significant differences between recovery methods offered on different TLDs.

\onecolumn

\newcolumntype{L}{>{\centering\arraybackslash}l{1cm}}

\begin{table*}[h!] 

    \caption{Codes used in our research. \textit{\mfa{} Methods} and \textit{Recovery Methods} coded within our procedure categorization described in Section~\ref{sec_meth_deploymentmeth}}
    \label{tab_codebook_deployment}
	\setlength{\tabcolsep}{2pt}
    \centering
\footnotesize

\begin{tabularx}{\textwidth}{lrl}

\toprule
\textbf{Code} & \textbf{Frequency} & \textbf{Description}\\ 

\midrule
\textbf{\mfa{} Methods} & &\\

SMS  & \var{AllCountMethSMS} (\var{AllPercentMethSMS}\%) & SMS (usually with OTP) sent to users' phone number.\\

Phone Call & \var{AllCountMethCall} (\var{AllPercentMethCall}\%) & Call used to convey information (\eg{}, spoken code, code hidden in caller ID). \\

Email & \var{AllCountMethMail} (\var{AllPercentMethMail}\%) & Email (usually with OTP) sent to users' associated email address. \\

Hardware Token  & \var{AllCountMethHWToken} (\var{AllPercentMethHWToken}\%) & Hardware device, \eg{}, U2F security key, or physical OTP generator. \\

Mobile App  & \var{AllCountMethSWToken} (\var{AllPercentMethSWToken}\%) & Mobile app to verify user. Usually TOTP apps, but also proprietary apps using, \eg{}, push notifications. \\

Other \mfa{}  & \var{AllCountMethOther} (\var{AllPercentMethOther}\%) & Other methods such as biometry, local files, printed codes via mail and others. \\

No \mfa{} & \var{PagesNoMFA} (\var{PagesPercentNoMFA}\%) & No \mfa{} present, \ie{}, the website never offered \mfa{}, removed it, or never mentions it on public help pages. \\

\midrule
\textbf{Recovery Methods} & &\\ 

Contact Website Support & \var{AllCountRecContactSupport} (\var{AllPercentRecContactSupport}\%) & Contact the service over varying mediums. Can include fixed wait or ID checks. \\

Backup Codes  & \var{AllCountRecCodes} (\var{AllPercentRecCodes}\%)  & Code provided by service. It can be both an OTP or secret knowledge, distributed number varies.\\

Contact Local Admin  & \var{AllCountRecContactAdmin} (\var{AllPercentRecContactAdmin}\%) &  Contact local admin or designated superuser to, \eg{}, reset \mfa{}. \\

Additional \mfa{} Method  & \var{AllCountRecAdd2FA} (\var{AllPercentRecAdd2FA}\%)  & Explicit mention of using multiple \mfa{} methods to have a backup. \\

Backup SMS/Phone Call  & \var{AllCountRecPhone} (\var{AllPercentRecPhone}\%)  & Send SMS or call user. Only when SMS/Call is not a \mfa{} method, \\ & & or if it is explicitly a separate backup number. \\

Backup Email  & \var{AllCountRecMail} (\var{AllPercentRecMail}\%)  & Send email. Only when email is not a method, or if it is explicitly a separate backup address. \\

Dedicated Account Recovery System  & \var{AllCountRecRecoveryService} (\var{AllPercentRecRecoveryService}\%) & Service offers a designated form to initiate recovery for lost \mfa{}. If this was just a standard contact form, \\ & & we chose \textit{Contact Service/Website} instead. \\

TOTP Seed & \var{AllCountRecSecret} (\var{AllPercentRecSecret}\%)  & Store the initial \mfa{} secret to be able to set it up again. Usually TOTP seeds. \\

Trusted Device  & \var{AllCountRecTrustedDevice} (\var{AllPercentRecTrustedDevice}\%) & \mfa{} not required after first authentication. Excludes cases in which this is limited to a certain time frame, \\ &  
& as it is only a backup during this time. \\

Photo/Official ID Proof  & \var{AllCountRecPhoto} (\var{AllPercentRecPhoto}\%) &  Requires upload of a selfie of user holding, \eg{}, ID cards and written notes as identity proof. \\

MFA Not Needed for Login  & \var{AllCountRecAccessNotLost} (\var{AllPercentRecAccessNotLost}\%)  &  \mfa{} not required for login, only for certain functions. Users can simply log in and update \mfa{}.\\

Security Question & \var{AllCountRecSecQuestion} (\var{AllPercentRecSecQuestion}\%)  & Answer a (set of) security questions to regain access. \\

Password Reset &  \var{AllCountRecPWReset} (\var{AllPercentRecPWReset}\%)  & Users can use the password reset function to deactivate \mfa{} as well, therefore avoiding real recovery.\\

Other Recovery  & \var{AllCountRecOther} (\var{AllPercentRecOther}\%) & Other methods or combinations of methods, \eg{}, Contact the phone provider for new SIM, letters with  \\ & & recovery codes, or using verification links via email and SMS and a photo proof. \\

Help Page Not Accessible  & \var{AllCountRecInaccesible} (\var{AllPercentRecInaccesible}\%) & Details cannot be verified due to unavailable help pages (\eg{}, geo-blocking, deleted pages) \\

No \mfa{} Recovery Available & \var{AllCountRecNone} (\var{AllPercentRecNone}\%)  & The help and documentation explicitly stresses that there is no recovery procedure. \\

No \mfa{} Recovery Documented  & \var{AllCountRecNoMention} (\var{AllPercentRecNoMention}\%) & \mfa{} is available, but no recovery is mentioned. \\

\bottomrule
\end{tabularx}
\end{table*}

\begin{table*}[hb!] 

    \caption{Codes used in our research. \textit{Detailed Information} describes the codes used within our \studytwo{} (cf. Section~\ref{sec_meth_recoverytestsmeth}) to account for additional interesting details not covered in the previous categorization.}

    \label{tab_codebook_detail}
	\setlength{\tabcolsep}{2pt}
    \centering
\footnotesize

\begin{tabularx}{\textwidth}{lrrl}

\toprule
\textbf{Code} & \multirow{2}{*}{\shortstack[l]{\textbf{Frequency} \\ \textbf{Experiment}}} & \multirow{2}{*}{\shortstack[l]{\textbf{Frequency} \\ \textbf{Documentation}}} &\textbf{Description}\\  & & & \\

\midrule
Recovery is Part Of MFA Setup  & \var{RecSetupExpCount} (\var{RecSetupExpPercent}\%) & \var{RecSetupDocCount} (\var{RecSetupDocPercent}\%) & Whether recovery is discussed during/alongside \mfa{} setup (yes, no)\\

Explicit Warning About Account Loss & \var{RecWarningsExpCount} (\var{RecWarningsExpPercent}\%) & \var{RecWarningsDocCount} (\var{RecWarningsDocPercent}\%) &Whether website warns users about losing access if \mfa{} is lost \\
 & & & and no(t enough) recovery methods were set. Only includes if the warning was explicit \\
 & & & and stresses that account loss is final. (emphasized, yes, no) \\

Vague Warning About Account Loss & \var{RecSoftWarningExpCount} (\var{RecSoftWarningExpPercent}\%) & \var{RecSoftWarningDocCount} (\var{RecSoftWarningDocPercent}\%) & Similar to \textit{Warning}, but also includes soft warnings in which alternative recoveries \\
 & & & are mentioned, or the account loss is only portrait as a vague possibility. (yes, no)\\
 
 Warning about SMS Usage & \var{RecSMSWarningExpCount} (\var{RecSMSWarningExpPercent}\%) & \var{RecSMSWarningDocCount} (\var{RecSMSWarningDocPercent}\%)&Whether website warns about using SMS OTPs due to their insecurity (yes, no).\\

Account Usage Mandates \mfa{} & \var{RecRestrictedExpCount} (\var{RecRestrictedExpPercent}\%) & \var{RecRestrictedDocCount} (\var{RecRestrictedDocPercent}\%) &Whether users are hindered to freely enable and disable \mfa{} (yes, no). \\

\mfa{} Enforces Recovery Setup & \var{RecForcedExpCount} (\var{RecForcedExpPercent}\%) & \var{RecForcedDocCount} (\var{RecForcedDocPercent}\%) &Whether users are forced to set up a backup (yes, no). \\

Mention of \mfa{} Trigger & \var{RecTriggerExpCount} (\var{RecTriggerExpPercent}\%) & \var{RecTriggerDocCount} (\var{RecTriggerDocPercent}\%) & List of triggers that cause the users \mfa{} to be requested (free text)\\

Number of Help Pages & - & \var{RecHelpNumDocCount} (\var{RecHelpNumDocPercent}\%) &Whether a service has more than three relevant help pages regarding \mfa{} \\
 & & & and its recovery (yes, no). \\

\bottomrule
\end{tabularx}
\end{table*}

\begin{table*}[htp] 
    \caption{List of websites examined during our recovery experience described in Section~\ref{sec_meth_recoverytests}, including their approximated Tranco rank, category or sample group. We chose to omit details on present \mfa{} and recovery methods, as well as the recovery result, within the public version of this work to ensure the anonymity of included websites.}

    \label{tab_examinedwebsites}
	\setlength{\tabcolsep}{5pt}
    \centering
\scriptsize

\newcommand{\req}[0]{{\cellcolor[HTML]{af2757}\color{white}\textasteriskcentered}}
\newcommand{\spec}[0]{{\cellcolor[HTML]{eeeeee}\textasteriskcentered}}
\newcommand{\nan}[0]{}
\newcommand{\reqbox}[0]{{\colorbox[HTML]{af2757}{\color{white}\textasteriskcentered}}}
\newcommand{\specbox}[0]{{\colorbox[HTML]{eeeeee}{\textasteriskcentered}}}

\begin{tabular}{llll|ccc|l}

\multirow{2}{*}{\shortstack{\specbox{} Specific purpose \reqbox{} Always required\\ \may{} offered \yes{} used}} 
& \multicolumn{3}{l|}{}  
& \multicolumn{3}{c|}{\textbf{Data}} 
& \multicolumn{1}{l}{}\\

\textbf{ID} & \textbf{Tranco} & \textbf{Category} & \textbf{Sample} & 
\textbf{{\rotatebox{90}{Financial}}}  & \textbf{{\rotatebox{90}{Address}}}  & \textbf{{\rotatebox{90}{Other PII}}}  & \textbf{Contact}\\

\midrule
1 & 10,001-50,000 & Cryptocurrencies & Handpicked & \req & \req & \nan & Email\\
2 & 10,001-50,000 & Hosting \& VPS & Random & \req & \req & \nan & Email\\
3 & 50,001-200,000 & Developers & Random & \req & \nan & \req & Email\\
4 & 50,001-200,000 & Communication & Random & \req & \req & \nan & Email\\
5 & 50,001-200,000 & Email & Handpicked & \spec & \nan & \nan & Email\\
6 & <1,000 & Security & Handpicked & \req & \req & \req & Email\\
7 & 10,001-50,000 & Domains & Handpicked & \spec & \spec & \nan & Email\\
8 & 50,001-200,000 & Identity Management & Handpicked & \req & \spec & \nan & Email\\
9 & <1,000 & Social & Top-Ranked & \spec & \nan & \nan & Email\\
10 & <1,000 & Gaming & Top-Ranked & \spec & \nan & \nan & Contact Form\\
11 & 5,001-10,000 & Communication & Random & \spec & \spec & \nan & Email\\
12 & 1,001-5,000 & Marketing \& Analytics & Handpicked & \spec & \spec & \nan & Email\\
13 & <1,000 & Social & Top-Ranked & \nan & \nan & \nan & Contact Form\\
14 & 5,001-10,000 & Developers & Random & \spec & \spec & \nan & Reset Form\\
15 & 1,001-5,000 & Other & Random & \spec & \nan & \nan & Contact Form\\
16 & 1,001-5,000 & Crowdfunding & Handpicked & \req & \spec & \nan & Email\\
17 & 1,001-5,000 & Security & Handpicked & \req & \req & \nan & Email\\
18 & <1,000 & Domains & Handpicked & \spec & \spec & \nan & Reset Form\\
19 & <1,000 & Other & Top-Ranked & \spec & \spec & \nan & Email\\
20 & <1,000 & Entertainment & Top-Ranked & \req & \nan & \nan & Email\\
21 & <1,000 & Hotels \& Accommodations & Top-Ranked & \req & \req & \req & Email\\
22 & <1,000 & Other & Top-Ranked & \req & \req & \nan & Email\\
23 & <1,000 & Education & Top-Ranked & \spec & \nan & \nan & Email\\
24 & <1,000 & Crowdfunding & Random & \req & \nan & \req & Email\\
25 & <1,000 & Domains & Random & \req & \req & \req & Email\\
26 & 1,001-5,000 & Other & Random & \req & \req & \nan & Email\\
27 & 5,001-10,000 & VPN Providers & Random & \req & \nan & \nan & Reset Form\\
28 & >200,000 & Cryptocurrencies & Random & \req & \nan & \nan & Email\\
29 & <1,000 & Communication & Top-Ranked & \spec & \spec & \spec & Contact Form\\
30 & <1,000 & Retail & Top-Ranked & \req & \req & \spec & Contact Form\\
31 & <1,000 & Other & Top-Ranked & \spec & \spec & \nan & Email\\
32 & <1,000 & Crowdfunding & Top-Ranked & \req & \spec & \nan & Contact Form\\
33 & 5,001-10,000 & Domains & Random & \req & \req & \req & Live Chat\\
34 & 1,001-5,000 & Cryptocurrencies & Random & \req & \nan & \nan & Contact Form\\
35 & 5,001-10,000 & Gaming & Random & \req & \req & \spec & Email\\
36 & >200,000 & Cloud Computing & Random & \req & \nan & \nan & Email\\
37 & 10,001-50,000 & Identity Management & Random & \req & \req & \nan & Live Chat\\
38 & 10,001-50,000 & Social & Random & \spec & \nan & \nan & Live Chat\\
39 & 10,001-50,000 & Cryptocurrencies & Handpicked & \req & \spec & \nan & Reset Form\\
40 & <1,000 & Social & Top-Ranked & \nan & \nan & \nan & Reset Form\\
41 & 50,001-200,000 & Betting & Random & \req & \nan & \nan & Live Chat\\
42 & <1,000 & Marketing \& Analytics & Top-Ranked & \spec & \spec & \nan & Reset Form\\
43 & <1,000 & Developers & Handpicked & \nan & \nan & \nan & Email\\
44 & 5,001-10,000 & Hosting \& VPS & Handpicked & \spec & \req & \req & Email\\
45 & 1,001-5,000 & Domains & Handpicked & \req & \req & \req & Email\\
46 & <1,000 & Social & Top-Ranked & \spec & \nan & \nan & Contact Form\\
47 & <1,000 & Social & Top-Ranked & \spec & \spec & \nan & Not Possible\\
48 & <1,000 & Backup \& Sync & Top-Ranked & \spec & \spec & \nan & Contact Form\\
49 & <1,000 & Developers & Top-Ranked & \nan & \nan & \nan & Contact Form\\
50 & <1,000 & Communication & Top-Ranked & \spec & \nan & \spec & Not Possible\\
51 & <1,000 & Domains & Top-Ranked & \spec & \spec & \spec & Email\\
52 & <1,000 & Communication & Top-Ranked & \spec & \spec & \nan & Email\\
53 & <1,000 & Remote Access & Handpicked & \spec & \spec & \nan & Not Possible\\
54 & 10,001-50,000 & Finance & Handpicked & \req & \req & \req & Not Possible\\
55 & <1,000 & Other & Top-Ranked & \spec & \spec & \nan & Not Possible\\
56 & <1,000 & Retail & Top-Ranked & \req & \req & \spec & Live Chat\\
57 & 50,001-200,000 & Food & Random & \req & \req & \req & Contact Form\\
58 & 10,001-50,000 & Communication & Random & \spec & \nan & \nan & Not Possible\\
59 & <1,000 & Other & Handpicked & \req & \req & \nan & Email\\
60 & <1,000 & Retail & Top-Ranked & \req & \req & \spec & Contact Form\\
61 & <1,000 & Domains & Top-Ranked & \req & \req & \req & Contact Form\\
62 & <1,000 & Investing & Top-Ranked & \req & \req & \nan & Reset Form\\
63 & <1,000 & Retail & Top-Ranked & \req & \req & \spec & Reset Form\\
64 & <1,000 & Security & Top-Ranked & \req & \req & \nan & Reset Form\\
65 & <1,000 & Hosting \& VPS & Random & \req & \req & \nan & Reset Form\\
66 & >200,000 & Cryptocurrencies & Random & \req & \spec & \spec & Email\\
67 & 10,001-50,000 & Domains & Random & \req & \req & \nan & Reset Form\\
68 & 10,001-50,000 & Cryptocurrencies & Random & \req & \req & \req & Contact Form\\
69 & 50,001-200,000 & Payment & Random & \req & \req & \req & Email\\
70 & 5,001-10,000 & Cryptocurrencies & Random & \req & \req & \req & Reset Form\\
71 & 10,001-50,000 & Domains & Random & \req & \req & \req & Email\\
\bottomrule
\end{tabular}

\end{table*}

\onecolumn{}
\footnotesize{}
\begin{longtable}{rlllll}
\caption{List of websites skipped during our recovery experience described in Section~\ref{sec_meth_recoverytests}, including their approximated Tranco rank, category, sample group and reason for skipping it.}\\

\toprule 
\textbf{ID} & \textbf{Tranco Range} & \textbf{Category} & \textbf{Sample} & \textbf{Skip Stage} & \textbf{Reason}\\ 
\midrule
\endfirsthead

\caption{List of websites skipped during our recovery experience continued.}\\

\toprule 
\textbf{ID} & \textbf{Tranco Range} & \textbf{Category} & \textbf{Sample} & \textbf{Skip Stage} & \textbf{Reason}\\ 
\midrule
\endhead

\bottomrule
\endfoot
\label{tab_skippedwebsites}
1 & 1,001-5,000 & Developers & Random & Registration & Blocked (eSIM usage) \\
2 & 1,001-5,000 & Investing & Random & Registration & Blocked (Location) \\
3 & 1,001-5,000 & Banking & Random & Registration & Blocked (Location) \\
4 & 50,001-200,000 & Betting & Random & Registration & Blocked (Location) \\
5 & 1,001-5,000 & Security & Random & Registration & Blocked (unknown reason) \\
6 & 10,001-50,000 & Gaming & Random & Registration & Blocked (unknown reason) \\
7 & 50,001-200,000 & Developers & Random & Registration & Blocked (unknown reason) \\
8 & <1,000 & Social & Handpicked & Registration & Blocked (unknown reason) \\
9 & <1,000 & Social & Top-Ranked & Registration & Blocked (unknown reason) \\
10 & >200,000 & Cryptocurrencies & Random & Registration & Blocked (unknown reason) \\
11 & 10,001-50,000 & Developers & Handpicked & Website Choice & Domain already within sample. \\
12 & <1,000 & Utilities & Top-Ranked & Website Choice & Domain already within sample. \\
13 & <1,000 & Entertainment & Top-Ranked & Website Choice & Domain already within sample. \\
14 & <1,000 & Backup \& Sync & Top-Ranked & Website Choice & Domain already within sample. \\
15 & <1,000 & Communication & Top-Ranked & Website Choice & Domain already within sample. \\
16 & <1,000 & Email & Top-Ranked & Website Choice & Domain already within sample. \\
17 & <1,000 & Payment & Random & Website Choice & Domain already within sample. \\
18 & <1,000 & Payment & Random & Website Choice & Domain already within sample. \\
19 & <1,000 & Backup \& Sync & Top-Ranked & Website Choice & Domain already within sample. \\
20 & <1,000 & Backup \& Sync & Random & Website Choice & Domain already within sample. \\
21 & <1,000 & Task Management & Random & Website Choice & Domain already within sample. \\
22 & <1,000 & IoT & Random & Website Choice & Domain already within sample. \\
23 & 50,001-200,000 & Gaming & Random & Website Choice & Maintained by volunteers \\
24 & 1,001-5,000 & Retail & Random & Website Choice & Method prohibited by ToS. \\
25 & 1,001-5,000 & Gaming & Random & Website Choice & Method prohibited by ToS. \\
26 & 1,001-5,000 & Payment & Random & Website Choice & Method prohibited by ToS. \\
27 & 10,001-50,000 & Cryptocurrencies & Random & Website Choice & Method prohibited by ToS. \\
28 & 10,001-50,000 & Other & Random & Website Choice & Method prohibited by ToS. \\
29 & <1,000 & Hosting  \& VPS & Top-Ranked & Website Choice & Method prohibited by ToS. \\
30 & <1,000 & IoT & Handpicked & Website Choice & Method prohibited by ToS. \\
31 & <1,000 & Social & Top-Ranked & Website Choice & Method prohibited by ToS. \\
32 & <1,000 & Social & Top-Ranked & Website Choice & Method prohibited by ToS. \\
33 & <1,000 & Developers & Top-Ranked & Website Choice & Method prohibited by ToS. \\
34 & <1,000 & Retail & Top-Ranked & Website Choice & Method prohibited by ToS. \\
35 & <1,000 & Email & Top-Ranked & Website Choice & Method prohibited by ToS. \\
36 & <1,000 & Entertainment & Top-Ranked & Website Choice & Method prohibited by ToS. \\*
37 & <1,000 & Gaming & Top-Ranked & Website Choice & Method prohibited by ToS. \\
38 & <1,000 & Gaming & Top-Ranked & Website Choice & Method prohibited by ToS. \\
39 & <1,000 & Domains & Top-Ranked & Website Choice & Method prohibited by ToS. \\
40 & <1,000 & Developers & Top-Ranked & Website Choice & Method prohibited by ToS. \\
41 & >200,000 & Cloud Computing & Random & Website Choice & Method prohibited by ToS. \\
42 & >200,000 & Gaming & Random & Website Choice & Method prohibited by ToS. \\
43 & >200,000 & Cryptocurrencies & Random & Website Choice & Method prohibited by ToS. \\
44 & 10,001-50,000 & Developers & Handpicked & MFA Setup & No MFA on base website. \\
45 & 10,001-50,000 & Gaming & Handpicked & MFA Setup & No MFA on base website. \\
46 & 10,001-50,000 & Marketing \& Analytics & Random & Registration & No MFA on base website. \\
47 & 10,001-50,000 & Payment & Random & Registration & No MFA on base website. \\
48 & 10,001-50,000 & Developers & Random & Registration & No MFA on base website. \\
49 & <1,000 & Communication & Top-Ranked & Registration & No MFA on base website. \\
50 & <1,000 & Social & Top-Ranked & Registration & No MFA on base website. \\
51 & <1,000 & Creativity & Top-Ranked & MFA Setup & No MFA on base website. \\
52 & <1,000 & Other & Top-Ranked & MFA Setup & No MFA on base website. \\
53 & <1,000 & Retail & Top-Ranked & MFA Setup & No MFA on base website. \\
54 & <1,000 & Cloud Computing & Top-Ranked & MFA Setup & No MFA on base website. \\
55 & <1,000 & Communication & Top-Ranked & MFA Setup & No MFA on base website. \\
56 & 10,001-50,000 & Investing & Random & Registration & Requirements (contract) \\
57 & 5,001-10,000 & Utilities & Random & Registration & Requirements (contract) \\
58 & 50,001-200,000 & Marketing \& Analytics & Random & Registration & Requirements (contract) \\
59 & 50,001-200,000 & Hosting \& VPS & Random & Registration & Requirements (contract) \\
60 & 50,001-200,000 & Banking & Random & Registration & Requirements (contract) \\
61 & 50,001-200,000 & Communication & Random & Registration & Requirements (contract) \\
62 & 50,001-200,000 & Email & Random & Registration & Requirements (contract) \\
63 & 50,001-200,000 & Developers & Random & Registration & Requirements (contract) \\
64 & 50,001-200,000 & Banking & Random & Registration & Requirements (contract) \\
65 & <1,000 & Cloud Computing & Random & Registration & Requirements (contract) \\
66 & <1,000 & Other & Random & Registration & Requirements (contract) \\
67 & >200,000 & Utilities & Random & Registration & Requirements (contract) \\
68 & 10,001-50,000 & Remote Access & Random & Registration & Requirements (device ownership) \\
69 & 5,001-10,000 & IoT & Handpicked & Registration & Requirements (device ownership) \\
70 & <1,000 & Remote Access & Top-Ranked & Registration & Requirements (device ownership) \\
71 & <1,000 & Social & Random & Registration & Requirements (device ownership) \\
72 & <1,000 & Communication & Top-Ranked & Registration & Requirements (device ownership) \\
73 & 10,001-50,000 & Finance & Random & Registration & Requirements (Foreign Address) \\
74 & 50,001-200,000 & IoT & Random & Registration & Requirements (Foreign Address) \\
75 & <1,000 & Payment & Top-Ranked & Registration & Requirements (Foreign Phone) \\
76 & 1,001-5,000 & Banking & Random & Registration & Requirements (ID) \\
77 & 10,001-50,000 & Banking & Random & Registration & Requirements (ID) \\
78 & 10,001-50,000 & Banking & Random & Registration & Requirements (ID) \\
79 & 10,001-50,000 & Finance & Random & Registration & Requirements (ID) \\
80 & 10,001-50,000 & Investing & Random & Registration & Requirements (ID) \\
81 & 10,001-50,000 & Banking & Random & Registration & Requirements (ID) \\
82 & 5,001-10,000 & Banking & Random & Registration & Requirements (ID) \\
83 & 50,001-200,000 & Investing & Handpicked & Registration & Requirements (ID) \\
84 & 50,001-200,000 & Banking & Random & Registration & Requirements (ID) \\
85 & 50,001-200,000 & Investing & Random & Registration & Requirements (ID) \\
86 & >200,000 & Cryptocurrencies & Random & Registration & Requirements (ID) \\
87 & 50,001-200,000 & Banking & Random & Registration & Requirements (ID/payment) \\
88 & <1,000 & Banking & Top-Ranked & Registration & Requirements (ID/payment) \\
89 & 1,001-5,000 & Universities & Random & Registration & Requirements (membership) \\
90 & <1,000 & Universities & Random & Registration & Requirements (membership) \\
91 & <1,000 & Universities & Top-Ranked & Registration & Requirements (membership) \\
92 & <1,000 & Universities & Top-Ranked & Registration & Requirements (membership) \\
93 & <1,000 & Universities & Top-Ranked & Registration & Requirements (membership) \\
94 & 10,001-50,000 & Finance & Handpicked & Registration & Requirements (payment) \\
95 & 50,001-200,000 & VPN Providers & Random & Registration & Requirements (payment) \\
96 & <1,000 & Cloud Computing & Handpicked & Registration & Requirements (payment) \\
97 & <1,000 & Cloud Computing & Top-Ranked & Registration & Requirements (payment) \\
98 & <1,000 & Cloud Computing & Top-Ranked & Registration & Requirements (payment) \\
99 & <1,000 & Hosting  \& VPS & Top-Ranked & Registration & Requirements (payment) \\
100 & >200,000 & Investing & Random & Registration & Requirements (payment) \\
101 & 1,001-5,000 & Hosting \& VPS & Handpicked & Registration & Requirements (payments) \\
102 & 1,001-5,000 & Other & Random & Website Choice & Time-limited free trial. \\
103 & 10,001-50,000 & Task Management & Random & Website Choice & Time-limited free trial. \\
104 & 10,001-50,000 & Finance & Handpicked & Website Choice & Time-limited free trial. \\
105 & 10,001-50,000 & Other & Handpicked & Website Choice & Time-limited free trial. \\
106 & 10,001-50,000 & Other & Random & Website Choice & Time-limited free trial. \\
107 & 10,001-50,000 & Security & Random & Website Choice & Time-limited free trial. \\
108 & 10,001-50,000 & Developers & Random & Website Choice & Time-limited free trial. \\
109 & 5,001-10,000 & Health & Random & Website Choice & Time-limited free trial. \\
110 & 50,001-200,000 & Hosting  \& VPS & Random & Website Choice & Time-limited free trial. \\
111 & 50,001-200,000 & Other & Random & Website Choice & Time-limited free trial. \\
112 & 50,001-200,000 & Cloud Computing & Random & Website Choice & Time-limited free trial. \\
113 & 50,001-200,000 & Security & Random & Website Choice & Time-limited free trial. \\
114 & <1,000 & Finance & Top-Ranked & Website Choice & Time-limited free trial. \\
115 & <1,000 & Communication & Top-Ranked & Website Choice & Time-limited free trial. \\
116 & <1,000 & Domains & Top-Ranked & Website Choice & Time-limited free trial. \\
117 & <1,000 & Other & Top-Ranked & Website Choice & Time-limited free trial. \\
118 & <1,000 & Other & Top-Ranked & Website Choice & Time-limited free trial. \\
119 & >200,000 & Hosting  \& VPS & Random & Website Choice & Time-limited free trial. \\
120 & >200,000 & Backup  \& Sync & Random & Website Choice & Time-limited free trial. \\
121 & >200,000 & Other & Random & Website Choice & Time-limited free trial. \\
122 & 50,001-200,000 & Cloud Computing & Random & Website Choice & Website not available. \\
123 & 50,001-200,000 & Backup  \& Sync & Random & Website Choice & Website not available. \\
124 & 50,001-200,000 & Health & Random & Website Choice & Website not available. \\
\end{longtable}
\twocolumn{}}

\clearpage

\section{Additional Figures}
\label{sec_app_figs}
 \begin{figure*}[htp!]
     \centering
     \includegraphics[width=1\textwidth]{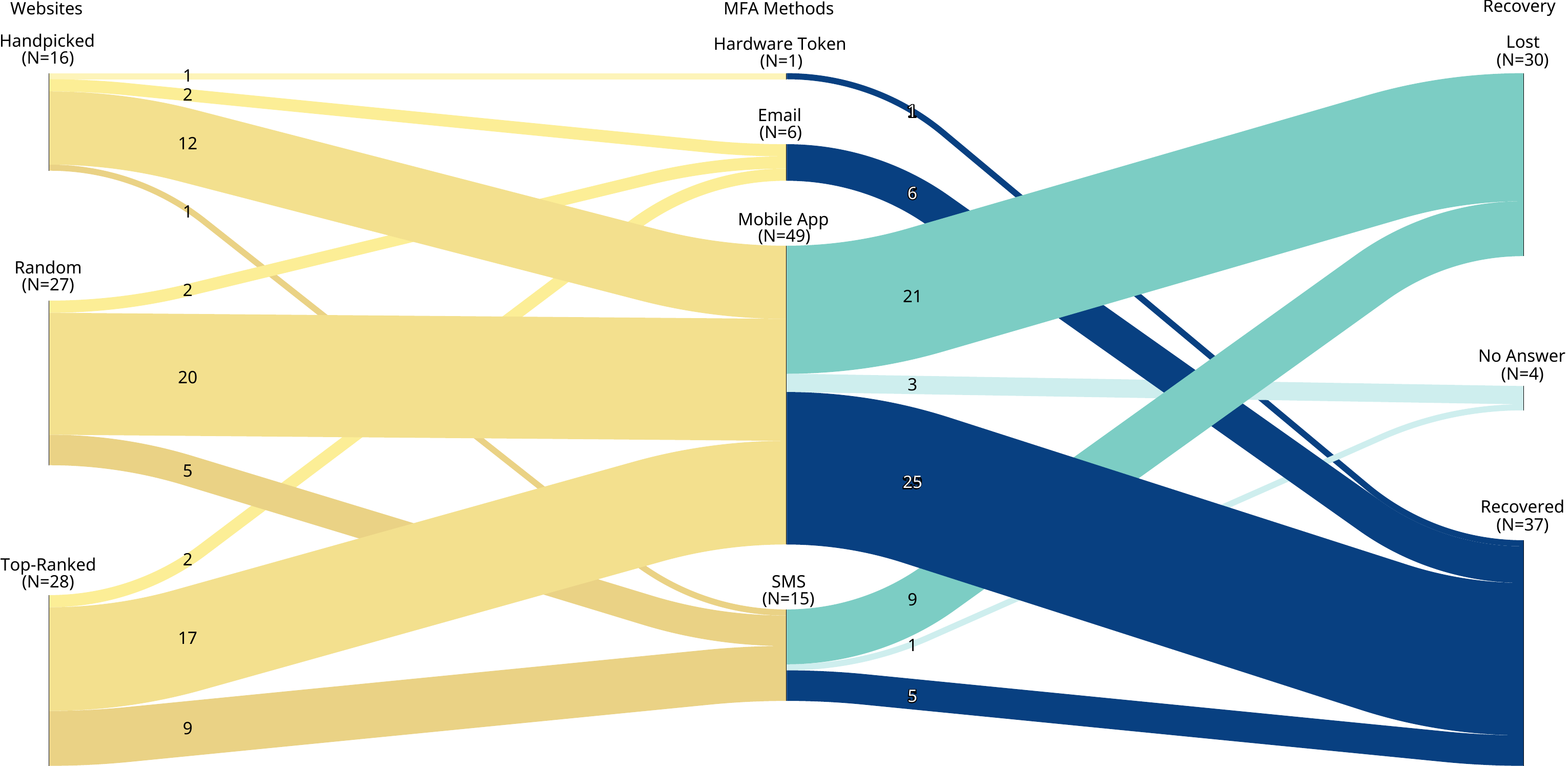}
    \caption{Illustration of our website sample, configured \mfa{} method and recovery result.}
    \label{fig_sankeyMFARecovery}
 \end{figure*}

\extendedversion{

\begin{figure*}[ht!]
	\centering
    \includegraphics[width=0.9\textwidth]{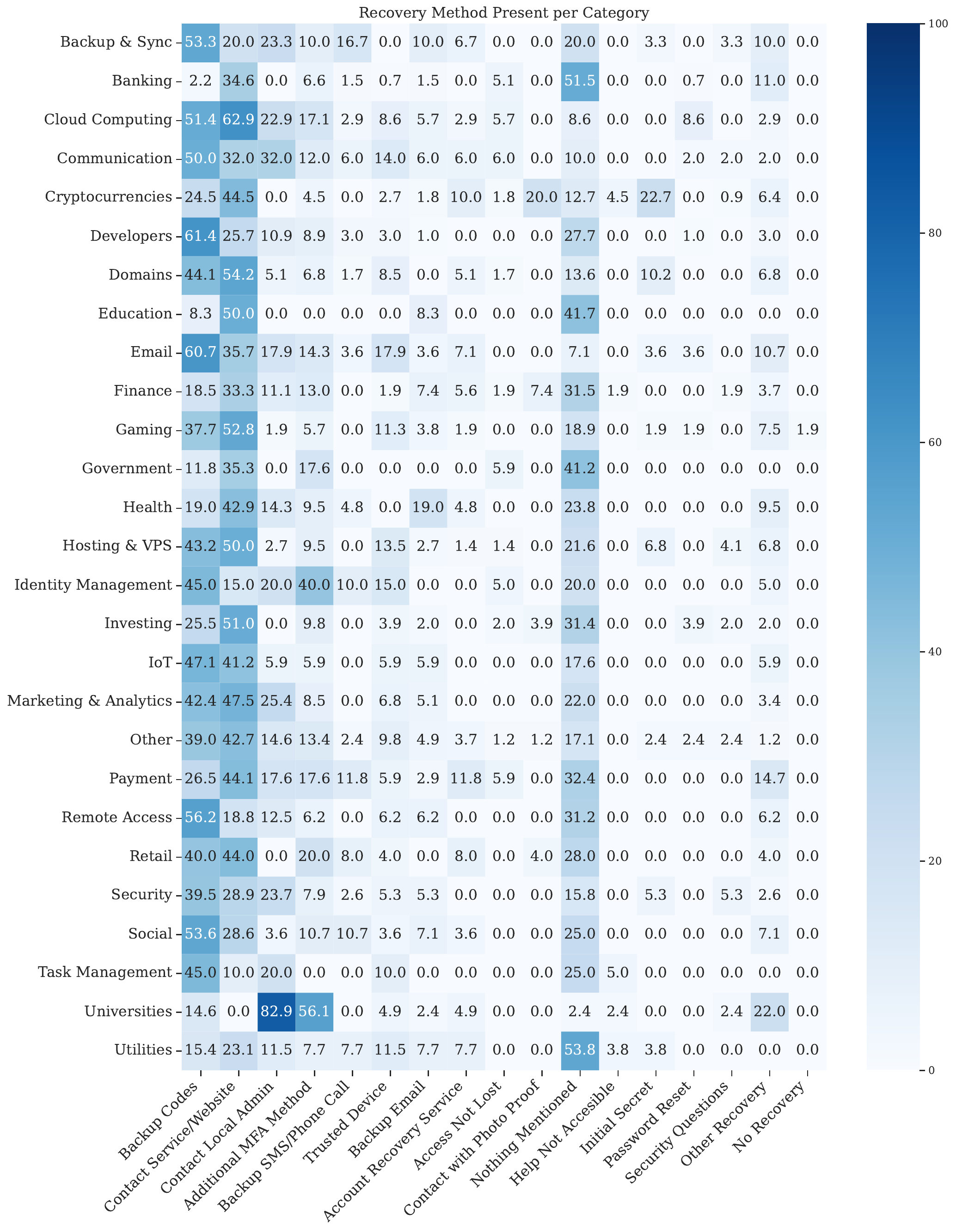}
     \caption{Presence of \mfa{} recovery methods within the initial \var{PagesMFA} websites over different categories on \url{2fa.directory} in \%. As websites can offer more than one recovery option, the numbers may not add up to 100\%.}
	\label{fig_recPresenceCategories}
\end{figure*}

}

\end{document}